\documentstyle[iopconf1,epsf]{article}
\newcommand{\be}[1]{\begin{equation} \label{(#1)}}
\newcommand{\ee}{\end{equation}}
\newcommand{\ba}[1]{\begin{eqnarray} \label{(#1)}}
\newcommand{\ea}{\end{eqnarray}}
\newcommand{\nn}{\nonumber}
\newcommand{\rf}[1]{(\ref{(#1)})}
\def \znbb {$0\nu\beta\beta$}
\def \tnbb {$2\nu\beta\beta$}
\def \Rpv{R_{P} \hspace{-0.9em}/\;\:}
\def\rp{$R_p \hspace{-1em}/\;\:$}
\def \emass {\langle m_{\nu} \rangle}

\begin{document}

\title{Double Beta Decay and Dark Matter Search - Window to New Physics now, 
and in future (GENIUS)}

\author{
H.V. Klapdor--Kleingrothaus}

\affil{Max--Planck--Institut f\"ur Kernphysik\\
P.O.Box 10 39 80, D--69029 Heidelberg, Germany \\
}

\beginabstract
Nuclear double beta decay provides an extraordinarily broad potential
to search for beyond Standard Model physics, probing already now the
TeV scale, on which new physics should manifest itself. These possibilities
are reviewed here.

First, the results of present generation experiments are presented.
The most sensitive one of them -- the Heidelberg--Moscow experiment in the 
Gran Sasso -- probes the electron mass now in the sub eV region and will reach 
a limit of $\sim$ 0.1 eV in a few years. Basing to a large extent on the theoretical
work of the Heidelberg Double Beta Group in the last two years, results
are obtained also for SUSY models (R--parity breaking, sneutrino mass),
leptoquarks (leptoquark-Higgs coupling), compositeness,
right--handed W boson mass and others. These results are comfortably 
competitive to corresponding results from high--energy accelerators like TEVATRON, HERA, etc.

Second, future perspectives of $\beta\beta$ research are discussed.  A new
Heidelberg experimental proposal (GENIUS) is presented which would allow to 
increase the sensitivity for Majorana neutrino masses from the present level
of at best 0.1 eV down to 0.01 or even 0.001 eV. Its physical potential would 
be a breakthrough into the multi-TeV range for many beyond standard models.
Its sensitivity for neutrino oscillation parameters would be larger than of 
all present terrestrial neutrino oscillation experiments and of those planned 
for the future. It could probe directly the atmospheric neutrino problem
and even the large angle solution of the solar neutrino problem. It would 
further, already in a first step, cover almost the full MSSM parameter 
space for 
prediction of neutralinos as cold dark matter, making the experiment 
competitive to LHC in the search for supersymmetry. 
\endabstract

\section{Introduction -- Motivation for the search for double beta decay
-- and a future perspective: GENIUS}
Double beta decay yields -- besides proton decay -- the most promising
possibilities to probe beyond standard model physics beyond accelerator 
energy scales. Propagator physics has to replace direct 
observations. That this method is very effective, is obvious from important
earlier research work and has been stressed, e.g. by \cite{4}, etc.. Examples
are the properties of $W$ and $Z$ bosons derived from neutral weak currents
and $\beta$--decay, and the top mass deduced from LEP electroweak radiative 
corrections. 

The potential of double beta decay includes information on the neutrino and
sneutrino mass, SUSY models, 
compositeness, leptoquarks, right--handed $W$ bosons and others 
(see Table 1). 
The recent results of the Heidelberg--Moscow experiment, 
which will be reported here,
have demonstrated 
that $0\nu\beta\beta$ decay probes already now the TeV
scale on which new physics should manifest itself according to present 
theoretical expectations. 
To give just one
example, inverse double beta decay $e^-e^-\rightarrow W^-W^-$ requires an 
energy of at least 4 TeV for observability, according to present constraints
from double beta decay \cite{14}. Similar energies are required to study,
e.g.
leptoquarks \cite{3,5,hir96a,126,127,128,129}.

To increase by a major step the 
present sensitivity for double beta decay and dark matter search, we present
here for the first time a new project which would operate one ton
of `naked' enriched {\bf GE}rmanium detectors in liquid 
{\bf NI}trogen as shielding in an {\bf U}nderground {\bf S}etup (GENIUS).
It would improve the sensitivity from the present potential of at best 
$\sim 0.1$ eV to neutrino masses down to 0.01 eV, a ten ton version even to
0.001 eV. The first version would allow to test 
a $\nu_e \rightarrow \nu_{\mu}$ explanation of
the atmospheric neutrino
problem, the second directly the large angle solution of the solar neutrino 
problem. The sensitivity for neutrino oscillation parameters would be larger
than for all present accelerator neutrino oscillation experiments, or those 
planned for the future. 
GENIUS would further allow to test the recent hypothesis of
a sterile neutrino and the underlying idea of a shadow world (see section 2).
Both versions of GENIUS would definitely be a breakthrough into the multi-TeV
range for many beyond standard models currently discussed in the literature,
and the sensitivity would be comparable or even superior to LHC for various
quantities such as right--handed W--bosons, R--parity violation, leptoquark
or compositeness searches.

Another issue of GENIUS is the search for Dark Matter in the universe.
The full
MSSM parameter space for predictions of neutralinos as cold 
dark matter could be 
covered already in a first step of the full experiment using only 100 kg
of $^{76}$Ge or even natural Ge, making the experiment competitive to LHC
in the search for supersymmetry.

\section{Double beta decay and particle physics}
We present a brief introductory outline of the potential of $\beta\beta$ decay
for some representative examples,
including some comments on the status of the required nuclear matrix
elements. The potential of double beta decay for probing neutrino oscillation
parameters will be addressed in section 4.2.

Double beta decay can occur in several decay modes
(Figs. 1--3) 
\be{1}
^{A}_{Z}X \rightarrow ^A_{Z+2}X + 2 e^- + 2 {\overline \nu_e}
\ee
\be{2}        
^{A}_{Z}X \rightarrow ^A_{Z+2}X + 2 e^- 
\ee
\be{3}
^{A}_{Z}X \rightarrow ^A_{Z+2}X + 2 e^- + \phi
\ee
\be{4}
^{A}_{Z}X \rightarrow ^A_{Z+2}X + 2 e^- + 2\phi
\ee
the last three of them violating lepton number conservation by $\Delta L=2$.
Fig. 3 shows the corresponding spectra, for the neutrinoless mode
(2) a sharp line at $E=Q_{\beta\beta}$, for the two--neutrino mode
and the various Majoron--accompanied modes classified by their spectral index,
continuous spectra.
Important for particle physics are the decay modes (2)--(4).

\begin{figure}
\setlength{\unitlength}{1in}                                                 
\begin{picture}(5,2)
\put(0.0,0.5){\includegraphics{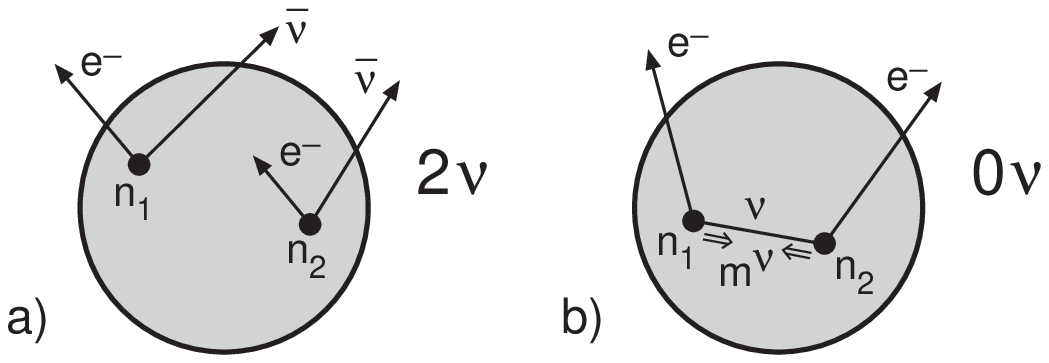}}
\end{picture}                                                                 
{\bf Fig. 1} {\it Schematic representation of $2\nu$ and $0\nu$ double beta 
decay.}
\end{figure}
\begin{figure}
\hspace*{10mm}
\epsfxsize=60mm
\epsfbox{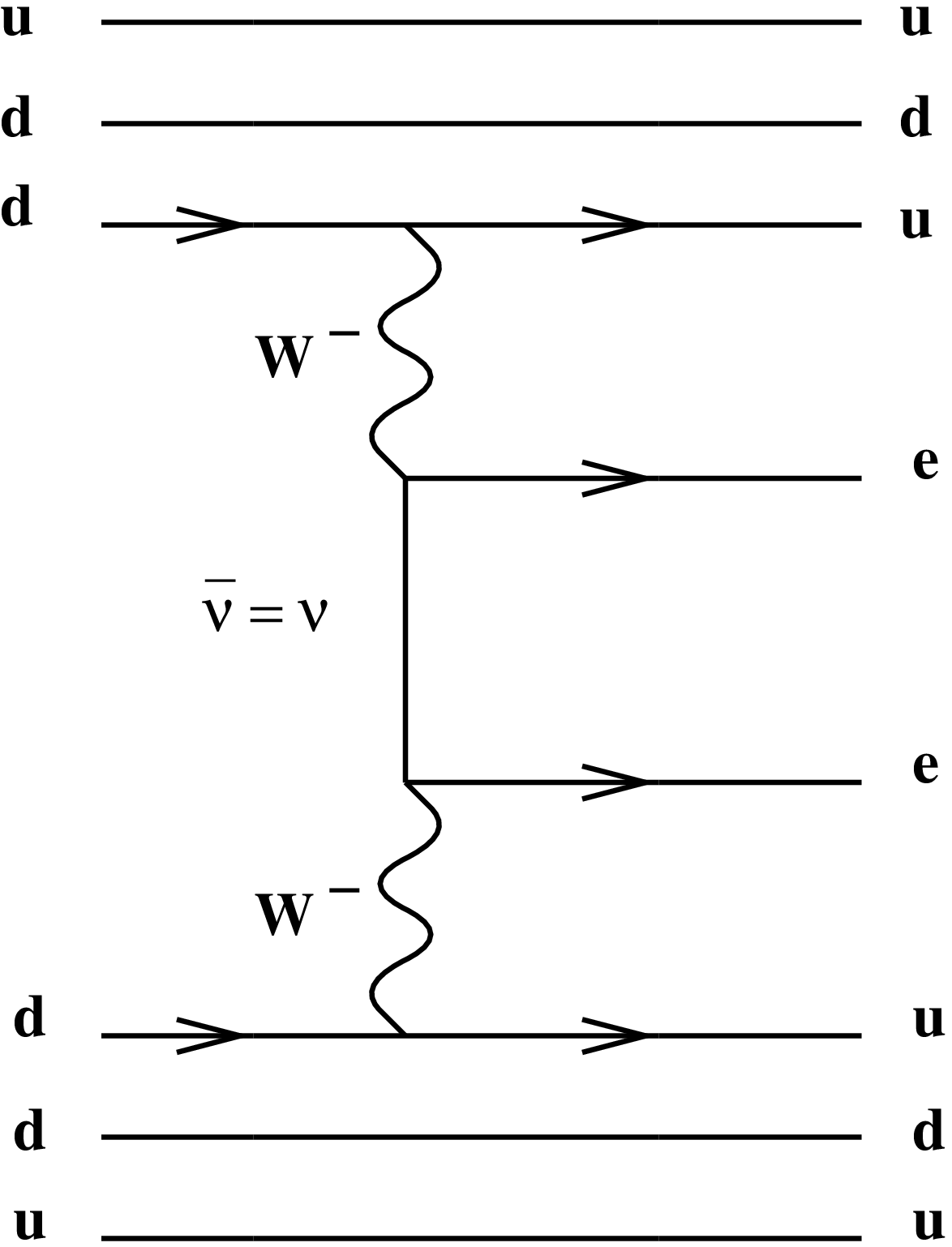}
\vspace*{5mm}

{\bf Fig. 2} {\it Feynman graph for neutrinoless double beta decay 
triggered  by exchange of a left--handed light or heavy neutrino}
\end{figure}

\begin{table}
{\footnotesize
\centerline{  
\begin{tabular}[!h]{|lll|}
\hline
Observable & Restrictions & Topics investigated\\
\hline
\hline
$0\nu$: &\underline{via $\nu$ exchange:} & 
Beyond the standard model and SU(5) model;\\
&Neutrino mass & early universe, matter--antimatter asymmetry\\
& \hskip 3mm Light Neutrino & Dark matter\\
& \hskip 3mm Heavy Neutrino & L--R --symmetric models (e.g. SO(10)),\\ 
&&compositeness\\
&Right handed weak currents & $ V+ A$ interaction, $W^{\pm}_{R}$ masses \\
&\underline{via photino, gluino, zino} & SUSY models: Bounds for parameter space\\
&\underline{(gaugino)  or sneutrino} & beyond the range of accelerators\\
&\underline{exchange:}& \\
&R-parity breaking, & \\
&sneutrino mass & \\
&\underline{via leptoquark exchange} & leptoquark masses and models\\
& leptoquark-Higgs interaction & \\
\hline
$0\nu\chi$: &existence of the Majoron & Mechanism of (B-L) breaking\\
&& 
-explicit\\ 
&& -spontaneous breaking of the local/global\\ 
&& B-L symmetry\\
&& new Majoron models\\
\hline
\end{tabular}
}}
{\bf Table 1} {\it $\beta\beta$ decay and particle physics}
\end{table}

The neutrinoless mode (2) needs not be necessarily connected with the 
exchange of a virtual neutrino or sneutrino. {\it Any} process violating 
lepton number can
in principle lead to a process with the same signature as usual 
$0\nu\beta\beta$
decay. It may be triggered by exchange of neutralinos, gluinos, squarks,
sleptons, leptoquarks,... (see below and \cite{Pae97}). This gives rise
to the broad potential of double beta decay for testing or yielding 
restrictions on
quantities of beyond standard model physics (see Table 1), realized and 
investigated to a large extent by the Heidelberg Double Beta Group in the last 
two years. 
There is, however, a generic relation between the amplitude of $0\nu\beta\beta$
decay and the $(B-L)$ violating Majorana mass of the neutrino. It has been 
recognized about 15 years ago \cite{Sch81} that if any of these two quantities
vanishes, the other one vanishes, too, and vice versa, if one of them is
non--zero, the other one also differs from zero. This Schechter-Valle-theorem 
is valid for
any gauge model with spontaneously broken symmetry at the weak scale,
independent of the mechanism of $0\nu\beta\beta$ decay. A generalisation
of this theorem to supersymmetry has been given recently \cite{Hir97,Hir97a}.
This Hirsch--Klapdor-Kleingrothaus--Kovalenko--theorem claims for the neutrino 
Majorana mass, the $B-L$ violating mass of the
sneutrino and neutrinoless double beta decay amplitude:
If one of them is non--zero, also the others are non--zero and vice versa,
independent of the mechanisms of $0\nu\beta\beta$ decay and (s-)neutrino
mass generation. This theorem connects double beta research with new processes
potentially observable at future colliders like NLC (next linear collider)
\cite{Hir97,Hir97a,Kol97}.

\begin{figure}
\epsfysize=70mm
\epsfbox{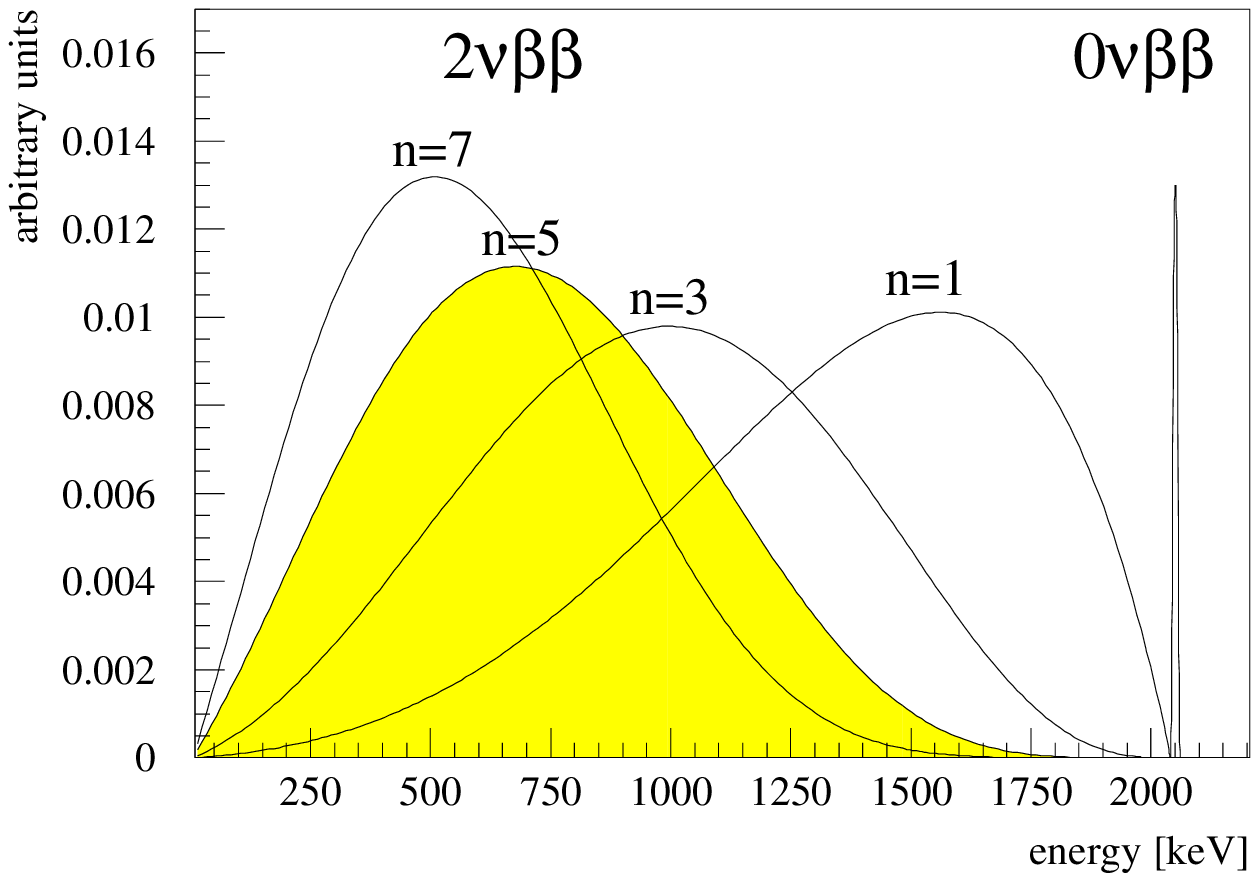}
{\bf Fig. 3} {\it Spectral shapes of the different modes of double beta 
decay,
n denotes the spectral index , n=5 for $2\nu\beta\beta$ decay (see text)} 
\end{figure}

\subsection{Mass of the (electron) neutrino}
The neutrino is one of the best examples for the merging of the different 
disciplines of micro-- and macrophysics. 
The neutrino plays, by its nature (Majorana or Dirac particle), and its
mass, a key role for the structure of modern particle physics theories
(GUTs, SUSYs, SUGRAs,...) \cite{1,Kla97,15,16,17,Kla97a,Kla97c}. 
At the same time it is
candidate for non--baryonic hot dark matter in the universe, and the neutrino mass
is connected -- by the sphaleron effect -- to the matter--antimatter asymmetry
of the early universe \cite{18}.
Neutrino physics has entered an era of new actuality in connection
with several possible indications of physics beyond the standard model
(SM) of particle physics: 
A lack of solar
($^7Be$) neutrinos, an atmospheric $\nu_{\mu}$ deficit and mixed dark matter 
models could all be explained simulaneously by non--vanishing neutrino masses.
Recent GUT models, for example an extended SO(10) scenario with $S_4$
horizontal symmetry could explain these observations by requiring 
degenerate neutrino masses of the order of 1 eV 
\cite{19,Moh94,20,21,22,23,12,13}.
For an overview see \cite{Smi96a}.
Such degenerate scenarios are the more general solution of
the well-known see-saw 
mechanism, of which the often discussed strongly hierarchical neutrino mass 
pattern is just a special solution (see \cite{Mohneu}). 
If the atmospheric neutrino data are excluded but LSND \cite{24,Ath96}, 
HDM and solar neutrino constraints are kept, they could be 
explained by an inverted 
mass texture \cite{26,Cal95}, where 
$m_{\nu_{e}} \simeq m_{\nu_{\tau}}
\simeq 2.4 eV >> m_{\nu_{\mu}}$. 
 
This brings double beta decay experiments into some key
position, since with some second generation $\beta\beta$ experiments like the 
HEIDELBERG--MOSCOW experiment using large amounts of enriched 
$\beta\beta$--emitter material the predictions of or assumptions in such
scenarios can now be tested. If the first of the above scenarios of neutrino
mass textures is ruled out by tightening the double beta limit on $m_{\nu_e}$,
then the only way to understand {\it all} neutrino results may 
require an 
additional sterile neutrino \cite{Cal93,Pel93}, coupling only extremely weakly 
to the Z--boson. Then the solar neutrino puzzle would be explained by the
$\nu_e--\nu_S$ oscillation, and atmospheric neutrino data by 
$\nu_{\mu}-\nu_{\tau}$ oscillations, and the $\nu_{\mu,\tau}$ would constitute
the hot dark matter (HDM) of the universe. The request for a light sterile
neutrino would naturally lead to the concept of a shadow world \cite{Ber95}.
This assumes exact duplication of the Standard Model in both the gauge and
the fermion content (the shadow sector), yielding three extra sterile
neutrinos $\nu^{'}$, the only interaction connecting known and shadow sector
being gravitation. Mixing of the $\nu$ and $\nu^{'}$ will occur by 
Planck scale 
effects. Such a scenario could explain all {\it four} present indications
for non-vanishing neutrino mass \cite{Mohneu}. The expectation for the effective
neutrino mass (see below) to be seen in double beta decay would be 
$ \langle m_{\nu_{e}} \rangle \simeq 0.02 eV$ \cite{Moh97a}. 
Thus it could
be checked by the new Genius project (see section 4.2.2). 
Interestingly in such a scenario
the $\nu^{'}_{\mu}$, $\nu_{\tau}^{'}$ having masses of $\sim$ 2 keV 
could act as
warm or cold dark matter in the universe \cite{Mohneu}.

At present neutrinoless double decay is the most
sensitive of the various existing methods to determine the mass of the
electron neutrino. It further provides a unique possibility
of deciding between a Dirac and a Majorana nature of the neutrino.
Neutrinoless double beta decay can be triggered by exchange of a
light or heavy left-handed Majorana neutrino  (Figs. 1,2).
For exchange of a heavy {\it right}--handed neutrino see section 2.3.
      The propagators in the first and second case show a different $m_{\nu}$
dependence: Fermion propagator $\sim \frac {m}{q^2-m^2} \Rightarrow$
\be{5}
a)\hskip5mm m\ll q \rightarrow \sim m \hskip5mm 'light' \hskip2mm neutrino
\ee
\be{6}
b)\hskip5mm m\gg q \rightarrow \sim \frac{1}{m} \hskip5mm 'heavy' \hskip2mm
neutrino
\ee   
The half--life for $0\nu\beta\beta$ decay induced by exchange of a light 
neutrino is given by \cite{27}
\ba{71}
[T^{0\nu}_{1/2}(0^+_i \rightarrow 0^+_f)]^{-1}= C_{mm} 
\frac{\langle m_{\nu} \rangle^2}{m_{e}^2}
+C_{\eta\eta} \langle \eta \rangle^2 + C_{\lambda\lambda} 
\langle \lambda \rangle^2 +C_{m\eta} \frac{m_{\nu}}{m_e} 
\nn
\ea
\be{ttt}
+ C_{m\lambda}
\langle \lambda \rangle \frac{\langle m_{\nu} \rangle}{m_e} 
+C_{\eta\lambda} 
\langle \eta \rangle \langle \lambda \rangle
\ee
or, when neglecting the effect of right--handed weak currents, by
\be{8}
[T^{0\nu}_{1/2}(0^+_i \rightarrow 0^+_f)]^{-1}=C_{mm} 
\frac{\langle m_{\nu} \rangle^2}{m_{e}^2}
=(M^{0\nu}_{GT}-M^{0\nu}_{F})^2 G_1 
\frac{\langle m_{\nu} \rangle^2}{m_e^2}
\ee
where $G_1$ denotes the phase space integral, $ \langle m_{\nu} \rangle$
denotes an effective neutrino mass
\be{9}
\langle m_{\nu} \rangle = \sum_i m_i U_{ei}^2,
\ee
respecting the possibility of the electron neutrino to be a mixed state
(mass matrix not diagonal in the flavor space)
\be{10}
|\nu_e \rangle = \sum_i U_{ei} |\nu_{i}\rangle
\ee

The effective mass $\langle m_{\nu} \rangle$ could be smaller than $m_i$
for all i for appropriate CP phases of the mixing coefficiants $U_{ei}$
\cite{Wol81}.
In general not too pathological GUT models yield 
$m_{\nu_e}=\langle m_{\nu_e}
\rangle$ (see \cite{15}).

$\eta$,$\lambda$ describe an admixture of right--handed weak currents, and
$M^{0\nu}\equiv M_{GT}^{0\nu}-M_{F}^{0\nu}$ denote nuclear matrix elements.

\subsection*{Nuclear matrix elements:}
A detailed discussion of $\beta\beta$ matrix elements for neutrino induced
transitions including the substantial (well--understood) differences
in the precision with which $2\nu$ and $0\nu\beta\beta$ rates can be 
calculated, can be found in \cite{16,27,28,29}. After the major step of 
recognizing the importance of ground state correlations for the calculation of 
$\beta\beta$ matrix elements \cite{30,31}, in recent years the main 
groups used the QRPA
model for calculation of $M^{0\nu}$. The different groups obtained very similar
results for $M^{0\nu}$ when using a realistic nucleon--nucleon interaction
\cite{28,29,32}, consistent with shell model approaches \cite{33,34}, 
where the latter are possible. Some deviation is found only when a 
non--realistic nucleon--nucleon interaction is used (e.g. $\delta$ force,
see \cite{35} and also \cite{36}). Also use of a by far too small configuration
space like in recent shell model Monte Carlo (SMMC) calculations \cite{37}
can hardly lead to reliable results. 
On the other hand refinements of the QRPA approach by going to higher order
QRPA (see \cite{38,39})lead only to minor changes for  the $0\nu\beta\beta$
ground state transitions. The most recent QRPA calculations 
including proton--neutron pairing \cite{Sim97}
do not fulfill the Ikeda sum rule by 30 \%. The calculated matrix elements
are (correspondingly ?) about 40 \% smaller than earlier calculations
fulfilling the sum rule properly \cite{28,29}.  
The consequences of high-lying
GT strength (in the GTGR and in the $\Delta$ resonance) have been studied
early already \cite{31}.

Since the usual QRPA approach does ignore deformation, some larger
uncertainty in these approaches may occur in deformed nuclei. This shows up
for example in different results obtained for $^{150}$Nd by QRPA and by a 
pseudo
SU(3) model as used by \cite{Hir95d}. 
Calculation of matrix elements of all double 
beta emitters have been published by \cite{43,29}. 
Typical uncertainties of calculated $0\nu\beta\beta$ rates originating 
from the limited knowledge of the particle--particle force, which is the main 
source of the uncertainty in those nuclei where this QRPA approach is   
applicable, are shown in \cite{29}. They are of the order of 
a factor of 2.

\subsection{Supersymmetry}
Supersymmetry (SUSY) is considered as prime candidate for a theory beyond the
standard model, which could overcome some of the most puzzling questions of
today's particle physics (see, e.g. \cite{44,45,Kan97}). 
Accelerator experiments 
have hunted
 for signs of supersymmetric particles so far without success. Lower limits on 
masses of SUSY particles are at present in the range of 20--100 GeV
\cite{PDG96}, mainly from experiments at LEP and TEVATRON.

Conservation of R--parity has been imposed ad hoc to the
minimal supersymmetric extension of the standard model (MSSM) to 
ensure
baryon number and lepton number conservation.
SUSY particles differ then from usual particles not only in their masses but also
in R--parity, assigned to be $R_P=1$ for usual particles and $R_P=-1$ for
SUSY particles. 
This assumption, however, is not guaranteed by supersymmetry or gauge 
invariance. Generally one can add the following R--parity violating terms  
to the usual superpotential \cite{hal84}.
\be{rpv}
W_{\Rpv}=\lambda_{ijk}L_{i}L_{j}\overline{E}_{k}+\lambda^{'}_{ijk}
L_i Q_j \overline{D}_k + \lambda^{''}\overline{U}_i \overline{D}_j 
\overline{D}_k,
\ee
where indices $i,j,k$ denote generations. $L$,$Q$ denote lepton and quark 
doublet superfields and $\overline{E}, \overline{U}, \overline{D}$ lepton and
up, down quark singlet superfields. Terms proportional to $\lambda$, 
$\lambda^{'}$
violate lepton number, those proportional to $\lambda^{''}$ violate baryon 
number. From proton decay limits it is clear that both types of terms cannot 
be present at the same time in the superpotential. On the other hand, once the
$\lambda^{''}$ terms being assumed to be zero, $\lambda$ and $\lambda^{'}$
terms are not limited. $0\nu\beta\beta$ decay can occur within the 
\rp-MSSM through Feynman graphs  such as those of Fig. 4. 
In lowest order 
there are alltogether six different graphs of this kind. \cite{6,47,hir96c}.    
 Attention has, therefore , been focused 
also on SUSY theories with R--parity violation, in which 
$0\nu\beta\beta$ decay can proceed by exchange of supersymmetric particles 
like gluinos, photinos,... 
Thus  $0\nu\beta\beta$ decay can be used to restrict R--parity violating
SUSY models \cite{6,hir96c,17,47,48}. From these graphs one derives \cite{6}
under some assumptions 
\be{7}
[T^{0\nu}_{1/2}(0^+ \rightarrow 0^+)]^{-1} \sim G_{01}
(\frac{\lambda_{111}'^2}{m^4_{{\tilde q},{\tilde e}}m_{{\tilde g}\chi}}M)^2
\ee
where $G_{01}$ is a phase space factor, 
$m_{{\tilde q}{\tilde e}{\tilde g}\chi}$
are the masses of supersymmetric particles involved: squarks, selectrons,
gluinos, or neutralinos. $\lambda'_{111}$ is the strength of an R--parity
breaking interaction (eq. 11), and $M$ is a nuclear matrix element. For the
matrix elements and their calculation see \cite{hir96c}.

\begin{figure}
\parbox{14cm}{
\epsfxsize=50mm
\epsfbox{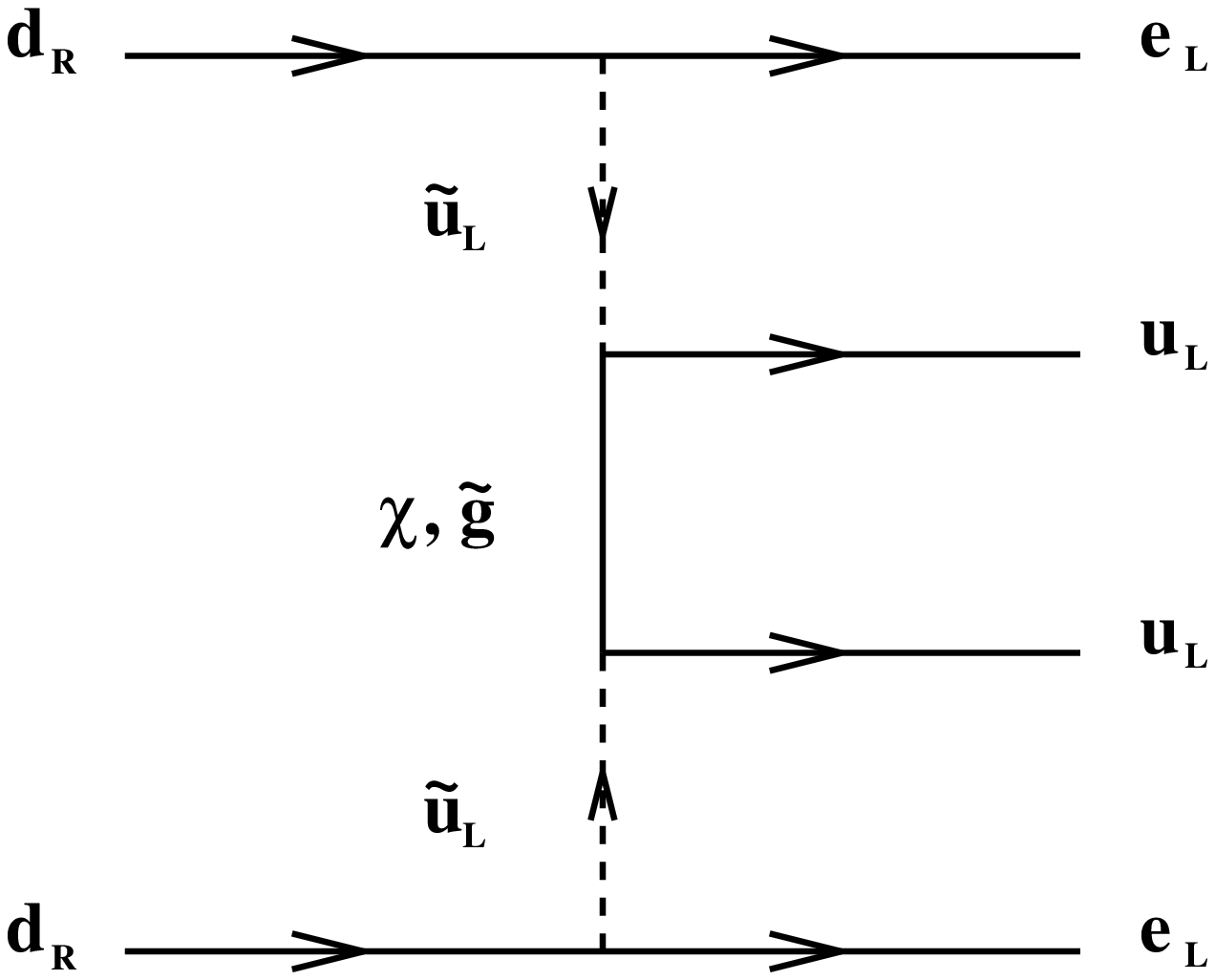}
\parbox{6cm}{
\vspace*{-45mm}
\hspace*{60mm}
\epsfxsize=50mm
\epsfbox{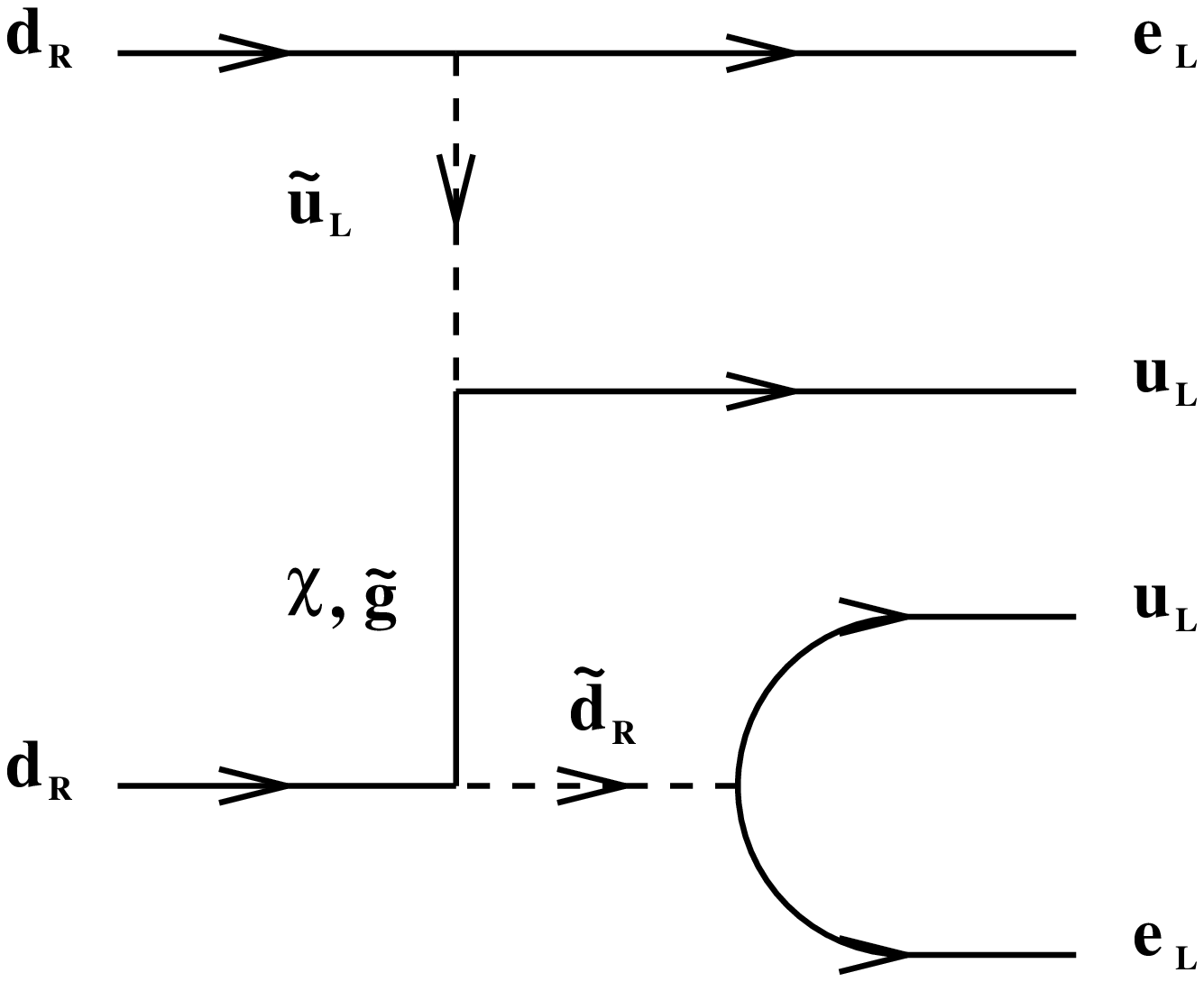}
}

{\bf Fig. 4} {\it Examples of Feynman graphs for $0\nu\beta\beta$ 
decay within R--parity \\
violating supersymmetric models (from [Hir95]).}}
\parbox{14cm}{
\epsfxsize=50mm
\epsfbox{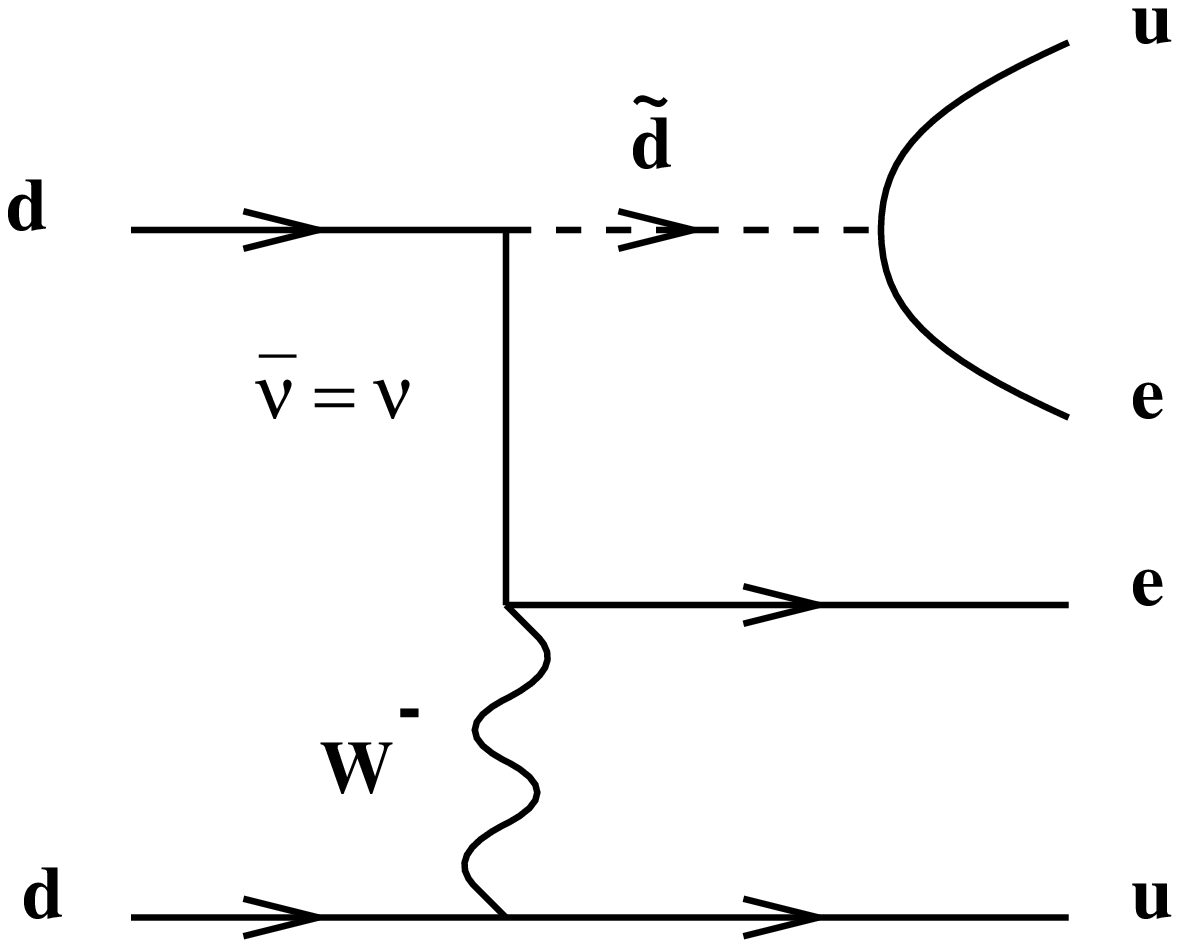}
\parbox{6cm}{
\vspace*{-45mm}
\hspace*{60mm}
\epsfxsize=50mm
\epsfbox{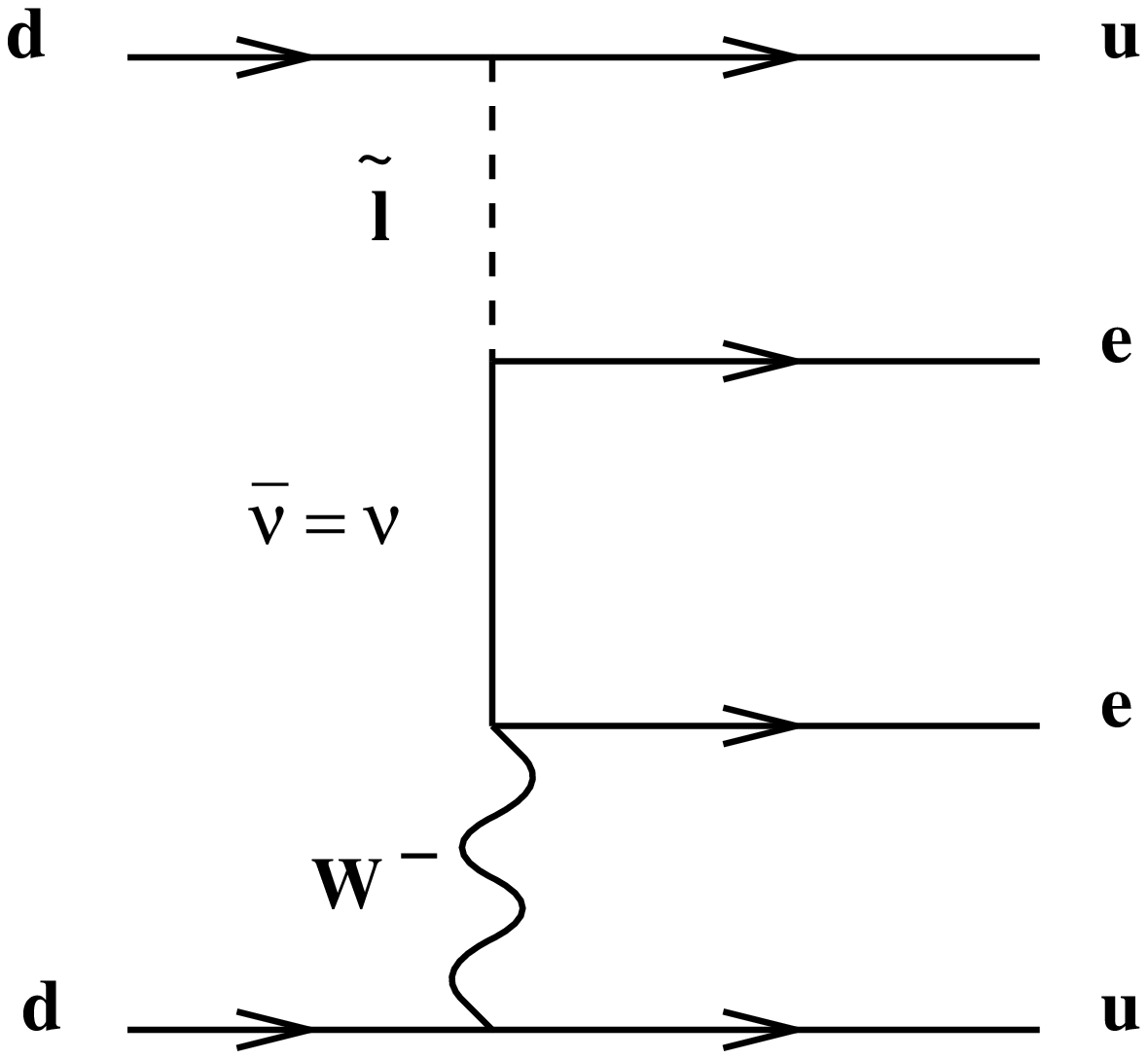}
}

{\bf Fig. 5 a)} {\it Feynman graph for the mixed 
SUSY--neutrino exchange \\
mechanism of 
$0\nu\beta\beta$ decay. R--parity violation occurs through scalar quark\\ 
exchange.} 
{\bf b)} {\it As figure 1, but for scalar lepton exchange
(from [Hir96]).}}
\parbox{14cm}{
\vspace*{6mm}
\epsfxsize=50mm
\epsfbox{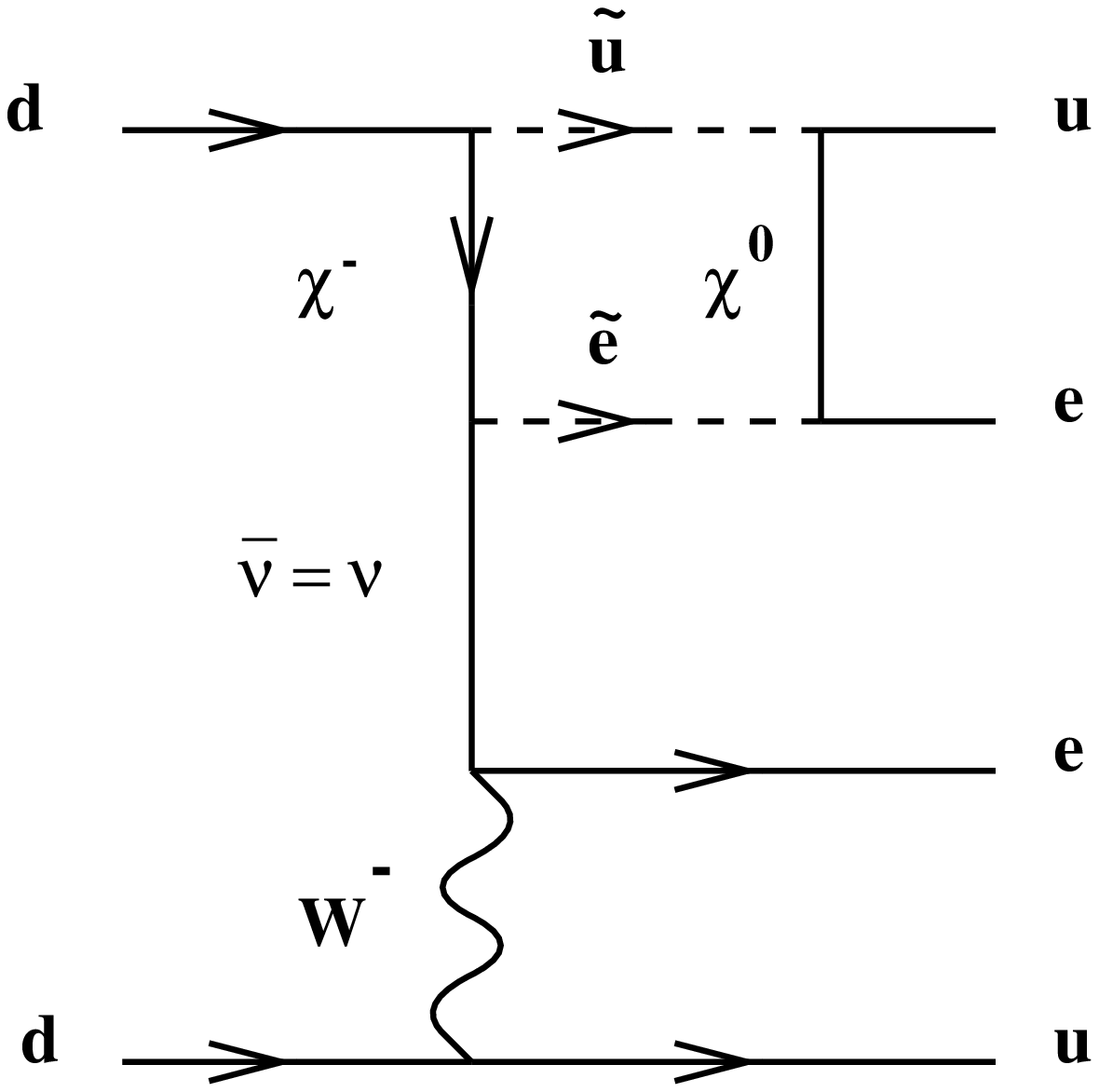}
\parbox{6cm}{
\vspace*{-51mm}
\hspace*{60mm}
\epsfxsize=50mm
\epsfbox{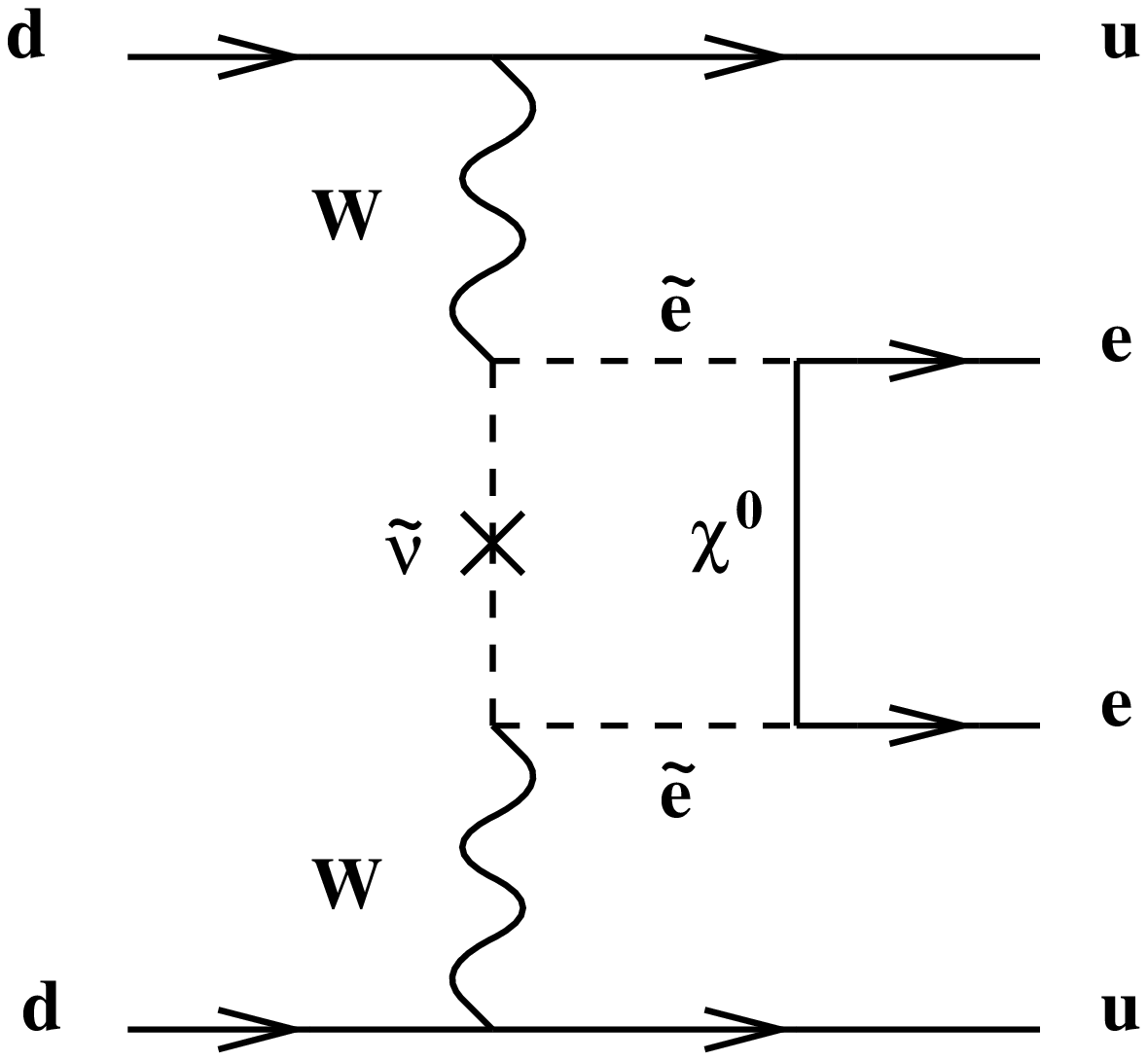}
}

{\bf Fig. 6} {\it Examples of $R_P$ conserving SUSY contributions
to $0\nu\beta\beta$ decay \\
(from [Hir97a]).}}   
\end{figure}

It is also worthwile to notice that $0\nu\beta\beta$ decay is not only 
sensitive to $\lambda^{'}_{111}$. Taking into account the fact that the SUSY
partners of the left and right--handed quark states can mix with each other, 
one can derive limits on different combinations of $\lambda^{'}$ 
\cite{hir96,7,bab95}.
Graphs allowing such information are as those shown in Fig. 5. The dominant 
diagram 
of this type is the one where the exchanged scalar particles are the
$\tilde{b}-\tilde{b}^C$ pair. Under some assumptions (e.g. the MSSM mass 
parameters to be approximately equal to the ``effective'' SUSY breaking scale 
$\Lambda_{SUSY}$), one obtains \cite{hir96}
\be{ncs1}
\lambda_{11i}^{'}\cdot \lambda_{1i1}^{'}\leq \epsilon_i^{'} 
\Big( \frac{\Lambda_{SUSY}}{100 GeV} \Big)^3
\ee
and
\be{ncs2}
\Delta_n \lambda^{'}_{311} \lambda_{n13} \leq \epsilon \Big(\frac
{\Lambda_{SUSY}
}{100 GeV}\Big)^3
\ee
With the known values of the SM quark and lepton masses follows 
$\epsilon^{'}_{1,2,3}=6.4 \cdot 10^{-5}$, $3.2\cdot 10^{-6}$ 
and $1.1 \cdot 10^{-7}$,
and $\epsilon = 6.5 \cdot 10 ^{-8}$. For an overview on our knowledge
on $\lambda^{'}_{ijk}$
from other sources we refer to \cite{Kol97a} and \cite{Bha97}.

Also R--parity {\it conserving} softly broken supersymmetry can give 
contributions to $0\nu\beta\beta$ decay, via the $B-L$--violating sneutrino 
mass term, the latter being a generic ingredient of any weak--scale SUSY
model with a Majorana neutrino mass \cite{Hir97,Hir97a}. 
These contributions are 
realized  at the level of box diagrams \cite{Hir97a} (fig. 6).
The $0\nu\beta\beta$ half-life for contributions from sneutrino exchange
is found to be \cite{Hir97a}
\be{rconv}
[{T_{1/2}^{0\nu\beta\beta}}]^{-1}=G_{01}\frac{4 m_p^2}{G^4_F} 
\Big|\frac{\eta^{SUSY}}{m^5_{SUSY}} M^{SUSY}\Big|,
\ee
where the phase factor $G_{01}$ is tabulated in \cite{74}, $\eta^{SUSY}$
is the effective lepton number violating parameter, which contains the
$(B-L)$ violating sneutrino mass $\tilde{m}_M$ and $M^{SUSY}$ is the nuclear 
matrix element \cite{11}.

\begin{figure}
\parbox{6cm}{
\epsfxsize=50mm
\epsfbox{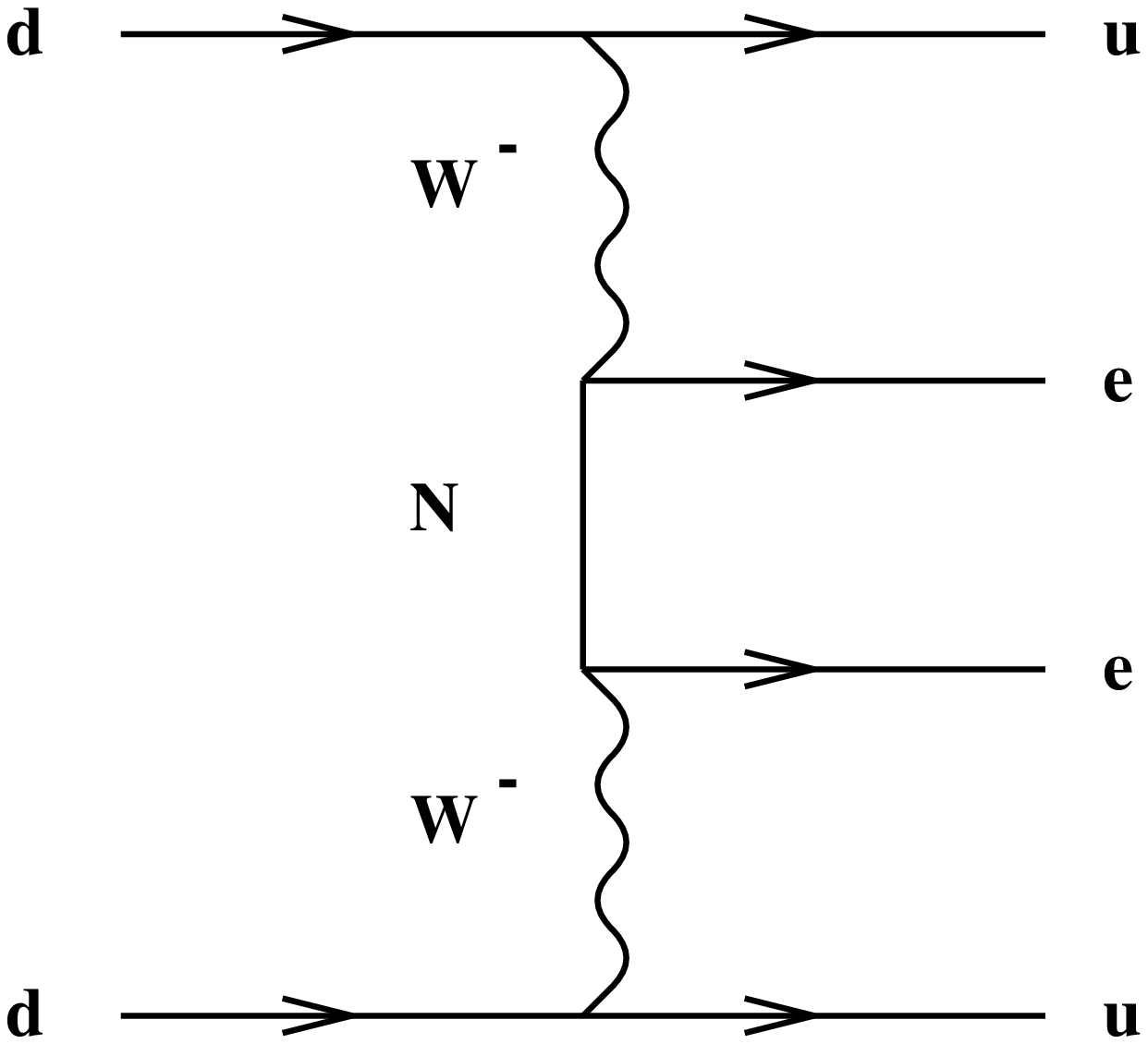}
}
\hspace*{6mm}
\parbox{6cm}{
\epsfxsize=50mm
\epsfbox{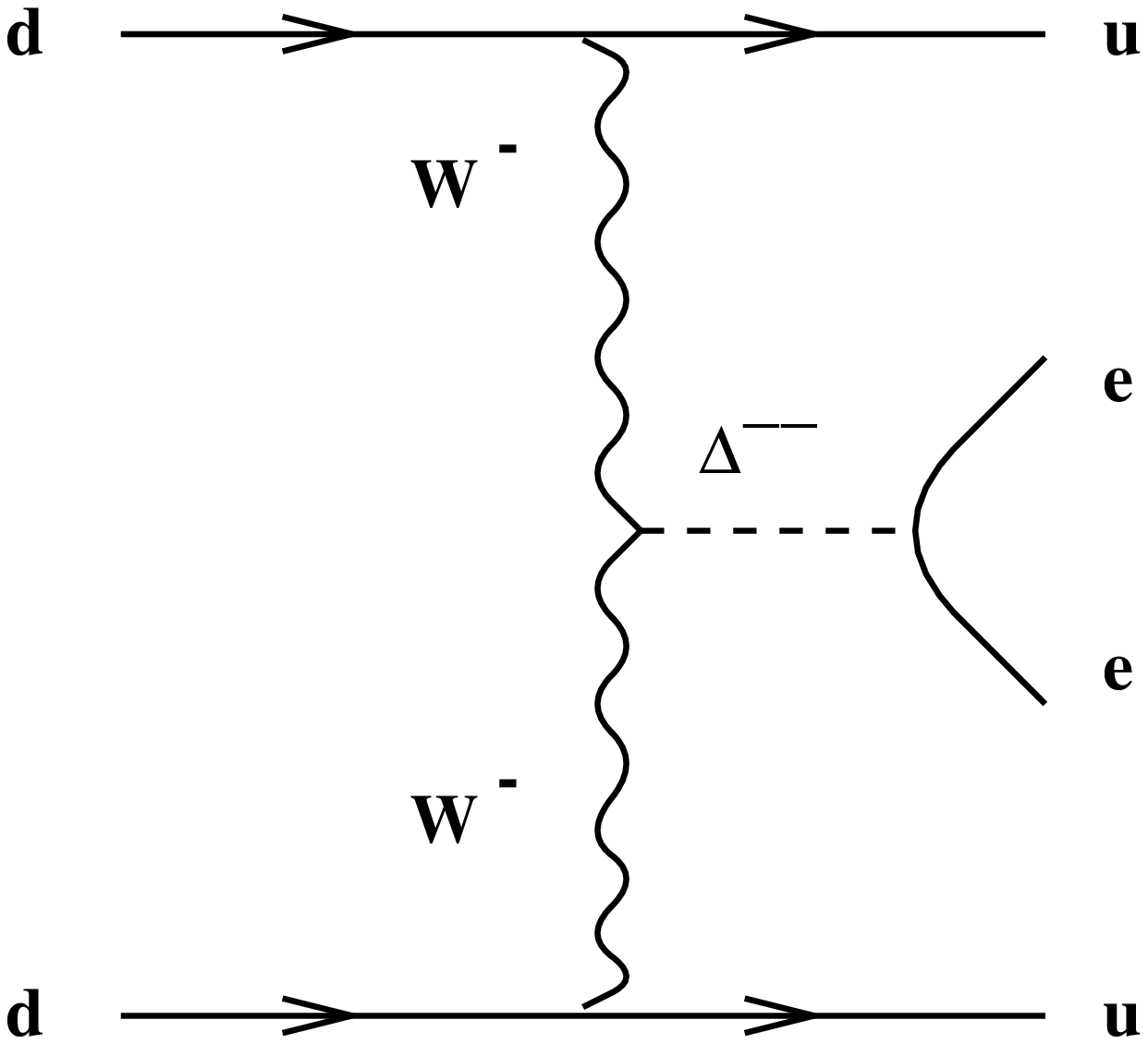}
\vspace*{2mm}
}
\vspace*{5mm}

{\bf Fig. 7 a)} {\it Heavy neutrino exchange contribution to neutrinoless 
double beta decay
in left right symmetric models, and} {\bf b)} {\it Feynman graph for 
the virtual exchange
of a doubly--charged Higgs boson, see text (from [Hir96d]).} 
\end{figure}

\subsection{Left--Right symmetric theories --
Heavy neutrinos and right--handed W Boson}
Heavy {\it right--handed } neutrinos appear quite naturally in left--right
symmetric GUT models. They offer in some natural way via the see--saw
mechanism explanation for the small neutrino masses compared to other fermions
and can explain also naturally parity violation. However the symmetry breaking
scale for the right--handed sector is not fixed by the theory and thus the
mass of the right--handed $W_R$ boson and the mixing angle between the mass 
eigenstates $W_1$, $W_2$ are free parameters. $0\nu\beta\beta$ decay taking
into account contributions from both, left-- and right--handed neutrinos
have been studied theoretically by \cite{11,49}. The former gives a more
general expression for the decay rate than introduced earlier by \cite{50}. 

$0\nu\beta\beta$ decay proceeds through the diagram shown in Fig. 7a, 
where 
$N$ denotes the heavy right--handed partner of the ordinary neutrino. 
In order to preserve the unitarity of the cross section in {\it inverse}
$0\nu\beta\beta$ decay, LR models must according to \cite{riz82}
include an additional Higgs triplet. This then gives rise to a second
contribution to $0\nu\beta\beta$ decay, shown in Fig. 7b. From the 
Feynman graphs of Fig. 7 it
is obvious that the amplitude will be proportional to \cite{11}
\be{ncs3}
\Big( \frac{m_{W_{L}}}{m_{W_R}}^4 \Big) \Big(\frac{1}{m_N}+\frac{m_N}
{m^2_{\Delta^{--}_R}}\Big)
\ee

Eq. \rf{ncs3} and the experimental lower limit of $0\nu\beta\beta$ decay leads 
to a constraint limit within the 3--dimensional parameter space
($m_{W_R}-m_N-m_{\Delta^{--}_R}$). The most conservative (weakest) limit
on $m_{W_R}$ is obtained in the limit, where the mass of the $\Delta^{--}$
goes to infinity (see section 3 below). If adding information on the 
vacuum 
stability, an absolute lower limit on the mass of the right--handed W--boson 
can be obtained.

\begin{figure}[htbp]
    \epsfxsize=90mm
    \epsfbox{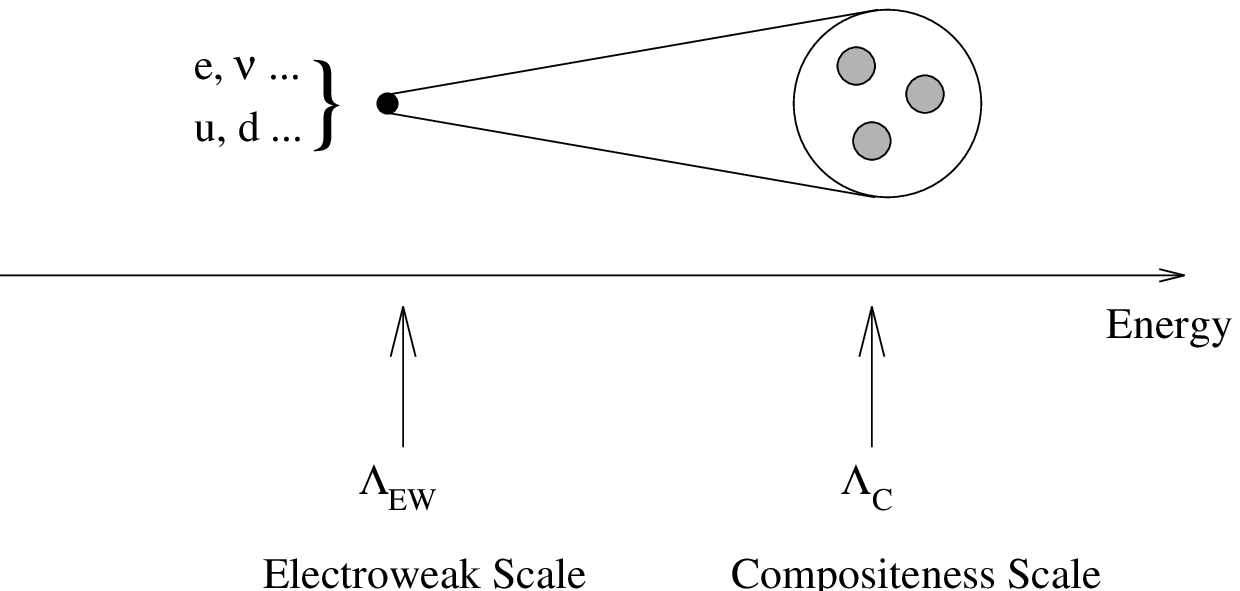}
    {\bf Fig. 8}
    {\it The idea of compositeness. At a (still unknown) energy scale 
             $\Lambda_{C}$ quarks and lepton might show an 
             internal structure}  
\end{figure}

\begin{figure}[htbp]
    \epsfxsize=90mm
    \epsfbox{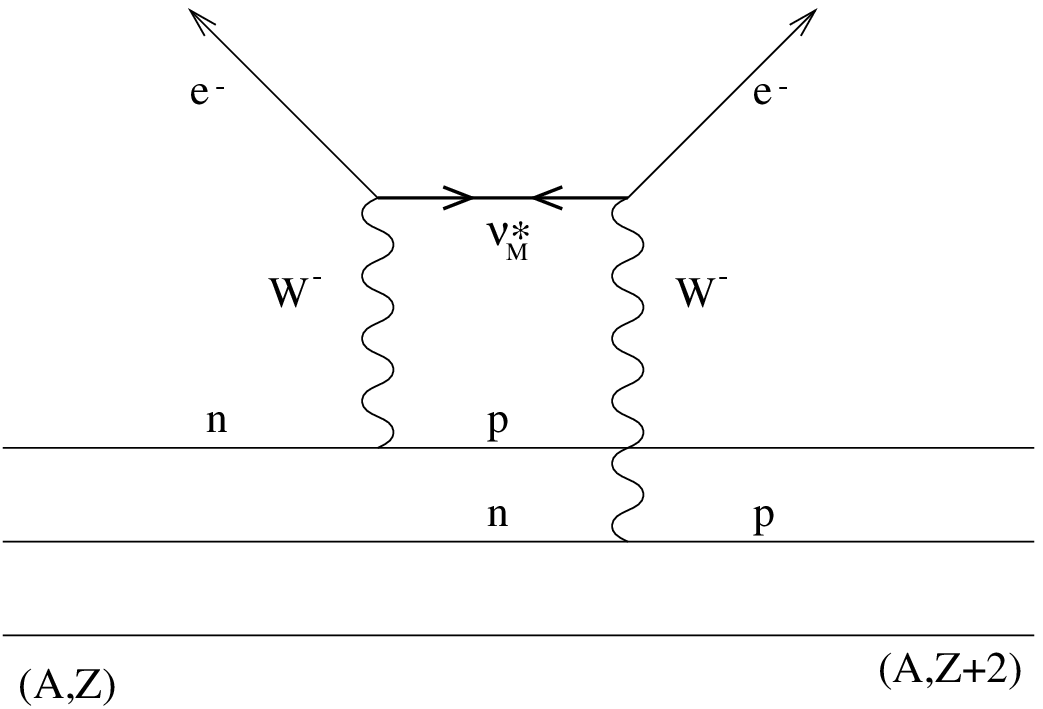}
{\bf Fig. 9}
{\it Neutrinoless double beta decay ($\Delta L = +2$ process)
             mediated by a composite heavy Majorana neutrino.}
\end{figure}

\subsection{Compositeness}
Although so far there are no experimental signals of a substructure of quarks 
and leptons, there are speculations that at some higher energy ranges beyond 1 
TeV or so there might exist an energy scale $\Lambda_C$ at which a
substructure of quarks and leptons (preons) might become visible 
\cite{8,45,51}(Fig. 8). 

The main consequences of compositeness of quarks and leptons are
(1) modifications to the gauge boson propagators and the interaction 
vertices with fermions, and additional effective four--fermion 
interactions through constituent exchange (2) highly massive excited states
which couple to the ordinary fermions through gauge interactions. This is 
discussed in detail in \cite{8}. Lower bounds on the compositeness scale
have been deduced from accelerator experiments at LEP \cite{52}, Fermilab
\cite{53}, HERA \cite{5} and from a theoretical analysis of the effect
of contact interactions in the leptonic $\tau$ decay \cite{54}.
They are all in the range of $\Lambda_C \ge 1.6$ TeV.

The {\it masses} of the excited leptons ($l^*$) and quarks ($q^*$) should not
be lower than the compositeness scale $\Lambda_C$. Already in 1982 it was shown
\cite{55} that precise measurements of the anomalous magnetic moment of the
electron give bounds on the masses of the excited states and thus the 
compositeness scale.

Limits on the masses of excited leptons from accelerators are $m_{e^*} > 127$ 
GeV \cite{56}, $m_{e^*,\nu^*} > 91$ GeV \cite{57}, $m_{\nu^*} > 180$ GeV
\cite{58,59} $m_{q^*} > 540$ GeV \cite{60}.

A possible low energy manifestation of compositeness could be neutrinoless
double beta decay, mediated by a composite heavy Majorana neutrino (Fig. 9), 
which then should be a Majorana particle.

Recent theoretical work shows (see \cite{8,9,Pan97,Tak97}) 
that the mass bounds for such an excited neutrino 
which can be derived from double 
beta decay are at
 least of the same order of magnitude as those coming from the
direct search of excited states in high energy accelerators 
(see also section 3).

\subsection{Majorons} 
The existence of new bosons, so--called Majorons, can play a significant
role in new physics beyond the standard model, in the history
of the early universe, in the evolution of stellar objects, in supernovae
astrophysics and the solar neutrino problem \cite{61,62,Kla92}.
In many theories of physics beyond the standard model neutrinoless 
double beta decay can occur with the emission of Majorons  
\be{phi1}
2n\rightarrow2p+2e^{-}+\phi
\ee
\be{phi2}
2n\rightarrow2p+2e^{-}+2\phi.
\ee 
In the classical Majoron model invented by Gelmini and Roncadelli 
in '81 \cite{63}, the Majoron is the Nambu--Goldstone boson 
associated with the spontaneous breaking of the $B-L$--symmetry 
and so generates Majorana masses of neutrinos. This was expected \cite{61}
to give a sizeable contribution to double beta decay. It was, however, ruled 
out, as also the doublet Majoron \cite{66} by LEP \cite{67} since it should
contribute the equivalent of two neutrino species to the width of the $Z^0$.
On the other hand, Majoron models in which the Majoron is an 
electroweak isospin singlet \cite{64,65} are still viable.
The drawback of the singlet Majoron is that it requires a severe finetuning   
in order to preserve existing 
bounds on neutrino masses and at the same time get an observable 
rate for Majoron accompanied $0\nu\beta\beta$ decay. 

To avoid such an unnatural fine--tuning in recent years several 
new Majoron models were proposed \cite{68,69,70}, 
where the term
Majoron denotes in a more general sense light or massless bosons 
with couplings to neutrinos. 

The main novel features of these ``New Majorons'' are that they
can carry leptonic charge, that they need not be 
Goldstone bosons and that emission of two Majorons 
can occur. 
The latter can be scalar--mediated 
or fermion--mediated. Table 2 shows some features of the 
different Majoron models according to \cite{69,70}. L denotes the 
leptonic charge, n the spectral index defining the phase space of the
emitting particles, M the nuclear matrix elements. For details we refer to
\cite{71,72}. 

The half--lifes are according to \cite{73,74} in some approximation given
by  
\be{14}
[T_{1/2}]^{-1}=|<g_{\alpha}>|^{2}\cdot|M_{\alpha}|^{2}\cdot G_{BB_{\alpha}}
\ee
for $\beta\beta\phi$-decays, or
\be{15}
[T_{1/2}]^{-1}=|<g_{\alpha}>|^{4}\cdot|M_{\alpha}|^{2}\cdot G_{BB_{\alpha}}
\ee
for $\beta\beta\phi\phi$--decays. The index ${\alpha}$ 
indicates that effective neutrino--Majoron coupling constants $g$, 
matrix elements $M$ and phase spaces $G$ differ for different models.

\newpage
\subsection*{Nuclear matrix elements:}
According to Table 2 there are five different nuclear matrix elements. Of
 these $M_{F}$ and $M_{GT}$ are the same which occur in $0\nu\beta\beta$ decay.
The other ones and the corresponding phase spaces have been calculated 
for the first time
by \cite{71,75}. The calculation of the matrix elements show 
that the new models predict, 
as consequence of the small matrix elements 
 very large half--lives and that unlikely large
coupling constants would be needed to produce observable decay rates
(see Table 3).

\begin{table}
\parbox{14cm}{   
\begin{tabular}{|cccccc|}
\hline
case & modus & Goldstone boson & L & n & Matrix element \\
\hline
\hline 
IB & $\beta\beta\phi$ & no & 0 & 1 & $M_{F}-M_{GT}$ \\
\hline
IC & $\beta\beta\phi$ & yes & 0 & 1 & $M_{F}-M_{GT}$ \\
\hline
ID & $\beta\beta\phi\phi$ & no  & 0 & 3 & 
$M_{F\omega^2}-M_{GT\omega^2}$ \\
\hline
IE & $\beta\beta\phi\phi$ & yes & 0 & 3 & 
$M_{F\omega^2}-M_{GT\omega^2}$ \\
\hline
IIB & $\beta\beta\phi$ & no  & -2 & 1 & $M_{F}-M_{GT}$ \\
\hline
IIC & $\beta\beta\phi$ & yes & -2 & 3 & $M_{CR}$ \\
\hline
IID & $\beta\beta\phi\phi$ & no  & -1 & 3 &
$M_{F\omega^2}-M_{GT\omega^2}$ \\
\hline
IIE & $\beta\beta\phi\phi$ & yes & -1 & 7 & 
$M_{F\omega^2}-M_{GT\omega^2}$ \\
\hline
IIF & $\beta\beta\phi$ & Gauge boson & -2 & 3 & $M_{CR}$ \\   
\hline
\end{tabular}}
\vskip2mm
{\bf Table 2} {\it Different Majoron models according to 
[Bam95].
The case IIF corresponds to the model 
of [Car93].}
\vskip3mm
\parbox{14cm}{
\begin{tabular}{|c|c|c|c|}
\hline
model & $T_{1/2}(<g>=10^{-4})$ & $T_{1/2}(<g>=1)$ & $T_{1/2 exp}$ \\ 
\hline
\hline
IB,IC,IIB  & $4\cdot10^{22}$ & $4\cdot10^{14}$ 
& $1.67\cdot10^{22}$ \\ 
\hline
ID,IE,IID  & $10^{38-42}$ & $10^{22-26}$ 
& $1.67\cdot10^{22}$ \\ 
\hline
IIC,IIF & $2\cdot10^{28}$ & $2\cdot10^{20}$ & $1.67\cdot10^{22}$ \\
\hline
IIE & $10^{38-42}$ & $10^{22-26}$ & $3.37\cdot10^{22}$\\
\hline
\end{tabular}}
\vskip2mm
{\bf Table 3}
{\it Comparison of half--lives calculated for 
different $<g>$--values for the new Majoron models with experimental 
best fit values, see section 3.1 (from [Hir96b])}
\end{table} 

\subsection{Sterile neutrinos}

Introduction of sterile neutrinos has been claimed to solve simultaneously the 
conflict between dark matter neutrinos, LSND and supernova nucleosynthesis
\cite{76} and light sterile neutrinos are part of popular 
neutrino mass textures
for understanding the various hints for neutrino
oscillations (see section 2.1) and \cite{Moh96,Mohneu,Moh97a}. 
Neutrinoless double beta decay can also
investigate several effects
of {\it heavy} sterile neutrinos \cite{77}.

\begin{figure}[!t]
\vspace*{20mm}
\setlength{\unitlength}{1in}                                                 
\begin{picture}(5,2)
\put(0.0,0.5){\includegraphics{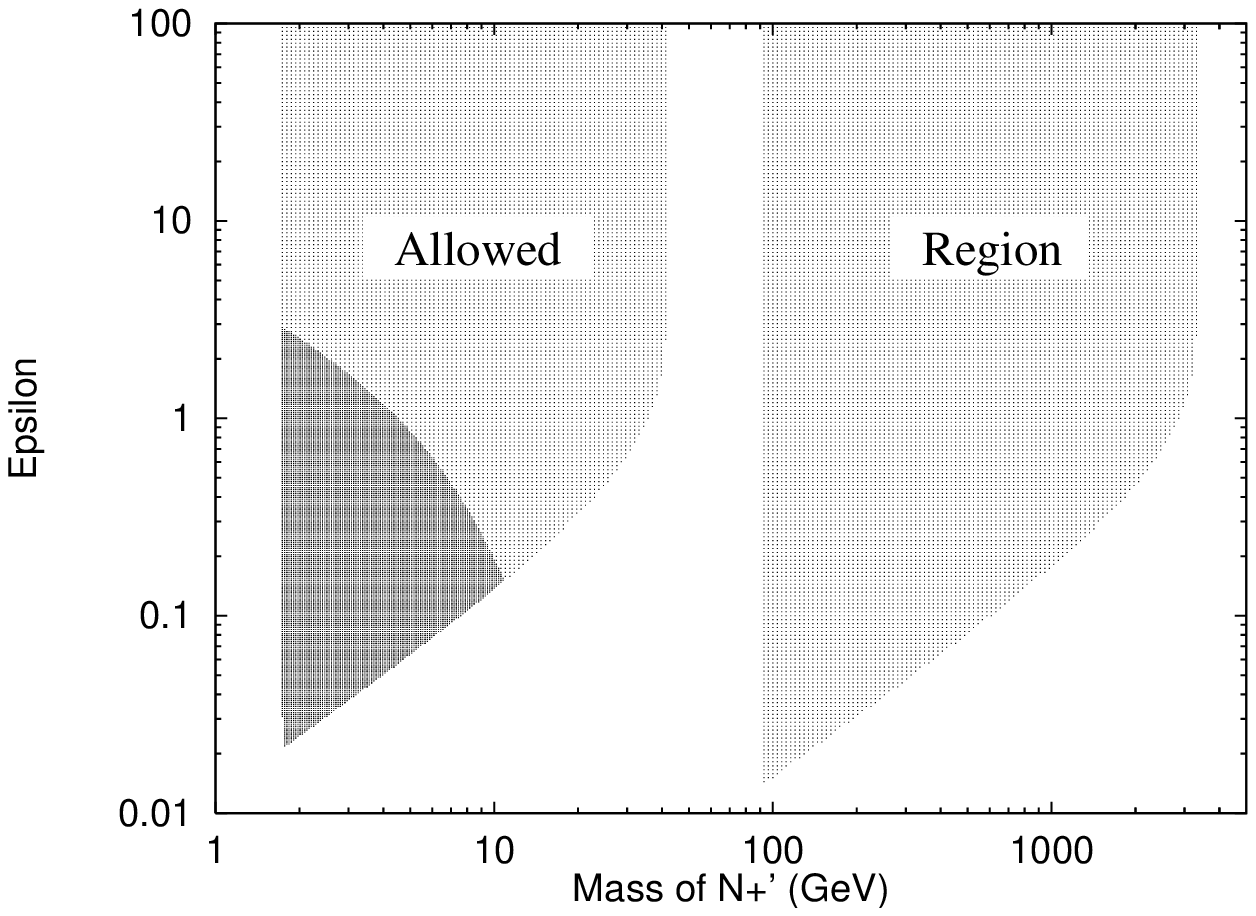}}
\end{picture}                                                                 
{\bf Fig. 10} {\it Regions of the parameter space 
($\epsilon-M_{N'_{+}}$ plane yielding
an observable signal (shaded areas) (from \cite{77}). Darker area: 'natural' 
region, lighter shaded: Finetuning needed, to keep $m_{\nu_e}$ below 1 eV.
$M_{N'_+}$: mass eigenstate, $\epsilon$: strength of lepton number violation in
mass matrix}
\end{figure}

If we assume having a light neutrino with a mass $\ll$ 1 eV, mixing with a much
 heavier (m $\ge$ 1 GeV) sterile neutrino can yield under certain conditions
a detectable signal in current $\beta\beta$ experiments.

In models with two (or more) sterile neutrinos, the sterile neutrinos can mix
appreciably even in the limit $m_{\nu_{e}} \rightarrow 0$ and so can be 
potentially visible in many processes \cite{78}. Neutrinoless double beta
decay proceeds in these models through the virtual exchange of the
heavier (i.e. GeV scale or higher) neutrinos. Fig. 10 shows the mass 
ranges leading to a $0\nu\beta\beta$ signal close to 
observability (shaded areas).     

\subsection{Leptoquarks}

Interest on leptoquarks (LQ) has been renewed during the last few years 
since ongoing collider experiments have good prospects for searching 
these particles \cite{Lagr1}. LQs are vector or scalar particles 
carrying both lepton and baryon numbers and, therefore, have a 
well distinguished experimental signature. Direct searches of LQs in 
deep inelastic ep-scattering at HERA \cite{H196} placed lower limits 
on their mass $M_{LQ} \ge 225-275$ GeV, depending on the LQ type and 
couplings. 

In addition to the direct searches on LQs, there are 
many constraints which can be derived from the study of low-energy 
processes \cite{DBC}. Effective 4-fermion interactions, induced 
by virtual LQ exchange at energies much smaller than their masses, 
can contribute to atomic parity violation, flavour-changing neutral 
current processes, meson decays, meson-antimeson mixing and some 
rare processes. 

\begin{figure}
\parbox{7cm}{
\epsfxsize=50mm
\epsfysize=50mm
\epsfbox{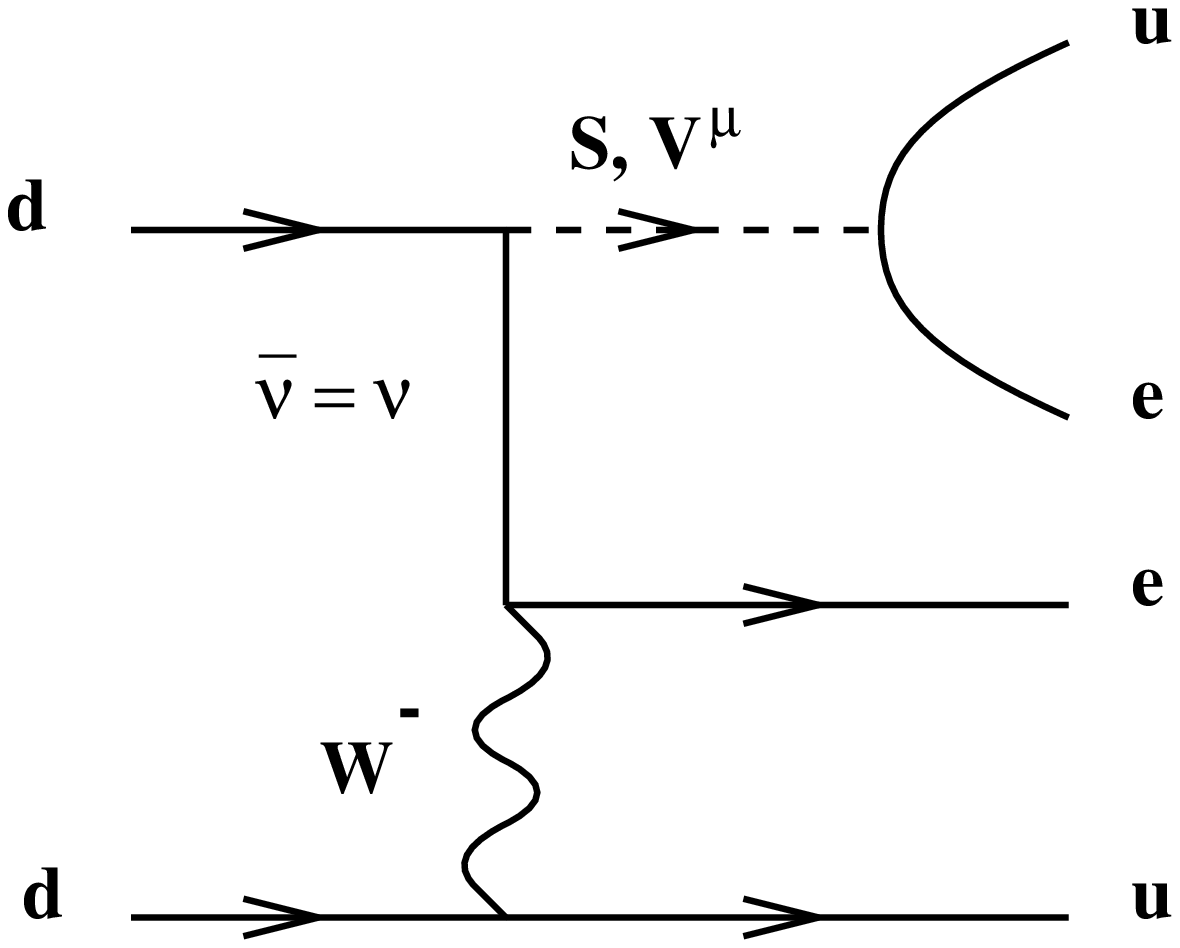}}
\parbox{7cm}{
\epsfxsize=50mm
\epsfysize=50mm
\epsfbox{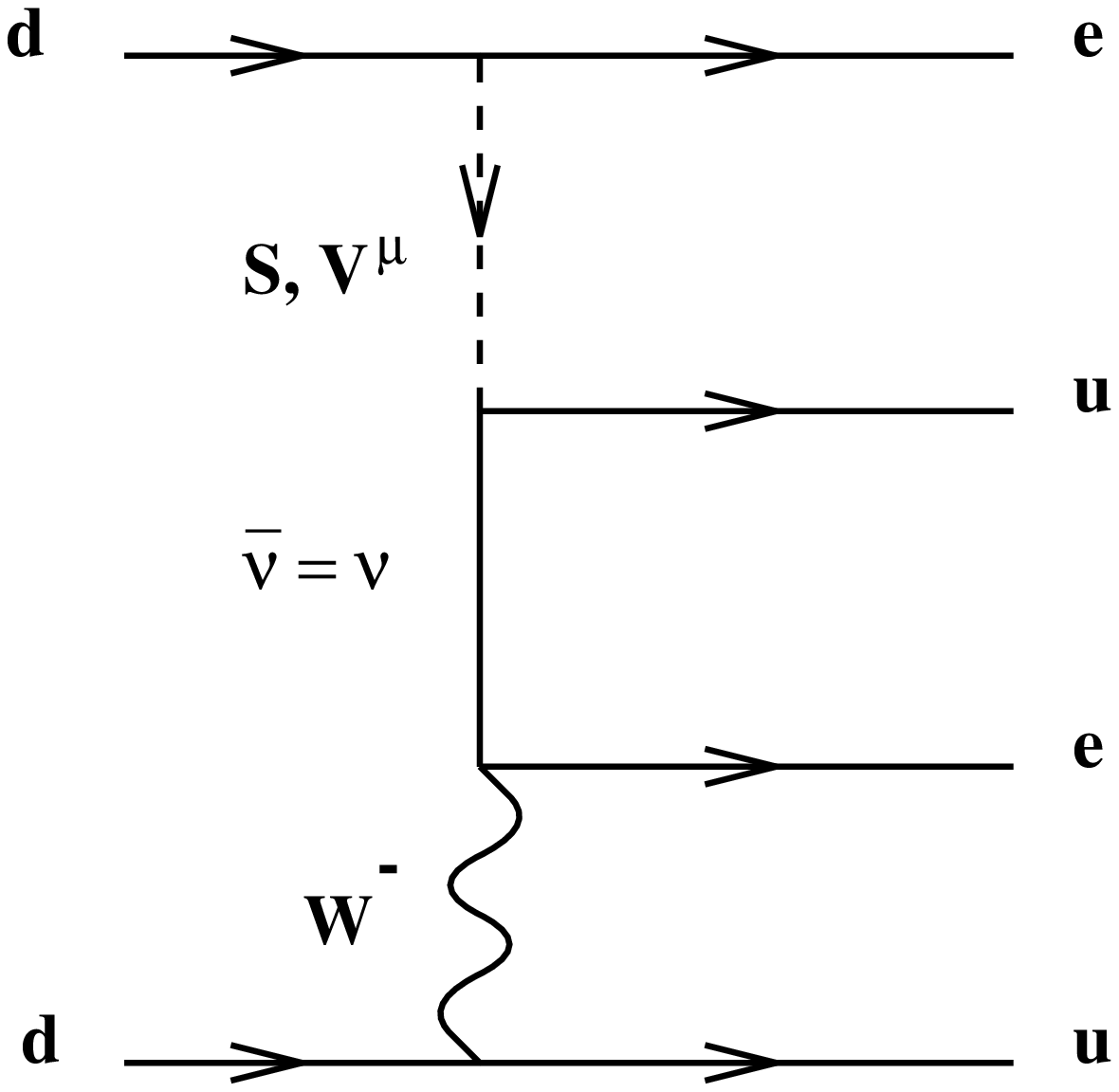}}
\vskip5mm
{\bf Fig. 11} {\it Examples of Feynman graphs for $0\nu\beta\beta$ decay 
within LQ models. $S$ and $V^{\mu}$ stand symbolically for scalar 
and vector LQs, respectively (from [Hir96a]).}
\end{figure}

To consider LQ phenomenology in a model-independent fashion one 
usually follows some general principles in constructing the Lagrangian 
of the LQ interactions with the standard model fields. In order to 
obey the stringent constraints from (c1) helicity-suppressed 
$\pi \rightarrow e\nu$ decay, from (c2) FCNC processes and 
from (c3) proton stability, the following assumptions are commonly adopted: 
(a1) LQ couplings are chiral, (a2) LQ couplings are generation 
diagonal, and (a3) there are no diquark couplings. 

Recently, however, it has been pointed out \cite{hir96a} that possible 
LQ-Higgs interactions spoil assumption (a1): Even if one assumes 
LQs to be chiral at some high energy scale, LQ-Higgs interactions 
introduce after electro-weak symmetry breaking mixing between 
LQ states with different chirality. Since there is no fundamental 
reason to forbid such LQ-Higgs interactions, it seems difficult 
to get rid of the unwanted non-chiral interactions in LQ models. 

In such LQ models there appear contributions to $0\nu\beta\beta$ 
decay via the Feynman graphs of Fig. 11. Here, $S$ and $V^{\mu}$ 
stand 
symbolically for scalar and vector LQs, respectively. 
The half--life for $0\nu\beta\beta$ decay arising from leptoquark 
exchange is given by \cite{hir96a}

\be{fuufu}
T_{1/2}^{0\nu}=|M_{GT}|^2 \frac{2}{G_F^2}[\tilde{C}_1a^2+C_4 b_R^2
+2 C_5 b_L^2].
\ee

with $a=\frac{\epsilon_S}{M_S^2}+\frac{\epsilon_{V}}{M_V^2}$, 
$b_{L,R}=\frac{\alpha_{S}^{(L,R)}}{M_S^2}+\frac{\alpha_V^{(L,R)}}{M_V^2}$,
$\tilde{C}_1=C_1 \Big(\frac{{\cal M}_1^{(\nu)}/(m_e R)}{M_{GT}-
\alpha_2 M_F}
\Big)^2$.

For the definition of the $C_n$ see \cite{74} and for
the calculation 
of the
matrix element ${\cal M}_{1}^{(\nu)}$ see \cite{hir96a}.
This allows to deduce information on leptoquark masses and leptoquark--Higgs
couplings (see section 3.2).

\begin{figure}
\parbox{10cm}{
\epsfxsize10cm
\epsffile{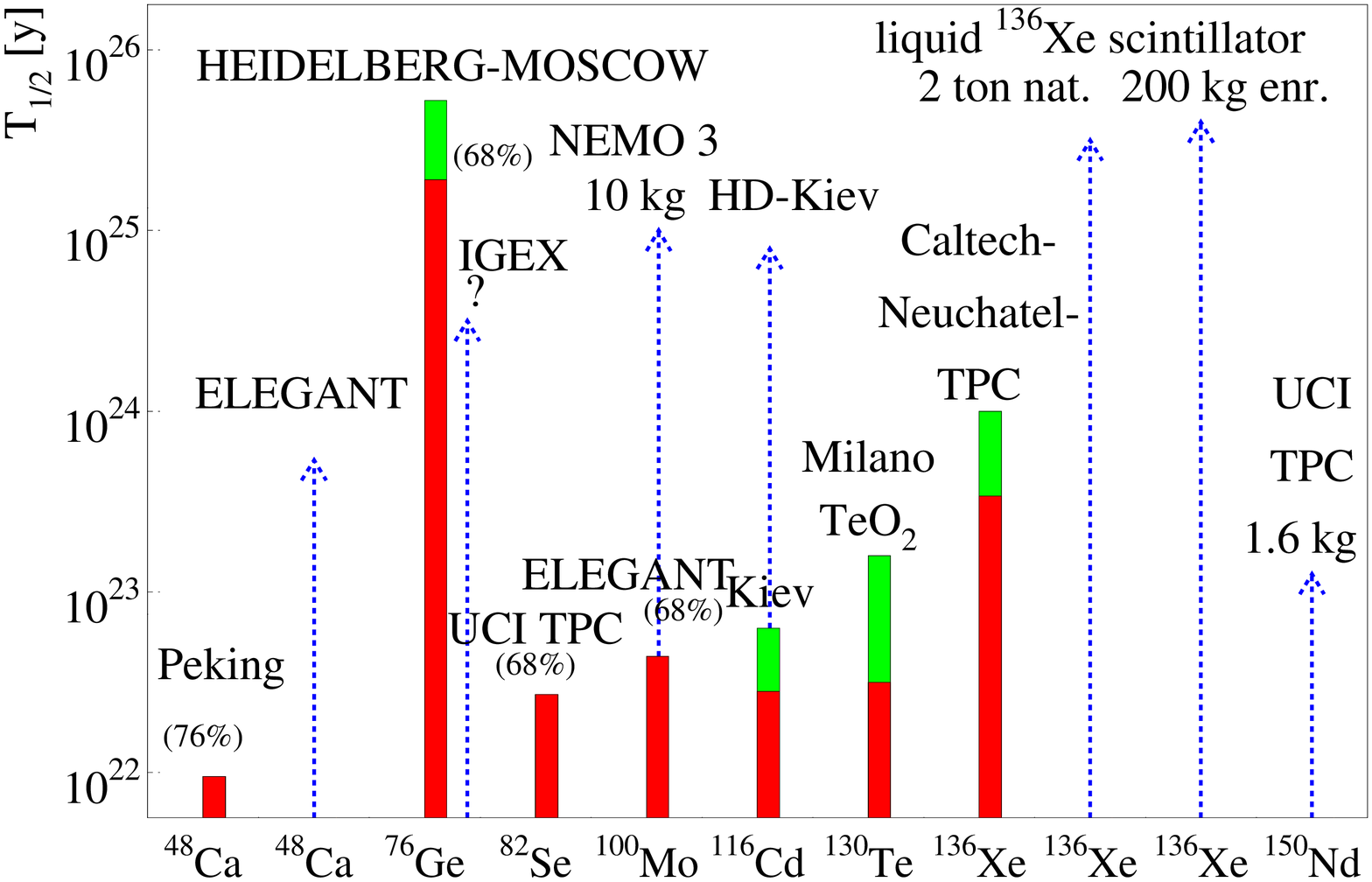}
}
\parbox{10cm}{
\epsfxsize10cm
\epsffile{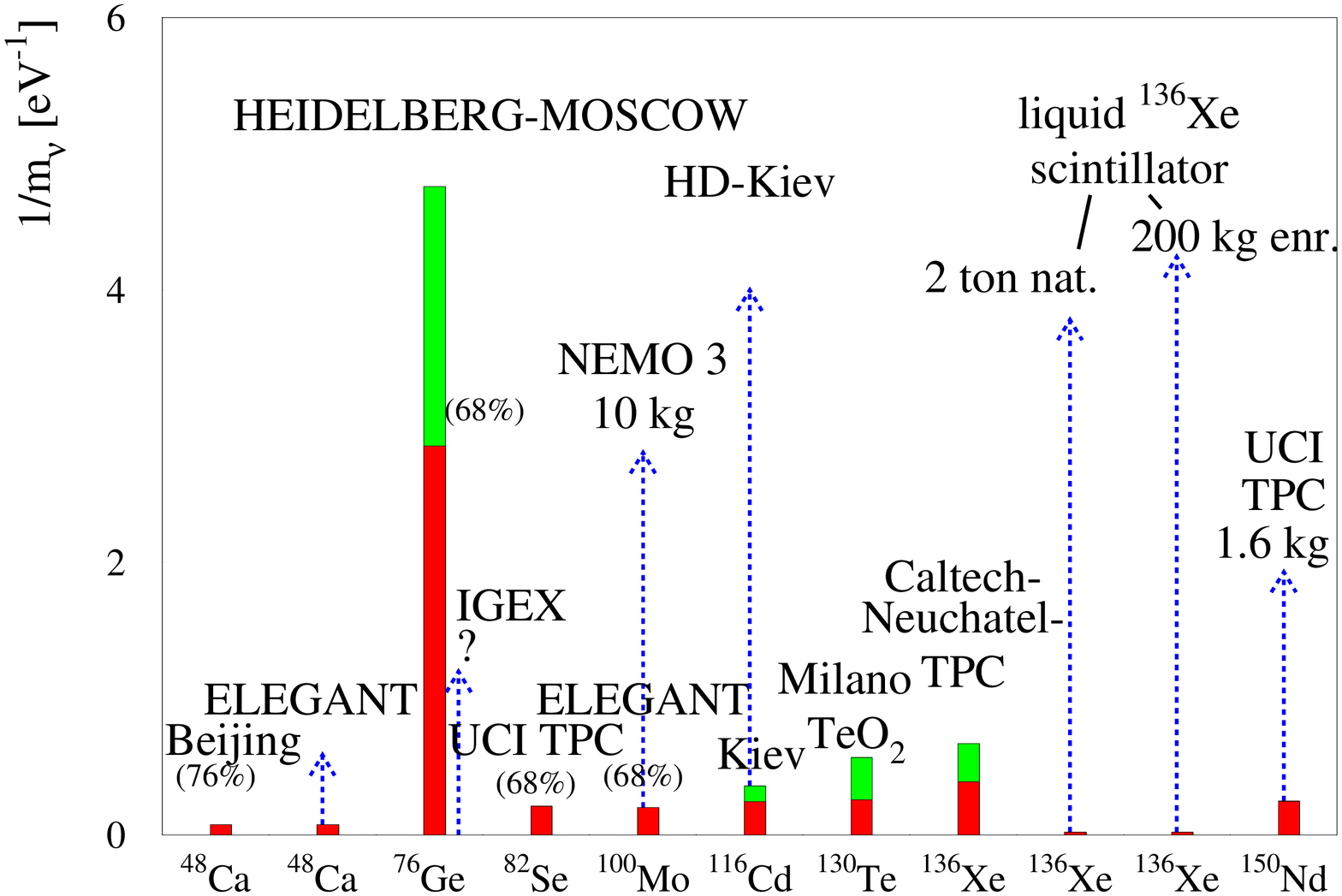}
}

{\bf Fig. 12} {\it 
Present situation, 1996, and expectation for the near future
 until the year 2000 and beyond, of 
the most promising $\beta\beta$-experiments concerning accessible half life 
(a) and neutrino mass limits (b). The filled bars correspond 
to the present status, open bars correspond to `safe' 
expectations for the year 2000 and dashed lines correspond to 
long--term planned or hypothetical 
experiments (from [Kla97]).}
\end{figure}

\section{Double Beta Decay Experiments: Present Status and Results}

\subsection{Present Experimental Status}
Fig. 12 shows an overview over measured 
$0\nu\beta\beta$ half--life limits and deduced mass limits. The largest 
sensitivity for $0\nu\beta\beta$ decay is obtained at present by active source 
experiments (source=detector), in particular $^{76}$Ge \cite{79,HM97,81,Kla97} 
and $^{136}$Xe \cite{82}.
The main reason is that large source strengths can be used (simultaneously
with high energy resolution), in particular when enriched $\beta\beta$
emitter materials are used. Geochemical experiments, though having
contributed important information to double beta decay, have no more
future in the sense that their inherent background from $2\nu\beta\beta$
decay cannot be eliminated.

Only a few of the present most sensitive experiments may probe the 
neutrino mass 
in the next years into
the sub--eV region, the 
Heidelberg--Moscow experiment being the by far 
most advanced and most sensitive one, see Fig. 12b. 
No one of them will pass below $\sim 0.1-0.2$ eV (see section
4.1)
A detailed discussion of the various experimental
possibilities can be found in \cite{1,tren,Kla95b}. 
A useful listing of existing 
data from the 
various $\beta\beta$ emitters is given in \cite{83}.

\subsection{Present limits on beyond standard model parameters}      
The sharpest limits from $0\nu\beta\beta$ decay are presently coming from
the Heidelberg--Moscow experiment \cite{84,79,81,HM97,Kla97}. 
They will be given in the following.
With five 
enriched (86\% of $^{76}$Ge) detectors of a total mass of 11.5 kg 
taking data in the Gran Sasso underground laboratory, and with a background
of at present 0.07 counts/kg year keV, 
the experiment has reached its final 
setup and is now 
exploring the sub--eV range for the mass of the electron neutrino.
Fig. 13 shows the spectrum taken in a measuring time of 31.8  kg y.

\subsection*{Half--life of neutrinoless double beta decay}
The deduced half--life limit for $0\nu\beta\beta$ decay is 

\be{t12}
T^{0\nu}_{1/2} > 1.2 \cdot 10^{25} y \hspace{2mm}(90\% C.L.)
\ee
\be{t13}
\hskip8mm     > 1.9 \cdot 10^{25} y \hspace{2mm}(68 \% C.L.)
\ee
\subsection*{Neutrino mass}

\begin{figure}[!t]
\hspace*{15mm}
\epsfxsize=90mm
\epsfbox{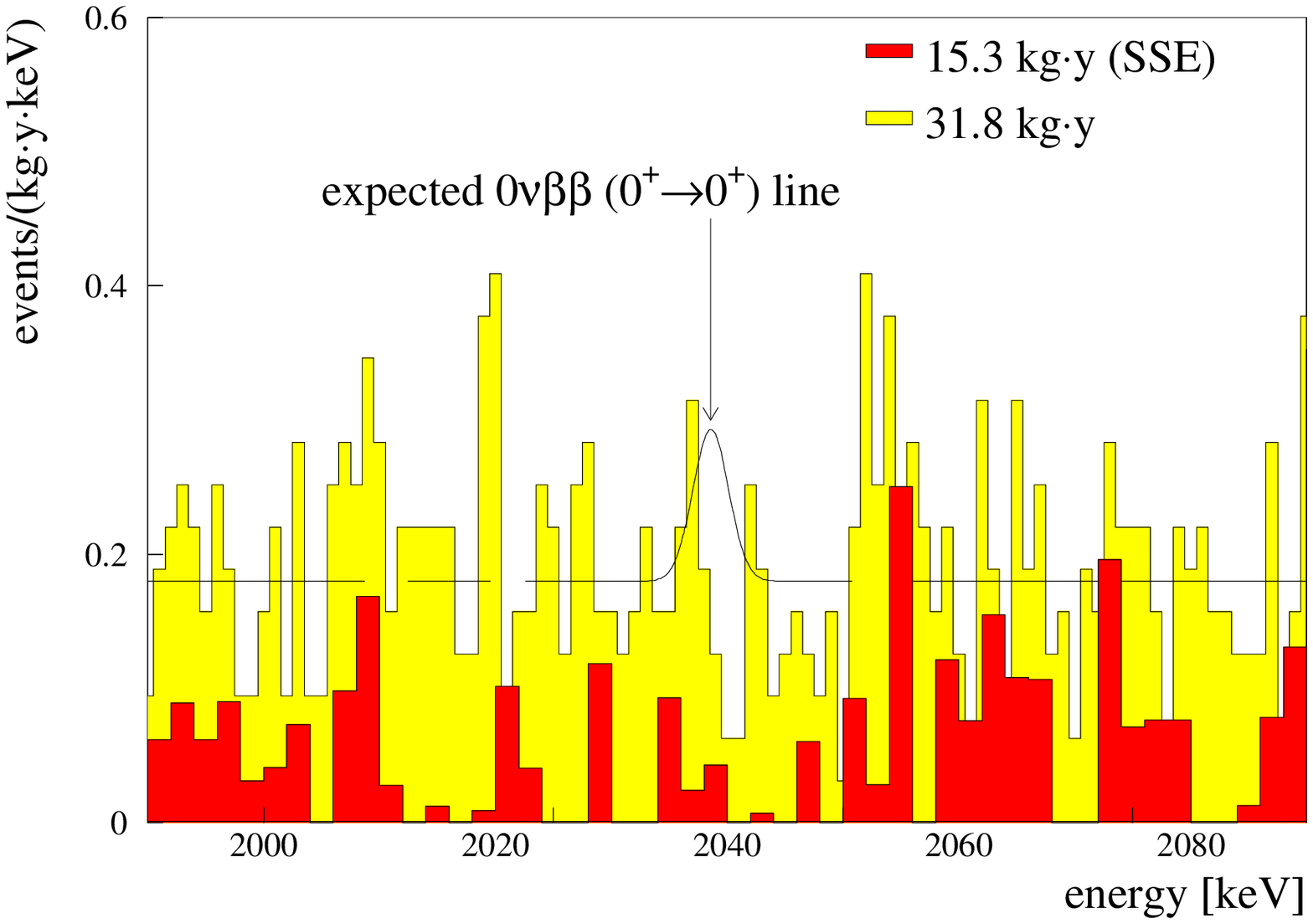}
\vskip-28mm
{\bf Fig. 13} {\it Integral spectrum in the region of interest after 
subtraction of the first 200 days of measurement of each detector, 
leaving 31.8
kg y of measuring time. The solid
 curve corresponds to the signal excluded
with $90 \% C.L.$ It corresponds to $T^{0\nu}_{1/2} > 1.2 \cdot 10^{25}$ y.
The darkened histogram corresponds to data accumulated meanwhile
using a new pulse shape analysis method \cite{Hel96} in a measuring time
of 15.3 kg y.}
\end{figure}

{\it Light neutrinos:} The deduced upper limit of an (effective) electron
neutrino 

Majorana mass is, with the matrix element from \cite{29}
\be{mnu}
\langle m_{\nu} \rangle < 0.45 eV \hspace{2mm}(90\% C.L.)
\ee
\be{mnu2}
\hskip10mm < 0.35 eV \hspace{2mm}(68 \% C.L.)
\ee
This is the sharpest limit for a Majorana mass of the electron neutrino so
far.

{\it Superheavy neutrinos:}     
For a superheavy {\it left}--handed neutrino we deduce 
\cite{79} exploiting the 
mass dependence of the matrix 
element (for the latter 
see \cite{28}) a lower limit 
\be{mh}
\langle m_{H} \rangle \ge 7 \cdot 10^7 GeV
\ee
For a heavy {\it right}--handed neutrino the relation obtained to the mass 
of the 
right--handed W boson is shown in Fig. 14 (see \cite{11}).

\begin{figure}[1t]
\hspace*{5mm}
\epsfxsize=85mm
\hspace*{4mm}
\epsfbox{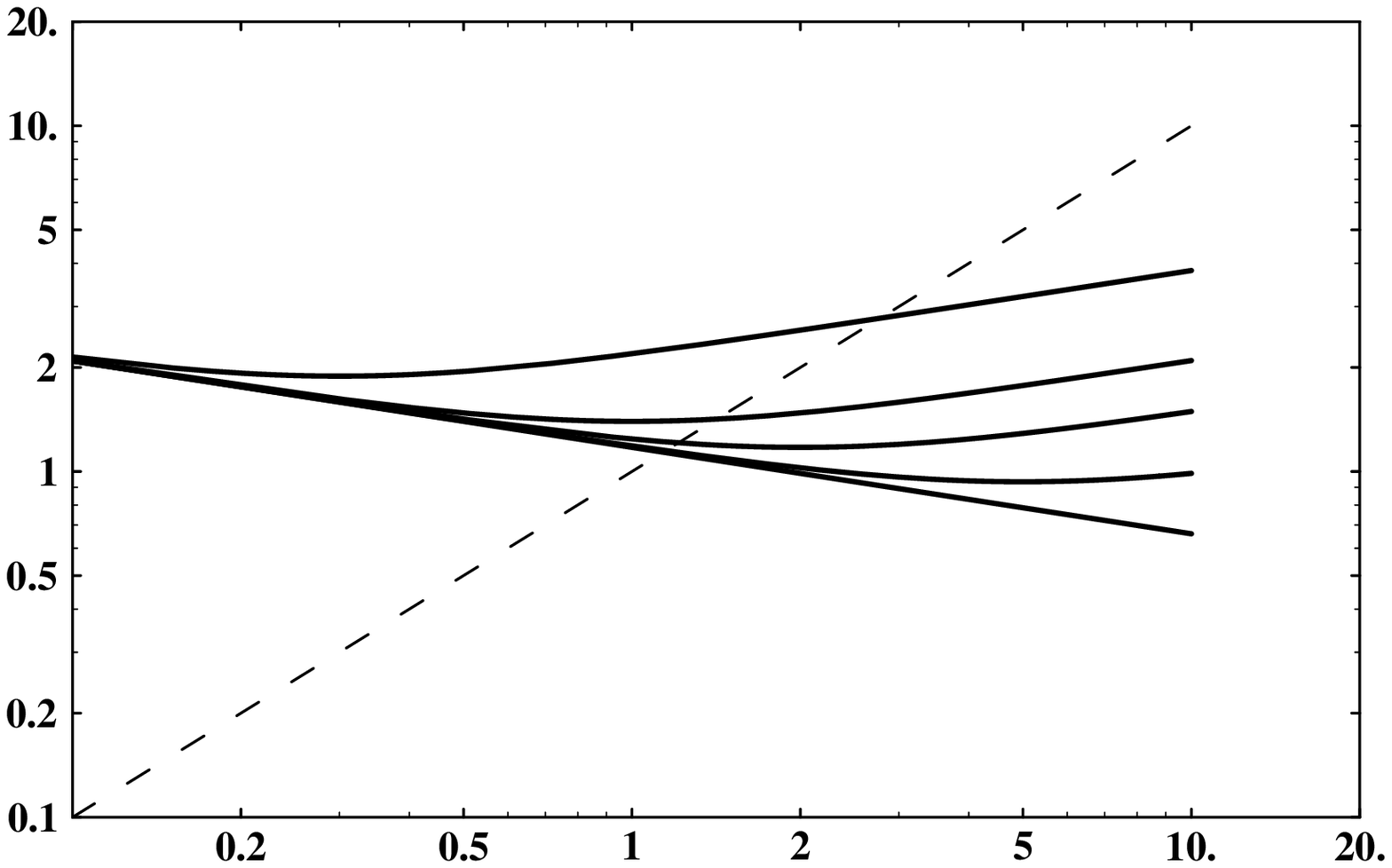}
\hspace*{-2mm}
\vskip-88mm
$m_{W_R}$

\hspace*{-4mm}
[TeV]
\vskip13mm

\vskip33mm
\hskip92mm
$\langle m_N \rangle$ [TeV]
\vskip5mm

\vskip-50mm
\hskip90mm
{\tiny
a)}

\vskip0mm
\hskip90mm
{\tiny
b)}

\vskip0mm
\hskip90mm
{\tiny
c)}

\vskip0mm
\hskip90mm
{\tiny
d)}

\vskip0mm
\hskip90mm
{\tiny
e)}

\vspace*{25mm}
{\bf Fig. 14} {\it Limits on the mass of the right-handed W-boson from 
neutrinoless double beta decay (full lines) and vacuum stability 
(dashed line). The five full lines correspond to the following 
masses of the doubly charged higgs, $m_{\Delta^{--}}$: a) 0.3, 
b) 1.0, c) 2.0, d) 5.0 and e) $\infty$ [TeV] (from \cite{11}).} 

\label{fig4}
\end{figure}

\subsection*{Right--handed W boson}
For the right--handed W boson a lower limit of (Fig. 14)
\be{mwr}
m_{W_R} \ge 1.1 TeV
\ee
is obtained \cite{11}.

\subsection*{SUSY parameters -- R--parity breaking and sneutrino mass}
The constraints on the parameters of the minimal supersymmetric standard model
 with explicit R--parity violation deduced \cite{6,hir96c,hir96} 
from the $0\nu\beta\beta$
half--life limit are more stringent than those from other
low--energy processes and from the largest high energy accelerators
(Fig. 15). The limits are
\be{mwr2}
\lambda^{'}_{111} \leq 3.9 \cdot 10^{-4} \Big(\frac {m_{\tilde{q}}}{100 GeV} 
\Big)^2 \Big(\frac {m_{\tilde{g}}}{100 GeV} \Big)^{\frac{1}{2}}
\ee
with  $m_{\tilde{q}}$ and  $m_{\tilde{g}}$ denoting squark and gluino masses,
respectively, and with the assumption $m_{\tilde{d_R}} \simeq m_{\tilde{u}_L}$.
This result is important for the discussion of new physics in the connection
with the high--$Q^2$ events seen at HERA. It excludes the possibility of 
squarks of first generation (of R--parity violating SUSY) being produced in the
high--$Q^2$ events \cite{Cho97,Alt97,Hir97b}.

We find further \cite{hir96} 
\be{mwr3}
\lambda_{113}^{'}\lambda_{131}^{'}\leq 1.1 \cdot 10^{-7}
\ee
\be{mwr4}
\lambda_{112}^{'}\lambda_{121}^{'}\leq 3.2 \cdot 10^{-6}.
\ee 
For the $(B-L)$ violating sneutrino mass $\tilde{m}_{M}$ the following limits 
are obtained \cite{Hir97a}
\ba{rconv2}
\tilde{m}_M &\leq& 2 \Big(\frac{m_{SUSY}}{100 GeV}\Big)^{\frac{3}{2}}GeV,
\hskip5mm \chi \simeq \tilde{B}\\
\tilde{m}_M &\leq& 11 \Big(\frac{m_{SUSY}}{100 GeV}\Big)^{\frac{7}{2}}GeV,
\hskip5mm \chi \simeq \tilde{H}
\ea
for the limiting cases that the lightest neutralino is a pure Bino $\tilde{B}$,
as suggested by the SUSY solution of the dark matter problem \cite{Jun96},
or a pure Higgsino. Actual values for $\tilde{m}_M$ for other choices of the
neutralino composition should lie in between these two values.
 
Another way to deduce a limit on the `Majorana' sneutrino mass $\tilde{m}_M$
is to start from the experimental neutrino mass limit, since the sneutrino 
contributes to the Majorana neutrino mass $m_M^{\nu}$ at the 1--loop level
proportional to $\tilde{m}^2_M$. This yields under some assumptions
\cite{Hir97a}
\be{ufu}
\tilde{m}_{M_{(i)}} \leq (60-125) \Big(\frac{m^{exp}_{\nu(i)}}{1 eV}\Big) ^{1/2}
MeV
\ee

Starting from the mass limit determined for the electron neutrino  by 
$0\nu\beta\beta$ decay this leads to 
\be{fu}
\tilde{m}_{M_{(e)}} \leq 22 MeV    
\ee
This result is somewhat dependent on neutralino masses and mixings. 
A non--vanishing `Majorana' sneutrino mass would result in new processes 
at future colliders, like sneutrino--antisneutrino oscillations.
Reactions at the Next Linear Collider (NLC) like the SUSY analog to inverse
neutrinoless double beta decay $e^-e^-\rightarrow \chi^-\chi^-$ (where $\chi^-$
denote charginos) or single sneutrino production, e.g. by 
$e^-\gamma \rightarrow \tilde{\nu}_e \chi^-$ could give information on the 
Majorana sneutrino mass, also. This is discussed by \cite{Hir97,Hir97a} and
by \cite{Hir97b,Kol97} in these proceedings. A conclusion is that future
accelerators can give information on second and third generation sneutrino
Majorana masses, but for first generation sneutrinos cannot compete with
$0\nu\beta\beta$--decay.

\begin{figure}[!t]
\vspace*{-60mm}
\hspace*{-20mm}
\epsfxsize=150mm
\epsfbox{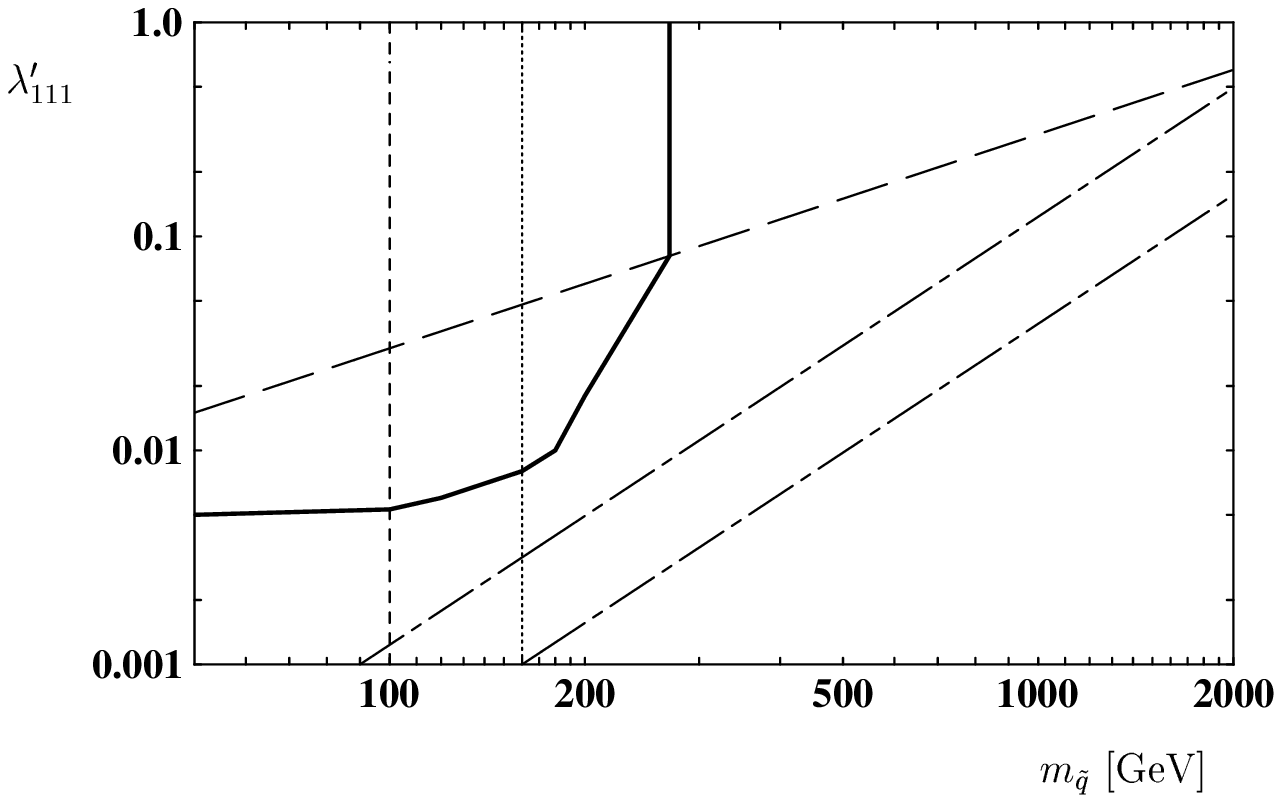}
\vspace*{-93mm}
\noindent
{\bf Fig. 15} {\it Comparison of limits on the R--parity violating 
MSSM parameters 
from different experiments in the $\lambda'_{111}$--$m_{\tilde{q}}$ plane. 
The 
dashed line is the limit from charged current universality according to 
\cite{113}. The vertical line is the limit from the data of Tevatron
\cite{114}. The thick full line is the region which might be explored by HERA
\cite{115}. The two dash--dotted lines to the right are the limits obtained
from the half--life limit for $0\nu\beta\beta$ decay of $^{76}$Ge, for
gluino masses of (from left to right) $m_{{\tilde{g}}}=$1TeV and 100 GeV,
respectively. The regions to the upper left of the lines are forbidden.
(from [Hir95])}
\end{figure}

\subsection*{Compositeness}
Evaluation of the $0\nu\beta\beta$ half--life limit assuming 
exchange of excited
Majorana neutrinos $\nu^*$ yields for the mass of the
 excited neutrino a lower bound of \cite{Pan97,Tak97}. 
\be{exc}
m_{N} \geq 3.4 m_W
\ee
for a coupling of order \cal{O}(1) and $\Lambda_c \simeq m_N$. Here,
$m_W$ is the W--boson mass.

\subsection*{Leptoquarks}
Assuming that either scalar or vector leptoquarks contribute
to $0\nu\beta\beta$ decay, the following constraints on the 
effective LQ parameters  (see section 2.7) can be derived \cite{hir96a}:
\ba{dbd_constraint}
\epsilon_I \leq 2.8 \times 10^{-9}
\left(\frac{M_I}{100\mbox{GeV}}\right)^2, \\
\alpha_I^{(L)} \leq 3.5 \times 10^{-10}
\left(\frac{M_I}{100\mbox{GeV}}\right)^2, \\
\alpha_I^{(R)} \leq 7.9  \times 10^{-8}
\left(\frac{M_I}{100\mbox{GeV}}\right)^2.
\ea

Here, different effective LQ couplings have been introduced. They are 
defined as: 
\ba{coeff}
\epsilon_I= 
2^{-\eta_I}
\left[\lambda_{I_1}^{(L)}\lambda_{\tilde{I}_{1/2}}^{(R)}
\left(\theta_{43}^I(Q_I^{(1)})
+ \eta_I \sqrt{2} \theta_{41}^I(Q_I^{(2)})\right) \right. \nn
\ea
\be{coeff2}
\left. \hskip5mm
 -\lambda_{I_0}^{(L)}\lambda_{\tilde{I}_{1/2}}^{(R)}
\theta_{13}^I(Q_I^{(1)})\right] 
\ee
\be{coeff3}
\alpha_I^{(L)}= 
\frac{2}{3 + \eta_I} \lambda_{I_{1/2}}^{(L)}\lambda_{I_1}^{(L)}
\theta_{24}^I(Q_I^{(2)})
\ee
\be{coeff4}
\alpha_I^{(R)} = \frac{2}{3 + \eta_I}
\lambda_{I_0}^{(R)} \lambda_{\tilde{I}_{1/2}}^{(R)}
\theta_{23}^I(Q_I^{(1)}).
\ee
$\eta_{S,V}= 1,-1$ for scalar and vector LQs.
$\theta_{kn}^I(Q)$ is a mixing parameter defined by
\ba{mp}
\theta_{kn}^I(Q) = \sum_{l} {\cal N}_{kl}^{(I)}(Q) {\cal N}_{nl}^{(I)}(Q)
\left(\frac{M_I}{M_{I_l}(Q)}\right)^2,
\ea
where ${\cal N}^{(I)}(Q)$ are mixing matrix elements which diagonalize 
the LQ mass matrices for 
the scalar $I= S$ and vector $I= V$ LQ fields with electric charges
$Q = -1/3,-2/3$, for complete definitions see \cite{hir96a}. 
Common mass scales $M_S$ of scalar and $M_V$ of vector LQs are 
introduced for convenience.

Since the LQ mass matrices appearing in $0\nu\beta\beta$ 
decay are ($4\times4$) 
matrices \cite{hir96a}, it is difficult to solve their diagonalization 
in full generality algebraically. However, if one assumes that only 
one LQ-Higgs coupling is present at a time, the (mathematical) problem is 
simplified greatly and one can deduce from, for example, 
eq. \rf{dbd_constraint} that either 
the LQ-Higgs coupling must be smaller than $\sim 10^{-(4-5)}$ or there can not 
be any LQ with e.g. couplings of electromagnetic strength with masses below
$\sim 250 GeV$. These bounds from $\beta\beta$ decay are of interest in 
connection with recently discussed evidence for new physics from HERA
\cite{Hew97,Bab97,Kal97,Cho97}. Assuming that actually leptoquarks have
been produced at HERA, double beta decay (the Heidelberg--Moscow experiment)
would allow to fix the leptoquark--Higgs coupling to a few $10^{-6}$
\cite{Hir97b}. It may be noted, that after the first 
consideration of leptoquark--Higgs coupling in \cite{hir96a} recently
Babu et al. \cite{Bab97b} noted that taking into account 
leptoquark--Higgs coupling reduces the leptoquark mass lower bound deduced
by TEVATRON -- making it more consistent with the value of 200 GeV 
required by 
HERA.

\subsection*{Half--life of $2\nu\beta\beta$ decay}
The Heidelberg--Moscow experiment 
produced for the first time a high statistics $2\nu\beta\beta$
spectrum ($\sim 20000$ counts, to be compared with the 40 counts on which the 
first detector observation of $2\nu\beta\beta$ decay by \cite{Ell87} 
(for the decay of $^{82}$Se) had to rely. The deduced half--life is \cite{HM97}
\be{t2v}
T^{2\nu}_{1/2} = (1.77^{+0.01}_{-0.01}(stat.)^{+0.13}_{-0.11}(syst.)) 
\cdot 10^{21} y
\ee
This result brings $\beta\beta$ research for the first time into the region 
of `normal' nuclear spectroscopy and allows for the first time statistically
reliable investigation of Majoron--accompanied decay modes.

\subsection*{Majoron--accompanied decay}
From simultaneous fits of
the $2\nu$ spectrum and one selected Majoron mode, experimental limits 
for the half--lives of the decay modes of
the newly introduced Majoron models \cite{72} are given
for the first time \cite{71,HM96}.

The small matrix elements and phase spaces for these modes 
\cite{71,75} already determined that these 
modes by far cannot be seen
in experiments of the present sensivity if we assume typical values for the
neutrino--Majoron coupling constants around $\langle g \rangle = 10^{-4}$
(see table 3). 

\section{Double Beta Experiments: Future Perspectives -- 
the GENIUS Project}
\subsection{The known experiments and proposals}
Figs. 12a,b show in addition to the present status 
the future perspectives of the main existing 
$\beta\beta$ decay experiments and includes some ideas for the future
which have been published.
The HEIDELBERG--MOSCOW experiment will probe the neutrino mass within 5
years down to the order of 0.1 eV. This limit will be reached 
taking into account the current 
background of 0.1 counts/kg y keV in the $0\nu\beta\beta$ region and a further 
reduction by digital pulse shape analysis (DPSA) (see \cite{HM97}). 
The best presently existing limits besides the HEIDELBERG-MOSCOW 
experiment (filled bars in Fig. 12),
have been obtained with the isotopes: 
$^{48}$Ca \cite{87}, 
$^{82}$Se \cite{88}, 
$^{100}$Mo \cite{89}, 
$^{116}$Cd \cite{90},
$^{130}$Te \cite{91},
$^{136}$Xe \cite{92} and
$^{150}$Nd \cite{93}.
These and other double beta decay setups presently under construction or 
partly in operation 
such as NEMO \cite{94,Bar97}, 
the Gotthard $^{136}$Xe TPC experiment \cite{95}, 
the $^{130}$Te cryogenic experiment \cite{91},
a new ELEGANT $^{48}$Ca experiment using 30 g of $^{48}$Ca \cite{96},
a hypothetical experiment with an improved UCI TPC \cite{93} assumed to use 1.6 kg of $^{136}$Xe, 
 etc., will not reach or exceed the $^{76}$Ge limits.
The goal 0.3 eV aimed at for the year 2004 by the NEMO experiment 
(see \cite{98,Bar97}
and Fig. 12) 
may even be very optimistic if claims about the effect of proton-neutron 
pairing on the $0\nu\beta\beta$ nuclear matrix elements by 
\cite{Pan96} will
turn out to be true, and also if the energy resolution will not be improved
considerably 
(see Fig. 1 in \cite{83}). 
Therefore, the conclusion given by \cite{Bed97c} concerning the
future SUSY potential of NEMO has no serious basis. 
As pointed out by Raghavan \cite{97}, even use of an 
amount of about 200 kg of 
enriched $^{136}$Xe or 2 tons of natural Xe added to the scintillator of the 
KAMIOKANDE detector 
or similar amounts added to BOREXINO (both primarily devoted to solar neutrino 
investigation) 
would hardly lead to a sensitivity larger 
than the present $^{76}$Ge experiment.
This idea is going to be realized at present by the KAMLAND
experiment \cite{Suz97}. 
An interesting future candidate was for some time a $^{150}$Nd bolometer 
exploiting the relatively large
phase space of this nucleus (see \cite{93}).
The way outlined by \cite{99} proposing a TPC filled with 1 ton of 
liquid enriched $^{136}$Xe 
and identification of the daughter by laser fluorescence seems
 not be feasible in
 a straight-forward way. However, another way of using liquid $^{136}$Xe
may be more promising \cite{CLI96}.

It is obvious that, from the experiments and proposals, 
the HEIDELBERG-MOSCOW 
experiment will give the sharpest limit for the electron neutrino 
 mass for the next decade. It is also obvious from Fig. 12 that {\it none}
of the present experimental approaches, or plans or even vague ideas has a
chance to surpass the border of 0.1 eV for the neutrino mass to lower values
(see also \cite{Nor97}).
At present there is only one way visible to reach the domain of lower 
neutrino masses,
suggested by the author of this report and meanwhile investigated in some
detail concerning its experimental realization and and physics potential in
\cite{Kla97d,Hel97}.

\subsection{Genius -- A Future Large Scale Double Beta and Dark Matter
Experiment}

The idea of GENIUS is to use a large amount of `naked' enriched 
{\bf GE}rmanium detectors in liquid {\bf NI}trogen as shielding in an 
{\bf U}nderground {\bf S}etup. Use of 1 (in an extended version 10) tons of
enriched $^{76}$Ge will increase the source strength largely, removing all
material from the vicinity of the detectors and shielding by liquid nitrogen
will lead to a drastic background reduction compared to the present level.
Using Ge detectors in liquid nitrogen has been discussed already earlier 
\cite{Heu95}.
That Ge detectors can be operated in liquid nitrogen has been demonstrated
recently in the Heidelberg low level laboratory \cite{Hel97}. The natural
site for GENIUS would be the Gran Sasso underground laboratory. The cost of 
the project would be a minor fraction of detectors prepared for LHC physics 
as CMS or ATLAS. We give in the next two subsections some results of Monte 
Carlo simulations of the setup \cite{Hel97} and some estimates of the physics 
potential \cite{Kla97d}. 

\begin{figure}[!t]
\epsfxsize10cm
\epsffile{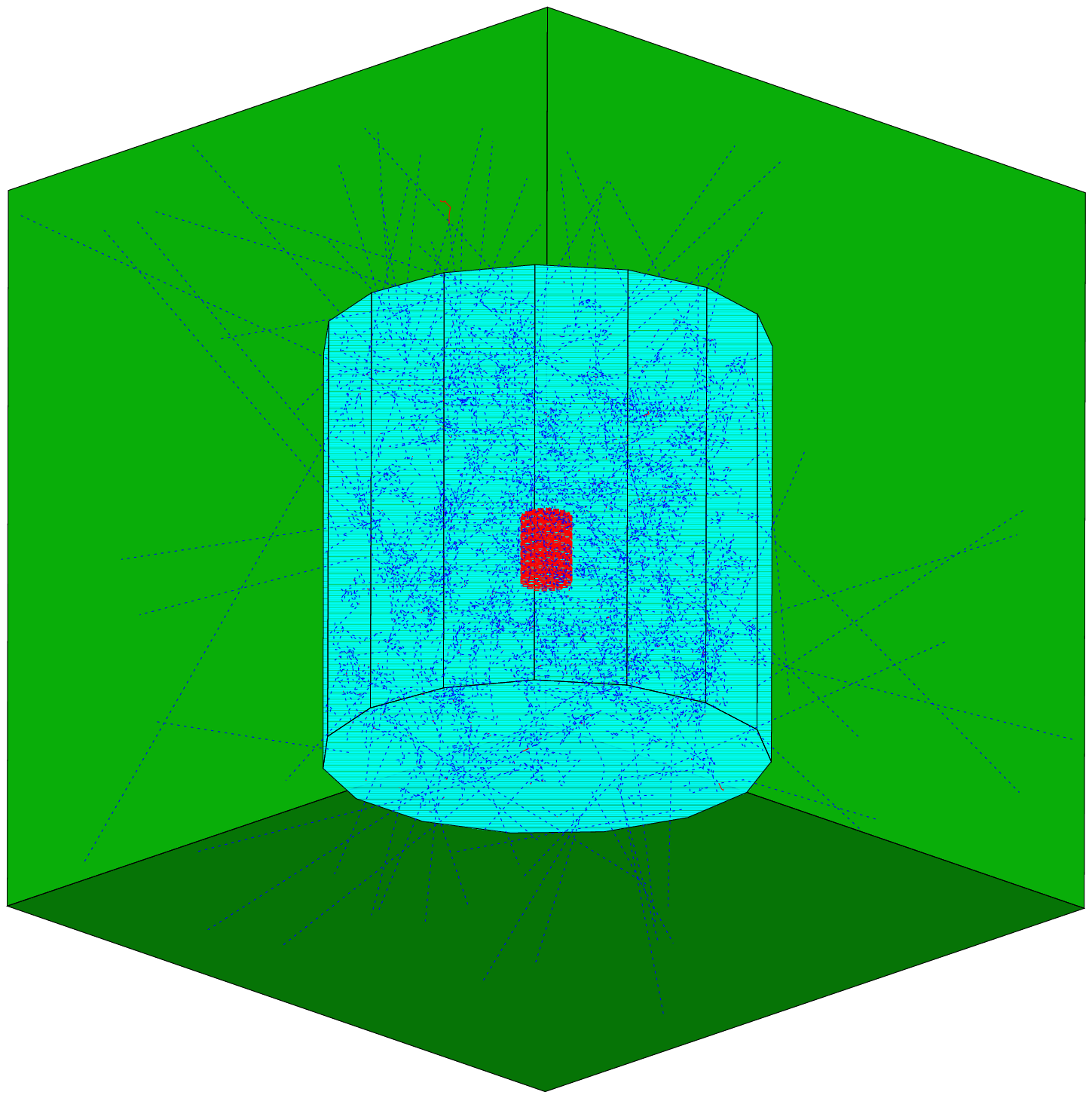}
{\bf Fig. 16}
{\it Simplified model of the GENIUS experiment: 288 enriched
$^{76}$Ge detectors
with a total of one ton mass in the center of a 9 m high liquid nitrogen 
tank with 9 m diameter; GEANT Monte Carlo simulation of 1000  2.6 MeV 
photons randomly
distributed in the nitrogen is also shown.}
\end{figure}

\subsubsection{Realization and Sensitivity of GENIUS}
A simplified model of GENIUS is shown in Fig. 16 consisting of about 300
enriched $^{76}Ge$ detectors with a total of one ton mass in the center of a 
9 m high liquid nitrogen tank with 9 m diameter. Figs. 17, 18 show the results
of Monte Carlo simulations, using the CERN GEANT code, of the background
\cite{Hel97}, starting from purity levels of the nitrogen being in general 
an order of magnitude less stringent than those already achieved in the CTF for the
BOREXINO experiment. The influence of muons penetrating the Gran Sasso
rock on the background can be reduced comfortably through coincidences
between the Germanium detectors from the muon induced showers. The count rate 
in the region of interest for neutrinoless double beta decay is 0.04 
counts/ keV $\cdot$ y $\cdot$ ton (Fig. 17). Below 100 keV the background 
count rate is about 10 counts/keV $\cdot$ y $\cdot$ ton.
Two neutrino double beta decay would dominate the spectrum with $4\cdot 10^6$
events per year.

\begin{figure}[!t]
\epsfxsize10cm
\epsffile{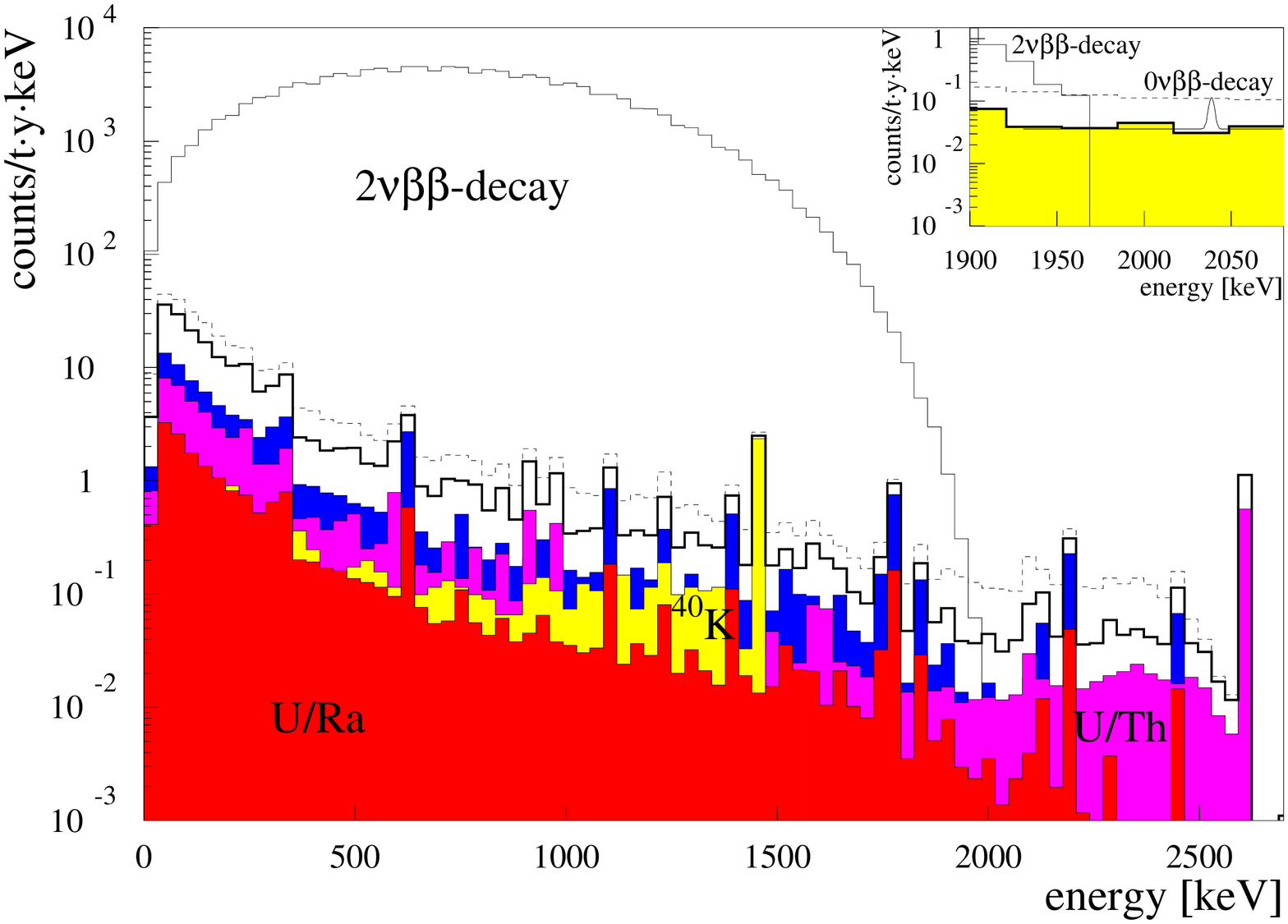}
{\bf Fig. 17}
{\it Monte Carlo simulation of the background of GENIUS.
Simulations of U/Ra, U/Th and $^{40}$K
(shaded), $^{222}$Rn (black histogram) activities in the liquid
nitrogen; the sum of the activities is shown with anticoincidence
between the 288 detectors (thick line) and without (dashed line); the
\tnbb -decay dominates the spectrum with 4 million events per year.
(from [Hel97])}
\end{figure}

\begin{figure}[!t]
\epsfxsize10cm
\epsffile{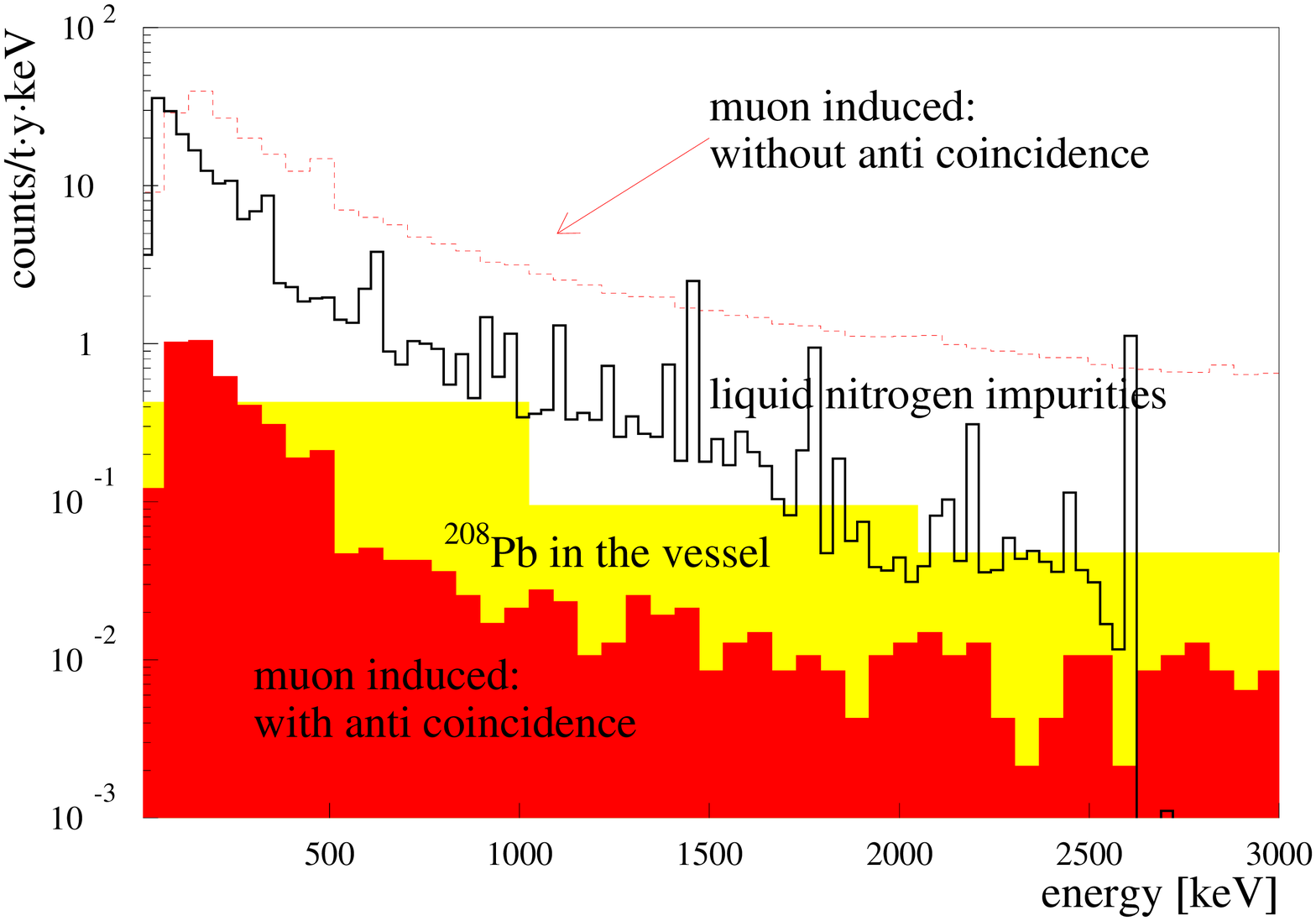}
{\bf Fig. 18}
{\it Background from outside the nitrogen: 200 GeV muons induced events 
(dashed line) and single hit events (filled histogram); decay of
$^{208}$Tl in the steel vessel (light shaded histogram) and the background 
originating from the nitrogen impurities for comparison (thick line)
(from [Hel97])}
\end{figure}

Starting from these numbers, a lower half--life limit of
\be{ofo}
T_{1/2}^{0\nu} \geq 5.8 \cdot 10^{27} \hskip8mm (68 \% C.L.)
\ee
can be reached within one year of measurement (following the 
highly conservative procedure for
analysis recommended by \cite{46}, which has been used also in the 
derivation of the results given in section 3.2, but is not used in the analysis 
of several other $\beta\beta$ experiments). This corresponds 
-- with the matrix
elements of \cite{29} -- to an upper limit on the neutrino mass of

\be{zofo}
\langle m_{\nu}\rangle \leq 0.02 eV \hskip8mm (68 \% C.L.)
\ee 

Figure 19 shows the obtainable limits on the neutrino mass in the case
of zero background. This assumption might be justified since our assumed
impurity concentrations
 are still more conservative than proved already now for example by Borexino. 
The final sensitivity of the experiment can be defined by the limit, 
which would
be obtained after 10 years of measurement. For the one ton experiment this 
would be:
\medskip
\be{lim3}
T^{0\nu}_{1/2}\quad\ge\quad 6.4 \cdot 10^{28}\, y \quad \mbox{(with
68\% C.L.)}\
\ee
\medskip
and
\medskip
\be{lim4}
\langle m_{\nu}\rangle \quad \le 0.006 eV \quad \mbox{(with 68\% C.L.)}\
\ee
\medskip
\begin{figure}[!t]
\epsfxsize10cm
\epsffile{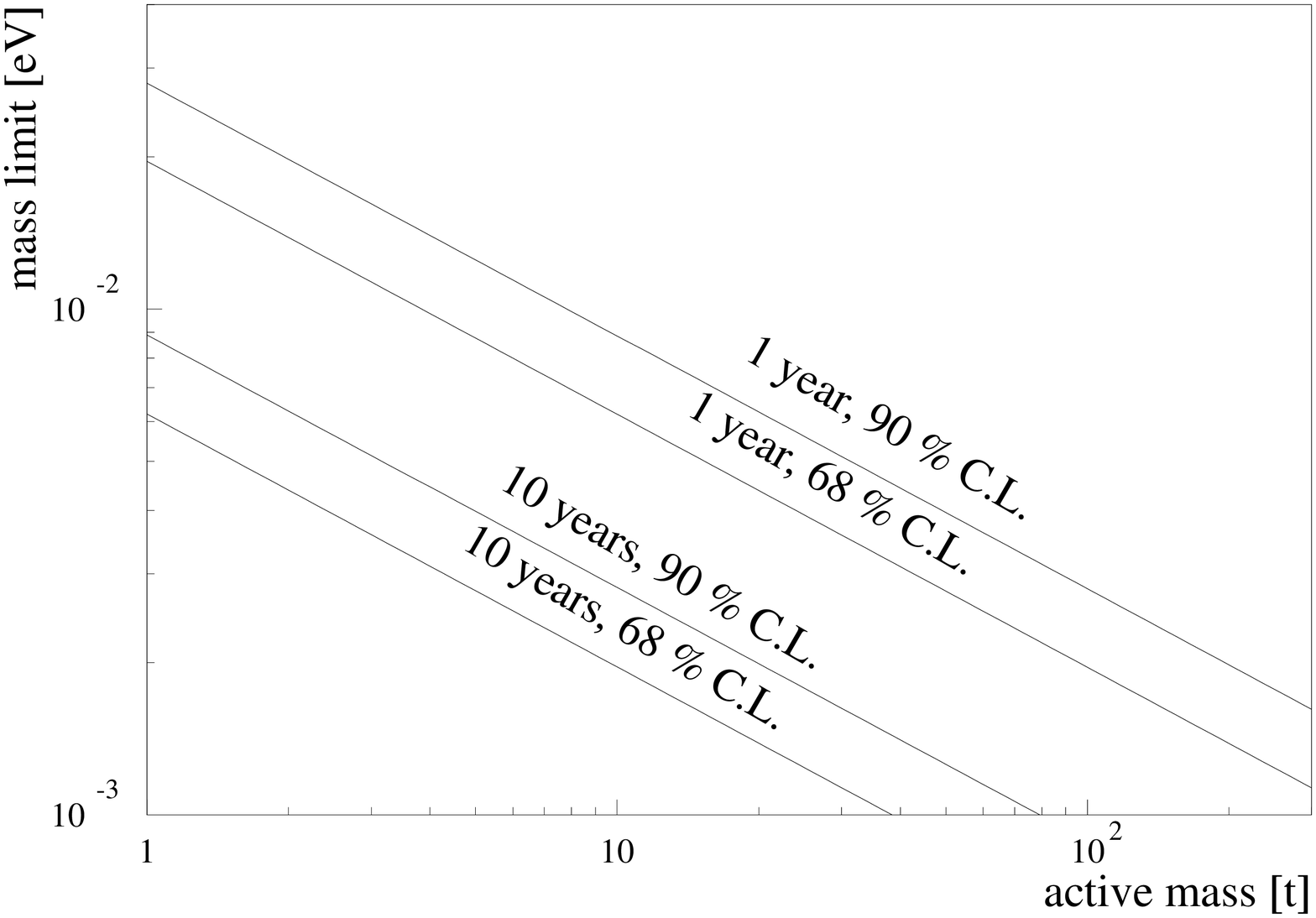}
{\bf Fig. 19}
{\it Mass limits on Majorana neutrino mass after one and ten years 
measuring time as function of the active detector mass; zero background 
is assumed (from [Hel97])}
\end{figure}
The ultimate experiment could test the $0\nu\beta\beta$ half life of
$^{76}$Ge up to a limit of 5.7$\cdot$10$^{29}$y 
and the neutrino mass down to 2$\cdot$10$^{-3}$eV using
10 tons of enriched Germanium.

\subsubsection{The Physics Potential of GENIUS} \hspace*{50mm}\\
{\it Neutrino mass textures and neutrino oscillations:}
GENIUS will allow a large step in sensitivity for probing the neutrino mass. 
It will allow to probe the neutrino mass down to 10$^{-(2-3)}$ eV, and thus 
surpass  the existing neutrino mass experiments by a factor of 50-500.
GENIUS will test the structure of the neutrino mass matrix and thereby also 
neutrino oscillation parameters 
\footnote{The double beta observable, the effective neutrino mass 
(eq. 10), can be expressed 
in terms of the usual neutrino oscillation parameters, once an assumption
on the ratio of $m_1/m_2$ is made. E.g., in the simplest two--generation case
\be{osc}
\langle m_{\nu} \rangle=|c_{12}^2 m_1 + s_{12}^2 m_2 e^{2 i \beta}|,
\ee
assuming CP conservation, i.e. $e^{2 i \beta}=\eta=\pm 1$, and 
$c_{12}^2 m_1 << \eta s_{12}^2 m_2$,
\be{osc2}
\Delta^2_{m_{12}}\simeq m_2^2=\frac{4 \langle m_{\nu} \rangle^2}{1-\sqrt{1-
sin^2 2 \theta}}
\ee
A little bit more general, keeping corrections of the order $(m_1/m_2)$ 
one obtains
\be{osc3}
m_2=\frac{ \langle m_{\nu} \rangle}{|(\frac{m_1}{m_2})+\frac{1}{2}
(1-\sqrt{1-sin^2 2 \theta})(\pm 1 - (\frac{m_1}{m_2}))|}.
\ee
For the general case see \cite{Kla97d}.}
superior in sensitivity to the best
proposed dedicated terrestrial neutrino oscillation experiments. Even in the 
first stage GENIUS will confirm or rule out degenerate or inverted neutrino 
mass scenarios, discussed in the literature as possible solutions of current 
hints to finite neutrino masses , and also test the $\nu_e \leftrightarrow
\nu_{\mu}$ hypothesis of the atmospheric neutrino problem. If the $10^{-3}$ eV
level is reached, GENIUS will even allow to test the large angle MSW 
solution of the solar neutrino problem. It will also allow to test the 
hypothesis of a shadow world underlying introduction of a sterile neutrino
mentioned in section 2.1.
The figures 20--24 show some examples of this potential. Fig 20 compares the
potential of GENIUS with the sensitivity of CHORUS/NOMAD and with the
proposed future experiments NAUSIKAA--CERN and NAUSIKAA--FNAL, looking for 
$\nu_e \leftrightarrow \nu_{\tau}$ oscillations, for different assumptions on
$m_1/m_2$. 
\begin{figure}[!t]
\setlength{\unitlength}{1in}                                                 
\begin{picture}(5,2)
\put(0.0,0.5){\includegraphics{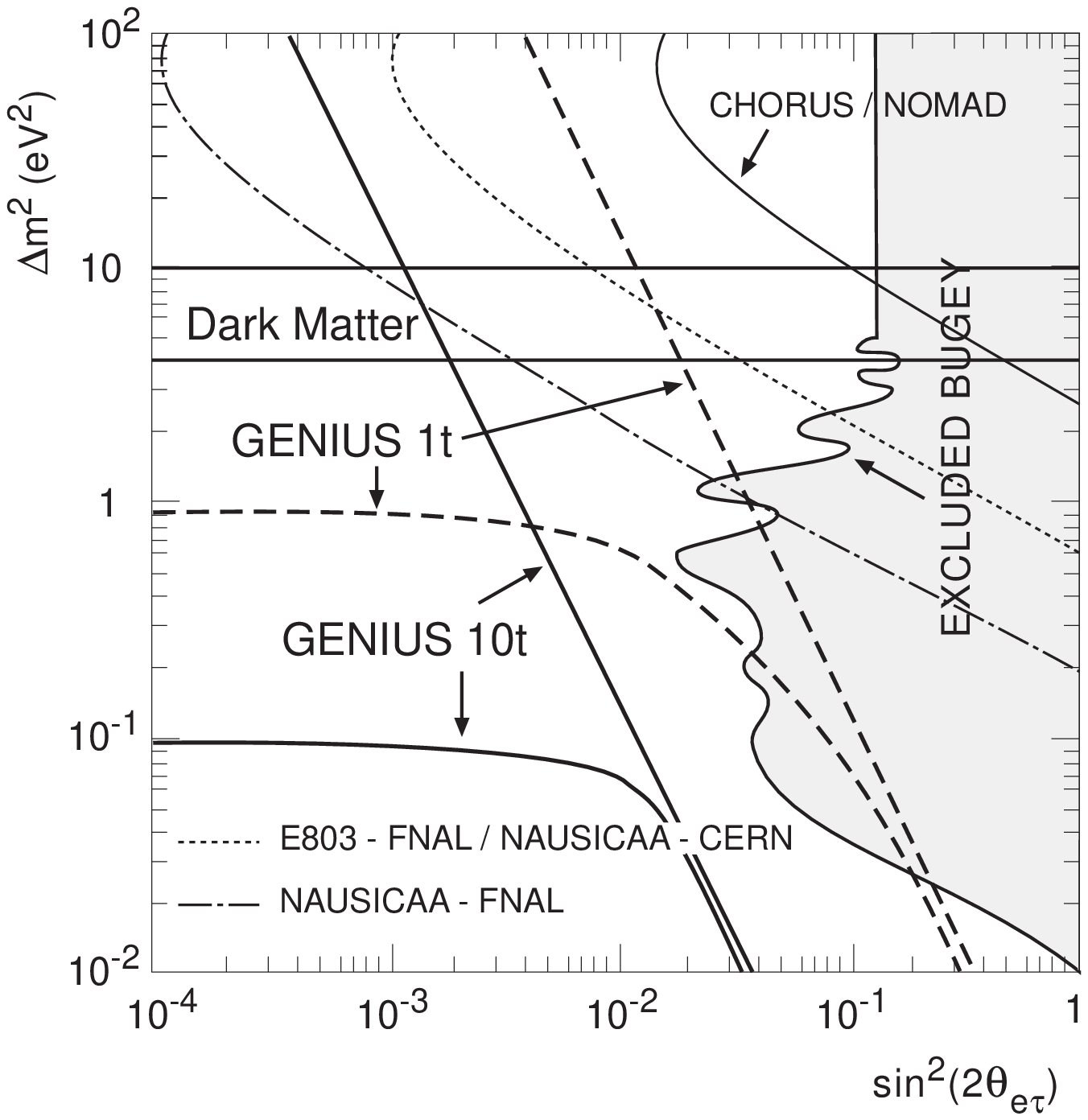}}
\end{picture}                                                                 

\vspace*{40mm}
{\bf Fig. 20}
{\it Current limits and future experimental sensitivity 
on $\nu_e - \nu_{\tau}$ oscillations. The shaded area is currently 
excluded from reactor experiments. The thin line is the estimated 
sensitivity of the CHORUS/NOMAD experiments. The dotted and dash-dotted 
thin lines are sensitivity limits of proposed accelerator experiments, 
NAUSICAA and E803-FNAL [Gon95]. 
The thick lines show the sensitivity of GENIUS (broken line: 
1 t, full line: 10 t), for two examples of mass ratios. The straight lines are 
for the strongly hierarchical case (R=0), while the lines bending to the left 
assume R=0.01.  
(from [Kla97d])}
\end{figure}

\begin{figure}[!t]
\vskip0mm 
\hskip0mm
\epsfxsize=100mm
\epsfysize=120mm
\epsfbox{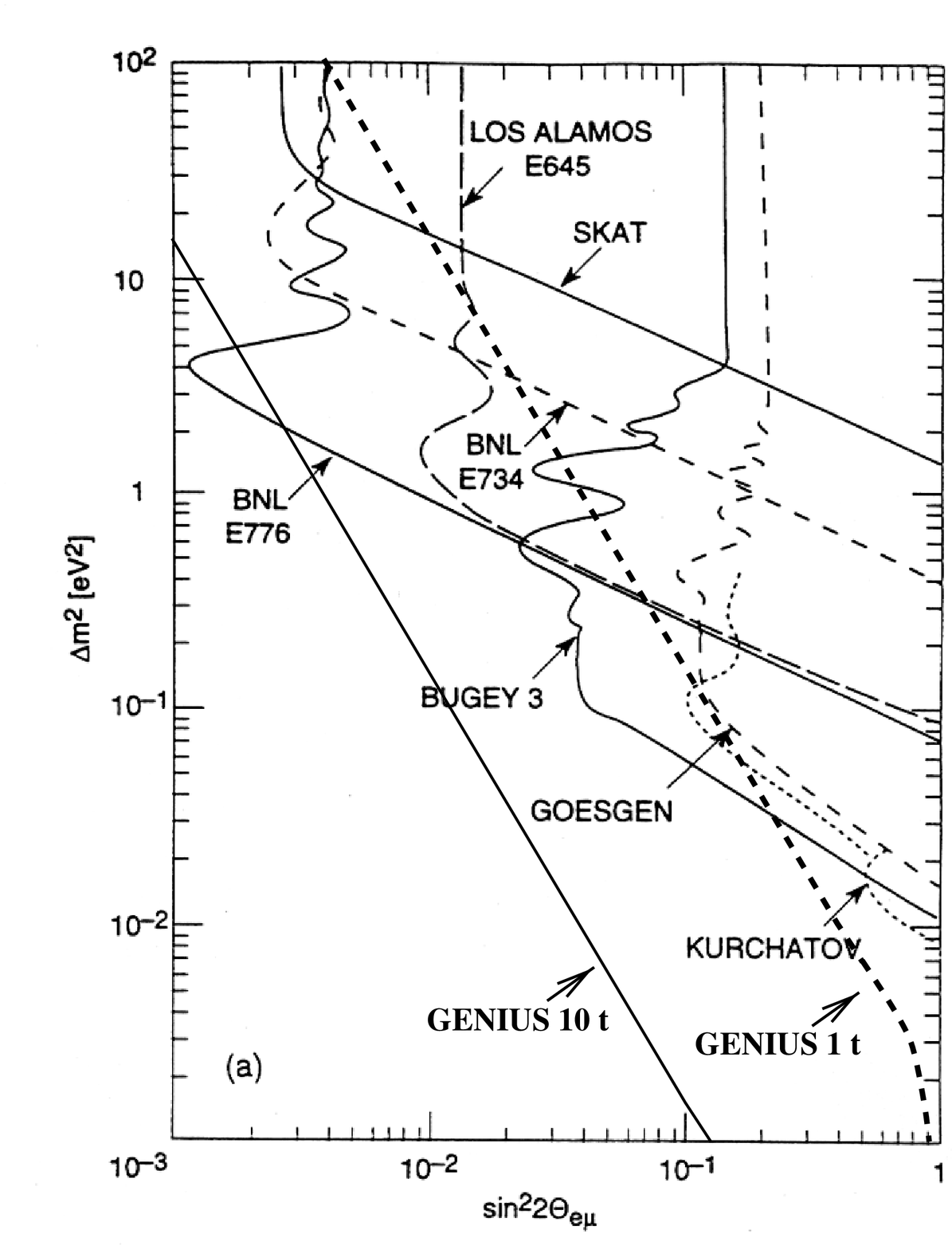}

\vskip0mm 
{\bf Fig. 21}
{\it Current limits on $\nu_e - \nu_{\mu}$ oscillations. 
Various existing experimental limits from reactor and accelerator 
experiments are indicated, as summarized in ref. [Gel95]. In addition, 
the figure shows the expected sensitivities for GENIUS with 1 ton (thick 
broken line) and GENIUS with 10 tons (thick, full line) 
(from [Kla97d])}
\end{figure}

\begin{figure}[!t]
\vskip5mm 
\hskip20mm
\epsfysize=120mm
\epsfbox{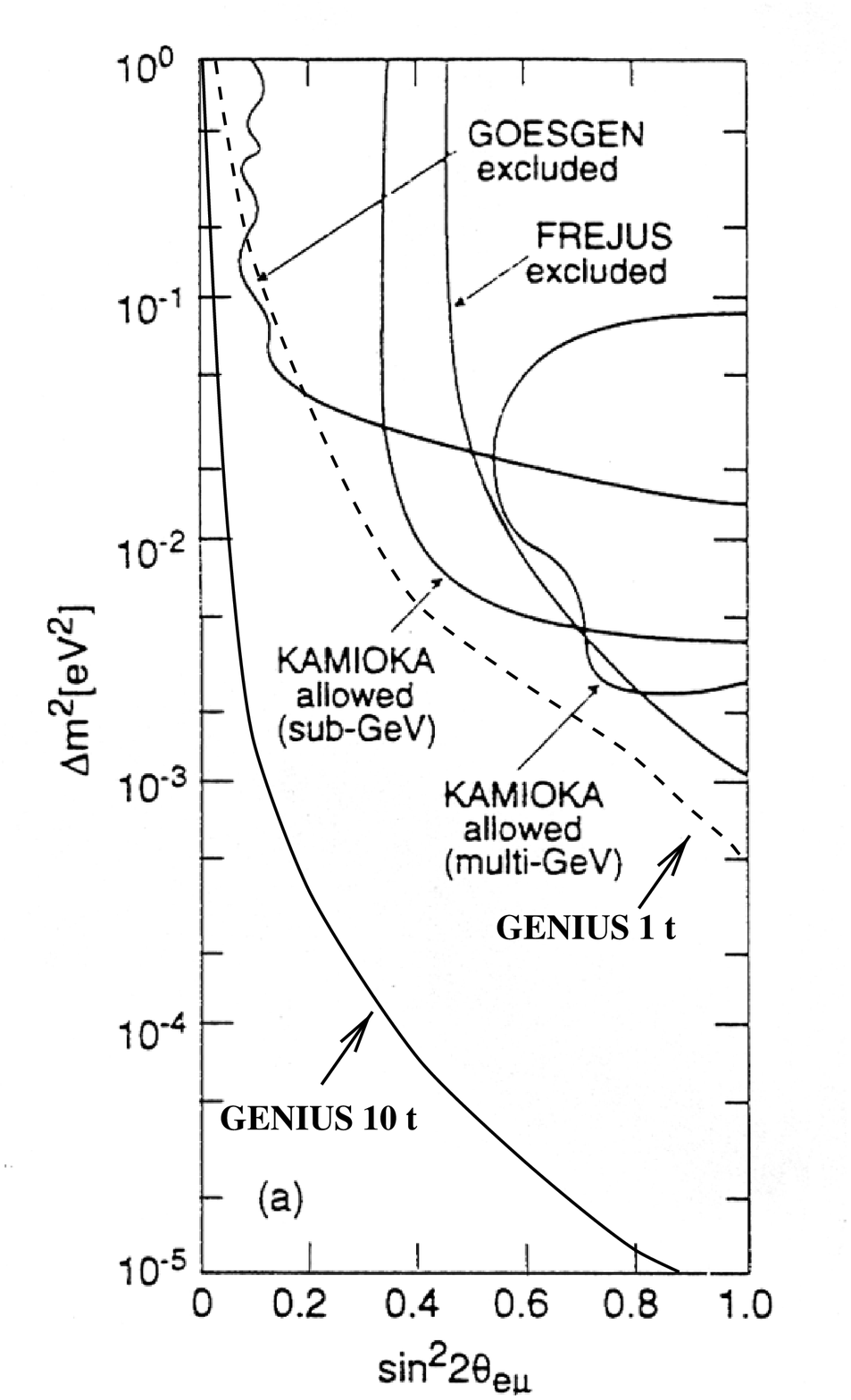}

\vskip0mm 
{\bf Fig. 22}
{\it Oscillation parameters which solve the 
atmospheric neutrino problem for $\nu_e \leftrightarrow \nu_{\mu}$ 
oscillations. In addition the best currently existing reactor 
constraints are shown. GENIUS would be able to test the 
atmospheric neutrino problem already with 1 ton, already in the shown,
worst strong hierarchy scenario ($m_1/m_2=0$)
(from [Kla97d]).}
\end{figure}

\begin{figure}[!t]
\vskip0mm 
\hskip5mm
\epsfxsize=58mm
\epsfysize=58mm
\epsfbox{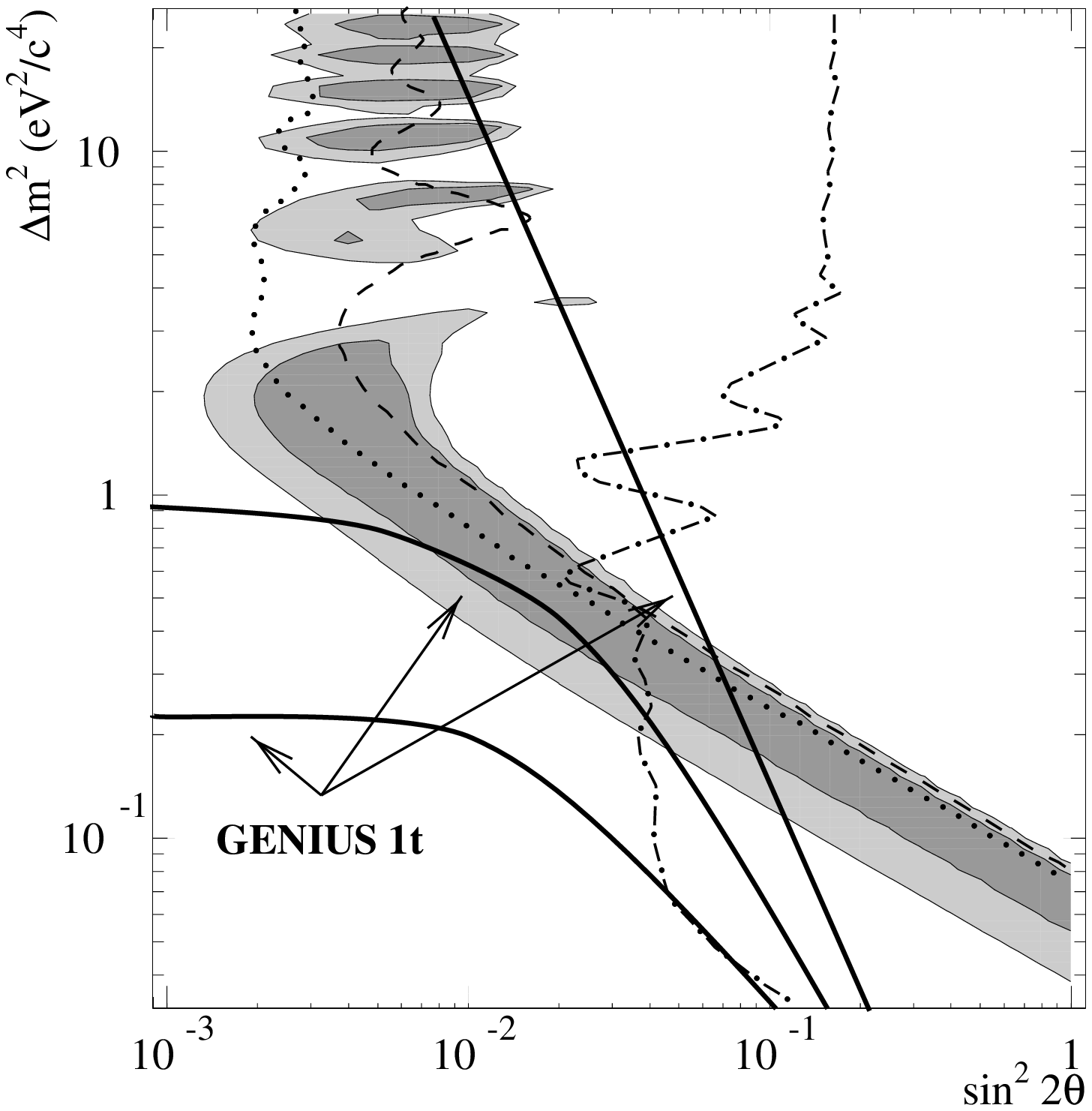}

{\bf Fig. 23}
{\it
LSND compared to the sensitivity of GENIUS 1t 
for $\eta^{CP} = +1$ and three ratios $R_{12}$, from top to bottom 
$R_{12}= 0, 0.01, 0.02$ (from [Kla97d]}
\end{figure}
Already in the worst case for double beta decay of $m_1/m_2=0$
GENIUS 1 ton is more sensitive than the running CERN experiments.
For quasi--degenerate models, for example $R=0.01$ already, GENIUS 1 ton would 
be more sensitive than all currently planned future accelerator neutrino experiments.

The situation of $\nu_e \leftrightarrow \nu_{\mu}$ oscillations (assuming
$sin^2\theta_{13}=0$) is shown in Fig. 21. The original figure is taken from
\cite{Gel95}. While the GENIUS 1 ton sensitivity is sufficient (even in the
worst case of $m_{\nu_e}<<m_{\nu_{\mu}}$) to extend to smaller values of
$\Delta m^2$ at large mixing angles, GENIUS 10 ton would have a sensitivity 
better than all existing or planned oscillation experiments, at least at large
$sin^2 2\theta$. In the quasi--degenerate models GENIUS would be much
more sensitive -- similar to the cases shown in Fig. 20.
Fig. 22 (background from \cite{Gel95}) compares the double beta worst case
of strong hierarchy ($m_1/m_2=0$), to the KAMIOKANDE allowed range for 
atmospheric neutrino oscillations. GENIUS 1 ton would already be able to test
the $\nu_e \leftrightarrow \nu_{\mu}$ oscillation hypothesis. 
Fig. 23 shows the potential of GENIUS for checking the LSND indication for
neutrino oscillations (original figure from \cite{Ath96}). 
Under the assumption 
$m_1/m_2 \geq 0.02$ and $\eta=1$, GENIUS 1 ton will be sufficient to find 
$0\nu\beta\beta$ decay if the LSND result is to be explained in terms of $\nu_e
\leftrightarrow \nu_{\mu}$ oscillations. This might be of particular interest 
also since the upgraded KARMEN will not completely cover \cite{Dre97} the full
allowed LSND range. Fig. 24 shows a summary of currently known constraints
on neutrino oscillation parameters (original taken from \cite{Hat94}), but 
including the $0\nu\beta\beta$ decay sensitivities of GENIUS 1 ton and GENIUS 
10 tons, for different assumptions on $m_1/m_2$ (and for $\eta^{CP}=+1$).
It is seen that already GENIUS 1 ton tests all degenerate or quasi--degenerate
($m_1/m_2 \geq \sim 0.01$) 
neutrino mass models in any range where neutrinos are 
interesting for cosmology, and also the atmospheric neutrino problem, if it is 
due to $\nu_e \leftrightarrow \nu_{\mu}$ oscillations. GENIUS in its 10 ton
version would directly test the large angle solution of the solar neutrino 
problem. 

\subsubsection*{GENIUS and left--right symmetry:}
If GENIUS is able to reach down to $\emass \le 0.01$ eV, it would at 
the same time be sensitive to right-handed $W$-boson masses up to 
$m_{W_R} \ge 8$ TeV (for a heavy right-handed neutrino mass of 
$1$ TeV) or $m_{W_R} \ge 5.3$ TeV (at $\langle m_N \rangle = m_{W_R}$). 
Such a limit would be comparable to the one expected for LHC, 
see for example \cite{Riz96}, which quotes a final sensitivity 
of something like $5-6$ TeV. Note, however that in order to 
obtain such a limit the experiments at LHC need to accumulate 
about $100 fb^{-1}$ of statistics. A 10 ton version of
 GENIUS 
could even reach a sensitivity of $m_{W_R} \ge 18$ TeV (for a heavy 
right-handed neutrino mass of

\newpage
\vspace*{-30mm} 
\hspace*{-30mm}
\epsfysize=200mm
\epsfbox{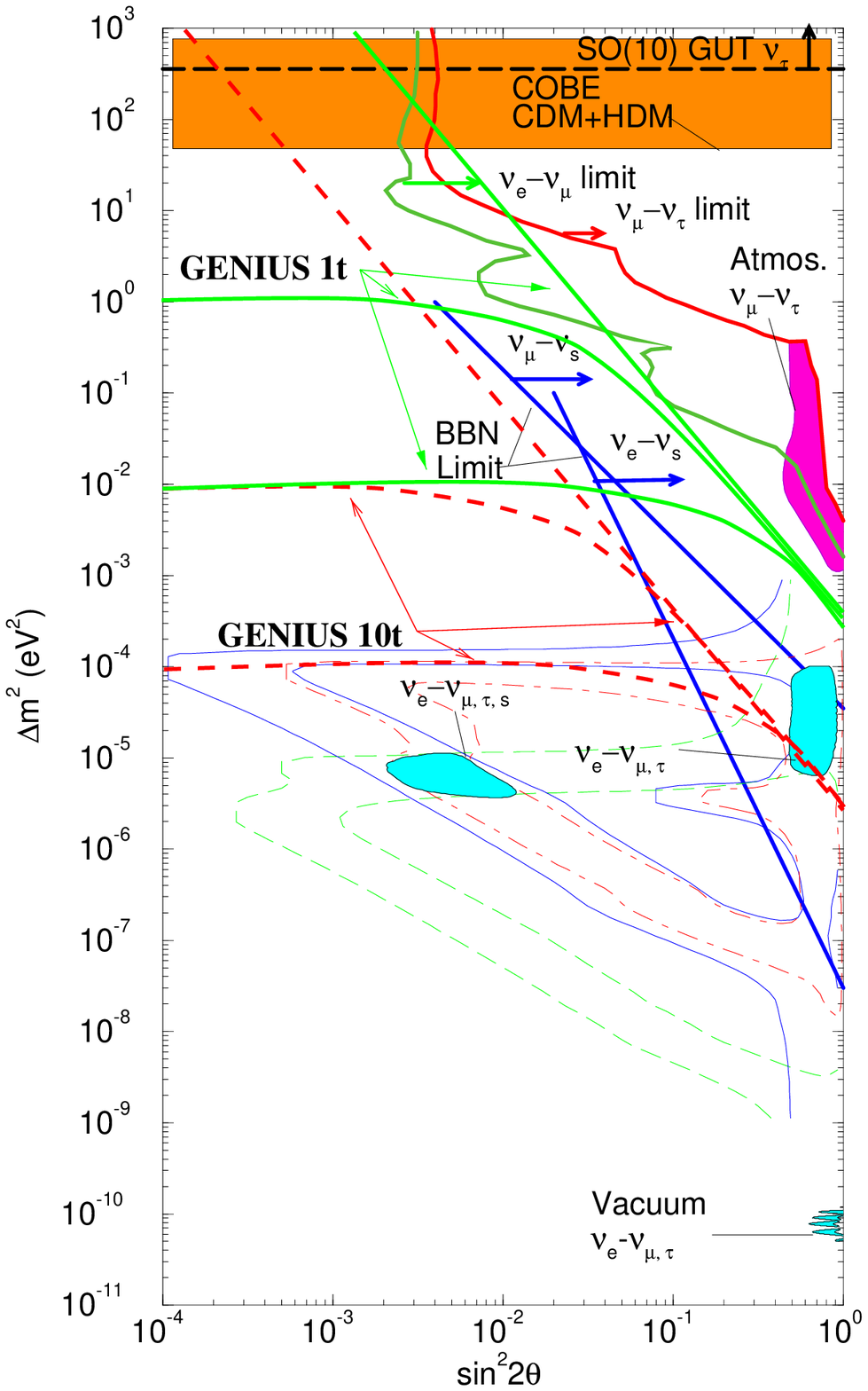}
\newpage
{\bf Fig. 24}
{\it Summary of currently known constraints on neutrino 
oscillation parameters. The (background) figure without the \znbb{} 
decay constraints can be obtained from\\ 
http://dept.physics.upenn.edu/~www/neutrino/solar.html. Shown are 
the vacuum and MSW solutions (for two generations of neutrinos) 
for the solar neutrino problem, 
the parameter range which would solve the atmospheric neutrino problem 
and various reactor and accelerator limits on neutrino oscillations. 
In addition, the mass range in which neutrinos are good hot dark matter 
candidates is indicated, 
as well as limits on neutrino oscillations into sterile states from 
considerations of big bang nucleosynthesis. Finally the 
thick lines indicate the sensitivity of GENIUS (full lines 1 ton, 
broken lines 10 ton) to neutrino oscillation parameters for three values 
of neutrino mass ratios $R = 0, 0.01$ and $0.1$ (from top to bottom). 
The region beyond the lines would be excluded.
While already the 1 ton GENIUS would be sufficient to constrain degenerate 
and quasi-degenerate neutrino mass models, and also would solve the 
atmospheric neutrino problem if it is due to 
$\nu_e \leftrightarrow \nu_{\mu}$ oscillations, the 10 ton version of 
GENIUS could cover a significant new part of the parameter space, 
including the large angle MSW solution to the solar neutrino problem, 
even in the worst case of $R=0$. } 
\bigskip

\begin{figure}[t]
\vskip-25mm 
\hskip10mm
\epsfxsize=100mm
\epsfysize=120mm
\epsfbox{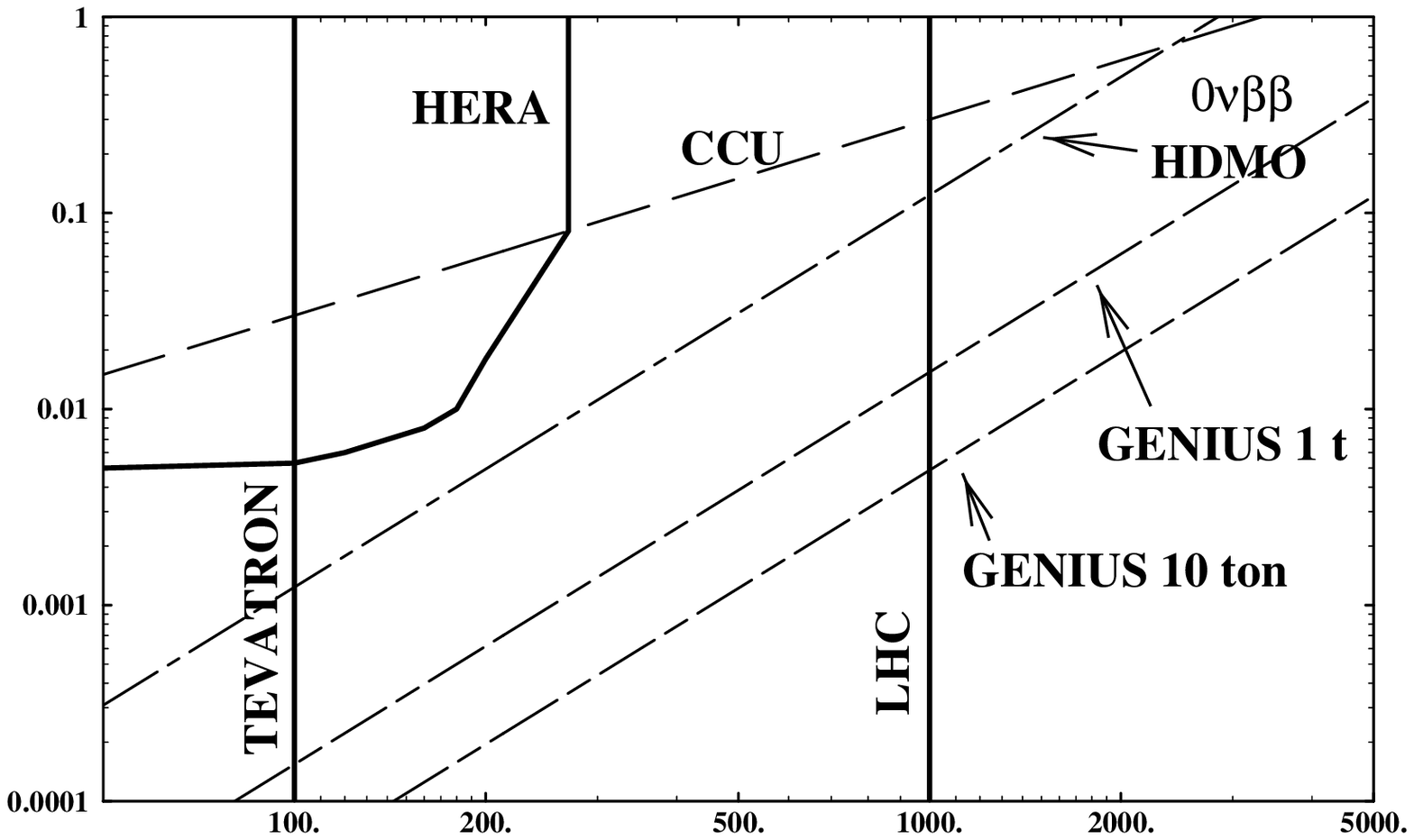}

\vskip-80mm
\noindent
$\lambda'_{111}$ 

\vskip45mm 
\hskip90mm $m_{\tilde q}$ [GeV] 

\bigskip

{\bf Fig. 25}
{\it Comparison of sensitivities of existing and future 
experiments on \rp SUSY models in the plane $\lambda'_{111}-m_{\tilde q}$. 
Note the double logarithmic scale! Shown are the areas currently excluded 
by the experiments at the TEVATRON, the limit from charged-current 
universality, denoted by CCU, and the limit from absence of \znbb{} 
decay from the Heidelberg-Moscow collaboration (\znbb{} HDMO). 
In addition, the estimated sensitivity of HERA and the LHC is compared to the 
one expected for GENIUS in the 1 ton and the 10 ton version. The figure 
is essentially an update of Fig. 15.}
\end{figure}

\noindent
$1$ TeV) or 
$m_{W_R} \ge 10.1$ TeV (at $\langle m_N \rangle = m_{W_R}$).

This means that already GENIUS 1 ton could be sufficient to definitely
test recent supersymmetric left--right symmetric models having the 
nice features of solving the strong CP problem without the need for an axion 
and having automatic R--parity conservation \cite{Kuc95,Moh96}.

\subsubsection*{GENIUS and $R_p$--violating SUSY:}
The improvement on the R--parity breaking Yukawa coupling $\lambda^{'}_{111}$
(see section 2.2) is shown in Fig. 25, which updates Fig. 15.
The full line to the right is the expected sensitivity of the 
LHC -- in the 
limit of large statistics. The three dashed--dotted lines denote (from top
to bottom) the current constraint from the Heidelberg--Moscow experiment
and the sensitivity of GENIUS 1 ton and GENIUS 10 tons, all
 for the 
conservative case of a gluino mass of 1 TeV. If squarks would be heavier than 
1 TeV, LHC could not compete with GENIUS. However, for typical squark masses  
below 1 TeV, LHC could probe smaller couplings.
However, one should keep in 
mind, that LHC can probe squark masses up to 1 TeV only with several years of 
data taking. 

\subsubsection*{GENIUS and $R_p$--conserving SUSY:}
Since the limits on a `Majorana--like' sneutrino mass $\tilde{m}_M$ scale
with $(T_{1/2})^{1/4}$, GENIUS 1 ton (or 10 tons)
would test `Majorana' sneutrino masses lower
by factors of about 7(20), compared with present constraints 
\cite{Hir97,Hir97a,Hir97b}. 

\subsubsection*{GENIUS and Leptoquarks:}
Limits on the lepton--number violating parameters defined in sections
2.7, 3.2
improve as $\sqrt{T_{1/2}}$. This means that for leptoquarks in the range
of 200 GeV LQ--Higgs couplings down to (a few) $10^{-8}$ could be explored. 
In other words, if leptoquarks interact with the standard model Higgs boson
with a coupling of the order ${\cal O}(1)$, either $0\nu\beta\beta$ must be 
found, or LQs must be heavier than (several) 10 TeV. 

\subsubsection*{GENIUS and composite neutrinos}
GENIUS in the 1(10) ton version would improve the limit on the excited
Majorana neutrino mass deduced from the Heidelberg--Moscow experiment
(eq. 32) to
\be{compri}
m_N\geq \sim 1.1 (2.3) \hskip3mm TeV
\ee

\begin{figure}[!h]
\vspace*{-13cm}
\hspace*{-1cm}
\epsfysize=130mm\centerline{\epsfbox{dmall.ps}}
\vspace*{10cm}

{\bf Fig. 26}
{\it WIMP--nucleon cross section
limits in pb for scalar interactions as
function of the WIMP--mass in 
GeV 
(hatched region: excluded by the Heidelberg--Moscow collaboration
[HM94]
and the UKDMC NaI experiment [Smi96];
solid line: DAMA result for NaI, see [Ber97]). 
Further shown are sensitivities of experiments under construction
 (dashed lines for
HDMS [Bau97,Kla97e], CDMS [Bar96] and for
GENIUS). These 
limits are compared to theoretical
expectations (scatter plot) for WIMP--neutralino cross sections calculated 
in the
MSSM framework with non--universal scalar mass unification
[Bed97b].  The 90~\% allowed region claimed by [Ber97a]
(light filled area), which is further
restricted by indirect dark matter searches [Bot97] (dark filled
area), could already be easily tested with a 100 kg version of the GENIUS experiment.}
\end{figure}

\subsubsection{The potential of GENIUS for Cold Dark Matter
Search}
 Weakly interacting massive particles (WIMPs) are candidates for the 
cold dark matter in the universe. The favorite WIMP candidate is 
the lightest supersymmetric particle, presumably the neutralino. 
The expected detection rates for neutralinos of typically less 
than one event per day and kg of detector mass \cite{Bed94,Bed97a,Bed97b,
Jun96}, 
however, make direct searches for WIMP scattering experimentally 
a formidable task. 

Fig. 26 shows a comparison of existing constraints and future 
sensitivities of cold dark matter experiments, together with the 
theoretical expectations for neutralino scattering rates 
\cite{Bed97b}.
Obviously, GENIUS could easily cover the range of positive 
evidence for dark matter
recently claimed by DAMA \cite{Ber97a,Bot97}. 
It would also be by far more
sensitive than all other dark matter experiments at present under construction
or proposed, like the cryogenic experiment CDMS. Furthermore,
obviously GENIUS will be the only experiment, 
which could seriously test the MSSM predictions over the whole 
SUSY parameter space. In this way, GENIUS could compete even 
with LHC in the search for SUSY, see for example the discussion 
in \cite{Bae97}.

It is interesting to note, that if WIMP scattering is found 
by GENIUS it could be used to constrain the amount of R-parity 
violation within supersymmetric models. The arguments are very 
simple \cite{Hir97c}. Due to the fact that neutralinos are 
abound in the 
galaxy even today, neutralino decays via R-parity violating 
operators would have to be highly suppressed. 

The details depend, of course, on the neutralino mass and composition. 
However, finding the neutralino with GENIUS would imply typical 
limits on R-parity violating couplings of the order of $10^{-(16-20)}$ 
for any of the $\lambda_{ijk}$, $\lambda'_{ijk}$ or $\lambda''_{ijk}$ 
in the superpotential (eq. 11). 
A positive result of the CDM search at hand, one could thus finally 
safely conclude that R-parity is conserved.

\section{Conclusion}
Double beta decay has a broad potential for providing important information
on modern particle physics beyond present and future high energy accelerator
energies which will be competitive for the next decade and more. 
This includes SUSY 
models, compositeness, left--right symmetric models, leptoquarks, and the
neutrino and sneutrino mass. 
Based to a large extent on the theoretical work of the Heidelberg
Double Beta group, results have been deduced from the HEIDELBERG--MOSCOW
experiment for these topics and have been presented here.
For the neutrino mass double beta decay now is particularly 
pushed into a key position by the recent possible indications of beyond 
standard model physics from the side of solar and atmospheric neutrinos,
dark matter COBE results and others. New classes of GUTs basing on
degenerate neutrino mass scenarios which could explain these observations,
can be checked by double beta decay in near future. The HEIDELBERG--MOSCOW 
experiment has reached a leading position among present $\beta\beta$
experiments and as the first of them now yields results in the sub--eV
range. We 
have presented a new idea and proposal of a future double beta experiment
(GENIUS) with highly increased sensitivity based on use of 1 ton or more
of enriched `naked' $^{76}$Ge detectors in liquid nitrogen.
This new experiment would be a breakthrough into the multi-TeV range for many 
beyond standard models. The sensitivity for the neutrino mass would reach
down to 0.01 or even 0.001 eV. The experiment would be competitive to LHC with 
respect to the mass of a right--handed W boson, in search for R--parity 
violation and others, and would improve the leptoquark and compositeness 
searches
by considerable factors. It would probe the Majorana electron
sneutrino mass more
sensitive than NLC (Next Linear Collider). It would yield constraints on 
neutrino oscillation parameters far beyond all present terestrial 
$\nu_e - \nu_x$ neutrino oscillation experiments and could test directly the
atmospheric neutrino problem and the large angle solution of the solar neutrino
problem. GENIUS would cover the full SUSY parameter space for prediction of 
neutralinos as cold dark matter and compete in this way with LHC in the search
for supersymmetry. Even if SUSY would be first observed by LHC, it would
still be fascinating to verify the existence and properties of neutralino
dark matter, which could be achieved by GENIUS. Concluding GENIUS
has the ability to provide a major tool for future particle-- and astrophysics.
    
 Finally it may be stressed that the technology of producing and using
enriched high purity germanium detectors, which have been produced for the 
first time for the Heidelberg--Moscow experiment, has found 
meanwhile applications also in pre-GENIUS dark matter search
\cite{101,102,Kla97e,Bau97} and in high--resolution $\gamma$-ray astrophysics,
using balloons and satellites \cite{Kla91,81,109,100,111,Kla97b}.


\begin{thebibliography}{200}


\bibitem[Abe94]{60}
F. Abe {\it et al.} (CDF Collab.), Phys. Rev. Lett. 72 (1994) 3004



\bibitem[Adr92]{56}
O. Adriani {\it et al.}, Phys. Lett. {\bf B 288} (1992) 404

\bibitem[ALE92]{57}
ALEPH Collab., Phys. Rep. 216 (1992) 343


\bibitem[ALE93]{52}
ALEPH Collab., Z.Phys. {\bf C 59} (1993) 215



\bibitem[Ale94]{91} A. Alessandrello et al., Phys. Lett {\bf B 335} (1994) 519
 



\bibitem[Als89]{89} M. Alston-Garnjost et al., Phys. Rev Lett. {\bf 7}1 (1993)
 831

\bibitem[Alt97]{Alt97}
G. Altarelli, J. Ellis, G.F. Guidice, S. Lola, M.L. Mangano,
preprint hep-ph/9703276

\bibitem[Ath95]{24}
C. Athanassopoulos {\it et al.} (LSND--Collaboration),
Phys. Rev. Lett. {\bf 75} (1995) 2650 and D. Smith, in \cite{tren}

\bibitem[Ath96]{Ath96}
C. Athanassopoulos {\it et al.}, LSND collab., Phys. Rev. {\bf C 54}
(1996) 2685, Phys. Rev. Lett. {\bf 77} (1996) 3082 

\bibitem[Aul82]{66}
C.S. Aulakh and R.N. Mohapatra, Phys. Lett. {\bf B 119} (1982) 136


\bibitem[Bab95]{bab95}
K.S. Babu, R.N. Mohapatra, Phys. Rev. Lett. {\bf 75} (1995) 2276

\bibitem[Bab97]{Bab97}
K.S. Babu et al., preprint hep-ph/9703299 (March 1997)

\bibitem[Bab97b]{Bab97b}
K.S. Babu {\it et al.}, preprint hep-ph/9705414v2 (1997)

\bibitem[Bae97]{Bae97}
H. Baer. M. Bhrlik, hep-ph/9706509

\bibitem[Bam95a]{77}
P. Bamert, C.P. Burgess, R.N. Mohapatra, Nucl. Phys. B 438 (1995) 3


\bibitem[Bam95]{69}   
P. Bamert, C.P. Burgess, R.N. Mohapatra, Nucl. Phys. {\bf B 449} (1995)
25 



\bibitem[Bar89]{113}
V. Barger, G.F. Guidice, T. Han. Phys.Rev. {\bf D 40} (1989) 2987


\bibitem[Bar93]{109}
S.D. Barthelmy {\it et al.}, AIP Conf. Proc. 280 (1993) 1166




\bibitem[Bar94]{100} S. D. Barthelmy et al., Astrophys. 
J. {\bf 427} (1994) 519 

\bibitem[Bar96]{Bar96}
P.B. Barnes {\it et al.}, Nucl. Instr. Meth. {\bf A 370} (1996) 233

   
\bibitem[Bar97]{Bar97}
A.S. Barabash, Proc. NEUTRINO 96, Helsinki, June 1996, World Scientific, 
Singapore (1997), p. 374

\bibitem[Bau97]{Bau97}
L. Baudis, J. Hellmig, H.V. Klapdor--Kleingrothaus, A. M\"uller, F. Petry,
Y. Ramachers, H. Strecker, Nucl. Inst. Meth.  {\bf A 385} (1997) 265

\bibitem[Bav95]{126} E. Baver, M. Leurer, Phys. Rev. {\bf D 51} (1995) 260

\bibitem[Bed94]{Bed94}
V. Bednyakov, H.V. Klapdor--Kleingrothaus, S.G. Kovalenko, Phys. Lett. 
{\bf B 329} (1994) 5 and Phys. Rev. {\bf D 50} (1994) 7128

\bibitem[Bed97a]{Bed97a}
V. Bednyakov, H.V. Klapdor--Kleingrothaus, S.G. Kovalenko, 
Phys. Rev. {\bf D 55} (1997) 503

\bibitem[Bed97b]{Bed97b}
V. Bednyakov, H.V. Klapdor--Kleingrothaus, S.G. Kovalenko, 
Y. Ramachers,
Z. Phys. {\bf A 357} (1997) 339

\bibitem[Bed97c]{Bed97c}
V. Bednyakov, V.B. Brudanin, S.G. Kovalenko, Ts. D. Vylov, Mod. Phys. Lett 
{\bf A 12} (1997) 233


\bibitem[Bel95]{14}
G. Belanger, F. Boudjema, D.London, H. Nadeau, Preprint hep-ph/9508317, 
August 1995

\bibitem[Ber92]{65}
Z.G. Berezhiani, A. Yu. Smirnov, J.W.F. Valle, Phys.Lett. {\bf B 291}
(1992) 99

\bibitem[Ber95]{Ber95}
Z. Berezhiani, R.N. Mohapatra, Phys. Rev. {\bf D 52} (1995) 6607

\bibitem[Ber97]{Ber97}
R. Bernabei {\it et al.}, Phys. Lett. {\bf B 389} (1997) 757


\bibitem[Ber97a]{Ber97a}
R. Bernabei {\it et al.}, ROM2F/97/33; P. Belli at TAUP, Gran Sasso, Sept. 7 
(1997)

\bibitem[Bha97]{Bha97}
G. Bhattacharyya, in \cite{Kla98}

\bibitem[Bl\"u94]{129} J. Bl\"umlein, E. Boos, A. Pukhov, preprint DESY 94-072
(April 1994)



\bibitem[Boc94]{111}
J. Bockholt, H.V. Klapdor--Kleingrothaus, Nucl. Phys. {\bf B}
(Proc. Suppl.) 35 (1994) 403

\bibitem[Bot97]{Bot97}
A. Bottino {\it et al.}, hep-ph/9709292

\bibitem[Buc87]{Lagr1} W. Buchm\"uller, R. R\"uckl and D. Wyler,
                       Phys.Lett. B191 (1987) 442.



\bibitem[Buc91]{3}
W. Buchm\"uller and G. Ingelman, Proc. Workshop Physics at HERA, Hamburg,
Oct. 29--30 (1991)

\bibitem[Bur93]{68}
C.P. Burgess, J.M. Cline, Phys. Lett. {\bf B 298} (1993) 141; 
Phys. Rev. {\bf D 49} (1994) 5925

\bibitem[Bur96]{72}
C.P.Burgess, in \cite{tren}

\bibitem[But93]{115}
J. Butterworth, H. Dreiner, Nucl. Phys. {\bf B397} (1993) 3
and H. Dreiner, P. Morawitz, Nucl. Phys. {\bf B428} (1994) 31




\bibitem[Cal93]{Cal93}
D.O. Caldwell, R.N. Mohapatra, Phys. Rev. {\bf D 48} (1993) 3259

\bibitem[Cal95]{Cal95}
D.O. Caldwell, R.N. Mohapatra, Phys. Lett. {\bf 354} (1995) 371

\bibitem[Car93]{70}
C.D. Carone, Phys. Lett. {\bf B 308} (1993) 85



\bibitem[CDF91]{53}
CDF Collab., Phys. Rev. Lett. {\bf 67} (1991) 2148




\bibitem[Chi81]{64}
Y. Chikashige, R.N. Mohapatra and R.D. Peccei, Phys. Lett. {\bf B 98}
(1981) 265; Phys. Rev. Lett. {\bf 45} (1980) 265


\bibitem[Cho94]{128} D. Choudhury, hep-ph/9408250 (1994)

\bibitem[Cho97]{Cho97}
D. Choudhury, S. Raychaudhuri, preprint hep-ph/9702392

\bibitem[Cli97]{CLI96} D. Cline, in: Proceedings of the {\it International 
Workshop Dark Matter in Astro-- and Particle Physics} (DARK96),
Eds. H.V. Klapdor--Kleingrothaus, Y. Ramachers, World Scientific 1997,
p. 479



\bibitem[Dan95]{90} F. A. Danevich et al., Phys. Lett. {\bf B 344} (1995) 72



\bibitem[Dav94]{DBC}  S. Davidson, D. Bailey and A. Campbell,
               Z.Phys. C61 (1994) 613;
               M. Leurer, Phys.Rev.Lett. 71 (1993) 1324; 
               Phys.Rev. D50 (1994) 536.



\bibitem[Der95]{58}
M. Derrik, {\it et al.} (ZEUS Collab.), Z. Phys. {\bf C 65} (1995) 627


\bibitem[Dia93]{54}
J.L. Diaz Cruz, O.A. Sampayo, Phys. Lett. {\bf B 306} (1993) 395;
Phys. Rev. {\bf D 49} (1994) R2149

\bibitem[Doi85]{74}
M. Doi, T. Kotani, E.Takasugi, Progr. Theor. Phys. Suppl. {\bf 83} (1985) 1



\bibitem[Doi93]{49}
M. Doi, T. Kotani, Progr. Theor. Phys. 89 (1993) 139

\bibitem[Dre97]{Dre97}
G. Drexlin, Proc. Internat. School on Neutrino Physics, Erice, Italy,
Sept. 1997, to be publ. in Plenum Press

\bibitem[Ell87]{Ell87}
S.R. Elliot, A.A. Hahn, M.K. Moe, Phys. Rev. Lett. {\bf 59} (1987) 1649

\bibitem[Ell92]{88} S. R. Elliott et al., Phys. Rev. {\bf C 46} (1992) 1535


\bibitem[Fal94]{102}
T. Falk, A. Olive, M. Srednicki, Phys. Lett. {\bf B339} (1994) 248

 

\bibitem[Fri88]{62}
J.Friemann, H.Haber, K.Freese, Phys. Lett. {\bf B 200} (1988) 115;
J. Bahcall, S. Petcov, S. Toshev and J.W.F. Valle, Phys.Lett. {\bf B 181}
(1986) 369; Z. Berezhiani and M. Vysotsky, Phys. Lett. {\bf B 199}
(1988) 281.



\bibitem[Fri95]{22}
H. Fritzsch and Zhi-zhong Xing, preprint hep-ph/9509389,
Phys. Lett. {\bf B 372} (1996) 265


\bibitem[Gel81]{63}
G.B. Gelmini and M. Roncadelli, Phys. Lett. {\bf B 99} (1981) 411

\bibitem[Gel95]{Gel95}
G. Gelmini, E. Roulet, Rep. Progr. Phys. {\bf 58}(1995) 1207

\bibitem[Geo81]{61}  
H.M. Georgi, S.L.Glashow and S. Nussinov, Nuc. Phys. {\bf B 193} (1981)
297


\bibitem[Ger96]{82}
G. Gervasio, in \cite{tren}

\bibitem[Gon95]{Gon95}
M. Gonzalez--Garcia, hep-ph/9510419


\bibitem[Gro85]{43}
K. Grotz, H. V. Klapdor, Phys. Lett. {\bf B 157} (1985) 242



\bibitem[Gro86]{31}
K. Grotz, H.V. Klapdor, Nucl. Phys {\bf A 460} (1986) 395


\bibitem[Gro90]{16}
K. Grotz, H.V. Klapdor, 'The Weak Interaction in Nuclear, Particle and 
Astrophysics' (Adam Hilger: Bristol, Philadelphia) 1990



\bibitem[H195]{5}
H1 Collab., Phys. Lett. {\bf B 353} (1995) 578, Z. Phys. {\bf C 64} (1994) 545 

\bibitem[H196]{H196}
H1 Collab., S. Aid et al., Phys. Lett. B 369 (1996) 173

\bibitem[Hab93]{44}
H. E. Haber, in Proc. on Recent Advances in the Superworld, Houston,
April 14-16 (1993), hep-ph/9308209


\bibitem[Hal84]{hal84}
L. Hall, M. Suzuki, Nuclear Physics {\bf B 231} (1984) 419

\bibitem[Hat94]{Hat94}
N. Hata, P. Langacker, Phys. Rev. {\bf D 50} (1994) 632

\bibitem[Hax84]{34}
W.C. Haxton, G.J. Stephenson, Progr. Part. Nucl. Phys. 12 (1984) 409

\bibitem[Hel96]{Hel96}
J. Hellmig, PhD Thesis, Univ. of Heidelberg, 1996

\bibitem[Hel97]{Hel97}
J. Hellmig, H.V. Klapdor--Kleingrothaus, Z. Phys. {\bf A}, in press 1997

\bibitem[Hew97]{Hew97}
J.L. Hewett, T.G. Rizzo, preprint hep-ph/9703337v3 (May1997)

\bibitem[Heu95]{Heu95}
G. Heusser, Ann. Rev. Nucl. Part. Sci. {\bf 45} (1995) 543

\bibitem[Hir95]{6}
M. Hirsch, H.V. Klapdor--Kleingrothaus, S.G. Kovalenko, Phys. Rev. Lett. 75
(1995) 17

\bibitem[Hir95b]{10}
M. Hirsch, H.V. Klapdor--Kleingrothaus, S.G. Kovalenko, in \cite{tren}



\bibitem[Hir95c]{47}
M. Hirsch, H.V. Klapdor--Kleingrothaus, S. Kovalenko, Phys. Lett. {\bf B 352}
(1995) 1

\bibitem[Hir95d]{Hir95d}
J. Hirsch, O. Castanos, P.O. Hess, Nucl. Phys. {\bf A 582} (1995) 124


\bibitem[Hir96]{hir96}
M. Hirsch, H.V. Klapdor--Kleingrothaus, S.G. Kovalenko,
Phys. Lett. {\bf B 372} (1996) 181, Erratum: Phys. Lett. {\bf B381} (1996) 488 


\bibitem[Hir96a]{hir96a}
M. Hirsch, H.V. Klapdor--Kleingrothaus, S.G. Kovalenko,
Phys. Lett. {\bf B 378} (1996) 17 and Phys. Rev. {\bf D 54}
(1996) R4207


\bibitem[Hir96b]{75}
M. Hirsch, H. V. Klapdor--Kleingrothaus, S. G. Kovalenko, H. P\"as,
Phys. Lett.  {\bf B 372} (1996) 8



\bibitem[Hir96c]{hir96c}
M. Hirsch, H.V. Klapdor--Kleingrothaus, S. Kovalenko, Phys. Rev. {\bf D 53}
(1996) 1329


\bibitem[Hir96d]{11}
M. Hirsch, H.V. Klapdor--Kleingrothaus, in \cite{tren};
M. Hirsch, H.V. Klapdor--Kleingrothaus, O. Panella, 
Phys. Lett. {\bf B 374} (1996) 7

\bibitem[Hir97]{Hir97}
M. Hirsch, H.V. Klapdor--Kleingrothaus, S.G. Kovalenko, 
Phys. Lett. {\bf B 398} (1997) 311 and {\bf 403}
(1997) 291


\bibitem[Hir97a]{Hir97a}
M. Hirsch, H.V. Klapdor--Kleingrothaus, S.G. Kovalenko, 
Phys. Rev. {\bf D} in press, 1997

\bibitem[Hir97b]{Hir97b}
M. Hirsch, H.V. Klapdor--Kleingrothaus, S. Kovalenko, in \cite{Kla98} 

\bibitem[Hir97c]{Hir97c}
M. Hirsch, H.V. Klapdor--Kleingrothaus, Proc. Int. Workshop on Dark 
Matter
in Astro-- and Particle Physics (DARK96), Heidelberg, Sept. 1996,
Eds. H.V. Klapdor--Kleingrothaus and Y. Ramachers (World Scientific,
Singapore) 1997, p. 640

\bibitem[HM94]{101}
HEIDELBERG--MOSCOW collab., Phys. Lett. {\bf B 336} (1994) 141



\bibitem[HM95]{79}
HEIDELBERG--MOSCOW collab., Phys. Lett. {\bf B 356} (1995) 450

\bibitem[HM96]{HM96}
HEIDELBERG--MOSCOW collab., Phys. Rev. {\bf D 54} (1996) 3641, 

\bibitem[HM97]{HM97}
HEIDELBERG--MOSCOW collab., Phys. Rev. {\bf D 55} (1997) 54 and Phys. Lett.
{\bf B 407} (1997) 219 



\bibitem[Ioa94]{21}
A. Ioanissyan, J.W.F. Valle, Phys. Lett {\bf B 322} (1994) 93

\bibitem[J\"or94]{95} V. J\"orgens et al., Nucl. Phys. (Proc. Suppl.) {\bf B  35} (1994) 378 

\bibitem[Jun96]{Jun96}
G. Jungmann, M. Kamionkowski, K. Griest, Phys. Rep. 267 (1996) 195


\bibitem[Kal97]{Kal97}
J. Kalinowski et al., preprint hep-ph/9703288v2 (March 1997)

\bibitem[Kan97]{Kan97}
G. Kane, in \cite{Kla98}

\bibitem[Kla84]{30}
H.V. Klapdor, K. Grotz, Phys.Lett. {\bf B142} (1984) 323

\bibitem[Kla87]{84}
H.V. Klapdor--Kleingrothaus, MPI--H 1987, proposal

\bibitem[Kla91]{Kla91}
H.V. Klapdor--Kleingrothaus, Proc. Int. Symposium on $\gamma$-Ray Astrophysics,
Paris 1990, AIP Conf. Proc. 232 (1991) 464

\bibitem[Kla92]{Kla92}
H.V. Klapdor--Kleingrothaus, K. Zuber, Phys. Bl. {\bf 48} (1992) 1017 

\bibitem[Kla94]{81}
H.V. Klapdor--Kleingrothaus, Progr. Part. Nucl. Phys. {\bf 32} (1994)261


\bibitem[Kla95]{1}
H. V. Klapdor--Kleingrothaus, A. Staudt, Non--Accelerator Particle Physics,
IOP Publ., Bristol, Philadelphia, 1995; and Teilchenphysik ohne Beschleuniger,
Teubner Verlag, Stuttgart, 1995

\bibitem[Kla96]{tren}
H.V. Klapdor--Kleingrothaus and S. Stoica (Eds.), Proc. {\it Int. Workshop
on Double Beta Decay and Related Topics}, Trento, 24.4.--5.5.95, 
World Scientific Singapore


\bibitem[Kla96a]{Kla95b}
H.V. Klapdor--Kleingrothaus, in Proc. 
{\it Int. Workshop
on Double Beta Decay and Related Topics}, Trento, 24.4.--5.5.95, 
World Scientific Singapore 1996, Ed.: H.V. Klapdor--Kleingrothaus and S. Stoica


\bibitem[Kla97]{Kla97}
H.V. Klapdor--Kleingrothaus, Invited talk at NEUTRINO 96, Helsinki, 
June 1996, World Scientific Singapore 1997, p. 317

\bibitem[Kla97a]{Kla97a}
H.V. Klapdor--Kleingrothaus,  
in Proc. Int. Workshop on Non-Accelerator New Physics (NANP-97), 
Dubna, Moscow region, 
Russia, July 7-11, 1997 and in Proc. Int. School on Neutrinos, 
Erice, Italy, Sept. 1997, to be publ. in Plenum Press

\bibitem[Kla97b]{Kla97b}
H.V. Klapdor--Kleingrothaus, M.I. Kudravtsev, V.G. Stolpovski,
S.I. Svertilov, V.F. Melnikov, I. Krivosheina, J. Moscow. Phys. Soc.
7 (1997) 41

\bibitem[Kla97c]{Kla97c}
H.V. Klapdor--Kleingrothaus and K. Zuber, Particle Astrophysics, IOP Publ., 
Bristol, Philadelphia 1997; and Teilchenastrophysik, Teubner Verlag, 
Stuttgart, 1997

\bibitem[Kla97d]{Kla97d}
H.V. Klapdor--Kleingrothaus, M. Hirsch, Z. Phys. {\bf A}, in press 1997

\bibitem[Kla97e]{Kla97e}
H.V. Klapdor--Kleingrothaus, Y. Ramachers, in: Proc. 
Int. Workshop on
Dark Matter in Astro-- and Particle Physics (DARK96) Sept. 1996,
Eds. H.V. Klapdor--Kleingrothaus and Y. Ramachers, Heidelberg, 
(World Scientific Singapore) 1997, p. 459

\bibitem[Kla98]{Kla98}
H.V. Klapdor--Kleingrothaus and H. P\"as (Eds.), Beyond the Desert --
Accelerator- and Non--Accelerator Approaches, IOP, Bristol, 1998

\bibitem[Kol97]{Kol97}
S. Kolb. M. Hirsch, H.V. Klapdor--Kleingrothaus, S. Kovalenko,
in \cite{Kla98}

\bibitem[Kol97a]{Kol97a}
S. Kolb, M. Hirsch, H.V. Klapdor--Kleingrothaus, Phys. Rev. {\bf D 56} (1997) 
4161

\bibitem[Kuc95]{Kuc95}
R. Kuchimanchi, R.N. Mohapatra, Phys. Rev. Lett. {\bf 75} (1995) 3939



\bibitem[Kum96]{96} K. Kume (ELEGANT collaboration), in \cite{tren}


\bibitem[Kuz90]{18}
V. Kuzmin, V. Rubakov, M. Shaposhnikov, Phys. Lett. {\bf B 185} (1985)
36; M. Fukugita, T. Yanagida, Phys. Rev. {\bf D 42} (1990) 1285;
G. Gelmini, T. Yanagida, Phys. Lett. {\bf B 294} (1992) 53; 
B. Campbell {\it et al.}, Phys. Lett. {\bf B 256} (91) 457



\bibitem[Lan88]{15}
P.Langacker, in `Neutrinos' (Springer: Heidelberg, New York), 1988,
ed. H.V. Klapdor, p. 71

\bibitem[Lee94]{19}
D.G. Lee, R.N. Mohapatra, Phys. Lett. {\bf B 329} (1994) 463

\bibitem[Leu94]{127} M. Leurer, Phys. Rev. {\bf D 49} (1994) 333


\bibitem[Moe91]{99} M. K. Moe, Phys. Rev. {\bf C 44} (1991) R931 

\bibitem[Moe94]{93} M. K. Moe, Prog. Part. Nucl. Phys. {\bf 32} (1994) 247; 
Nucl. Phys. (Proc. Suppl.) {\bf B 38} (1995) 36


\bibitem[Moh86]{48}
R.N. Mohapatra, Phys. Rev. {\bf D 34} (1986) 3457

\bibitem[Moh86b]{50}
R.N. Mohapatra, Phys. Rev. {\bf D 34} (1986) 909 


\bibitem[Moh88]{73}
R.N. Mohapatra and E. Takasugi, Phys. Lett. {\bf B 211} (1988) 192


\bibitem[Moh91]{17}
R.N. Mohapatra, P.B. Pal, 
'Massive Neutrinos in Physics and Astrophysics'
(World Scientific, Singapore) 1991

\bibitem[Moh92]{45}
R.N. Mohapatra, Unification and Supersymmetry (Springer: Heidelberg, New York)
1986 and  1992

\bibitem[Moh94]{Moh94}
R.N. Mohapatra, Progr. Part. Nucl. Phys. {\bf 32} (1994) 187

\bibitem[Moh95]{23}
R.N. Mohapatra, S. Nussinov, Phys. Lett. {\bf B 346} (1995) 75

\bibitem[Moh96]{Moh96}
R.N. Mohapatra, A. Rasin, Phys. Rev. Lett. {\bf 76} (1996) 3490 and Phys. Rev. 
{\bf D54} (1996) 5835

\bibitem[Moh96a]{7}
R.N. Mohapatra, in \cite{tren}


\bibitem[Moh97]{Mohneu}
R. N. Mohapatra, Proc. {\it Neutrino 96}, Helsinki, 1996, World 
Scientific Singapore, 1997, p. 290

\bibitem[Moh97a]{Moh97a}
R.N. Mohapatra, Proc. Int. School on Neutrinos, 
Erice, Italy, Sept. 1997, to be publ. in Plenum Press,
and priv. comm.


\bibitem[Mut88]{27}
K. Muto, H.V. Klapdor, in 'Neutrinos (Springer: Heidelberg, 
New York) 1988, ed. H.V. Klapdor, p. 183

\bibitem[Mut89]{28}
K. Muto, E. Bender, H.V. Klapdor, Z. Phys. {\bf A 334} (1989) 177,187;

\bibitem[Mut91]{33}
K. Muto, E.Bender, H.V. Klapdor--Kleingrothaus, Z. Phys {\bf A} 339 (1991) 435

\bibitem[NEM94]{94} NEMO Collaboration, Nucl. Phys. (Proc. Suppl.) {\bf B 35} 
(1994) 369 

\bibitem[Nor97]{Nor97}
D. Normile, Science 276 (1997) 1795 

\bibitem[PDG94]{46}
Particle Data Group, Phys. Rev. {\bf D 50} (1994)

\bibitem[PDG96]{PDG96}
Particle Data Group, Phys. Rev. {\bf D 54} (1996) 1

\bibitem[P\"as96]{71}
H. P\"as {\it et al.}, in \cite{tren};

\bibitem[P\"as97]{Pae97}
H. P\"as, M. Hirsch, H.V. Klapdor--Kleingrothaus, S.G. Kovalenko,
in \cite{Kla98}

\bibitem[Pan96]{8}
O.Panella, in \cite{tren}

\bibitem[Pan97]{Pan97}
O. Panella, in \cite{Kla98}

\bibitem[Pan96b]{Pan96}
G. Pantis, F. Simkovic, J.D. Vergados, A. Faessler, Phys. Rev. {\bf C 53} 
(1996) 695

\bibitem[Pel93]{Pel93}
J. Peltoniemi, J. Valle, Nucl. Phys. {\bf B 406} (1993) 409

\bibitem[Pel95]{76}
J.T. Peltoniemi, preprint hep-ph/9506228


\bibitem[Pet94]{20}
S.T. Petcov, A.Yu. Smirnov, Phys. Lett. {\bf B 322} (1994) 109



\bibitem[Pet96]{12}
S. Petcov, in \cite{tren}


\bibitem[Pil93]{78}
A. Pilaftsis, Phys. Rev. {\bf D 49} (1993) 2398

\bibitem[Piq96]{98} F. Piquemal {\it et al.}, in \cite{tren}  


\bibitem[Pri95]{25}
J.R. Primack, J. Holtzman, A. Klypin, D.O. Caldwell, Phys. Rev. Lett. 74 (1995)
2160




\bibitem[Rad95]{37}
P.B. Radha {\it et. al}, preprint nucl-th/9510052 (Oct. 1995)


\bibitem[Raf96]{26}
G. Raffelt, J. Silk, Phys. Lett. {\bf B 366} (1996) 429

\bibitem[Rag94]{97} R. S. Raghavan, Phys. Rev. Lett. {\bf 72} (1994) 1411 



\bibitem[Rau94]{59}
F. Raupach (H1 Collab.), Proc. Int. Europhys. Conf. on High En. Phys.,
Marseille, 1993, Edition Frontieres, Gif--Sur--Yvette, France 1994. Eds J. Carr, M. Perrothet



\bibitem[Ren82]{55}
F.M. Renard, Phys. Lett. {\bf B 116} (1982) 264

\bibitem[Riz82]{riz82}
T.G. Rizzo, Phys. Lett. {\bf B 116} (1982) 23

\bibitem[Riz96]{Riz96}
T.G. Rizzo, hep/ph/9612440

\bibitem[Roy92]{114}
 D.P. Roy, Phys.Lett. {\bf B 283} (1992) 270

\bibitem[Rub96]{4}
C. Rubbia, Proc. TAUP 95, Toledo, Sept. 17--21 (1995),
Nucl. Phys. {\bf B} (Proc. Supl.) 48 (1996)172

\bibitem[Sch81]{Sch81}
J. Schechter, J.W.F. Valle, Phys. Rev. {\bf D 25} (1982) 2951


\bibitem[Sim96]{41}
F. Simkovic et al., Proc. TAUP 95, Toledo, 17--21 Sept. 1995; Nucl. Phys.
{\bf B} (Proc. Supplement) 48 (1996) 257


\bibitem[Sim97]{Sim97}
F. Simkovic {\it et al.}, Phys. Lett. {\bf B 393} (1997) 267

\bibitem[Smi96]{Smi96}
P.F. Smith {\it et al.}, Phys. Lett. {\bf B 379} (1996) 299

\bibitem[Smi96a]{Smi96a}
A. Yu. Smirnov, Proc. Int. Conf. on High Energy Physics, 
Warsaw 1996, hep-ph/9611465v2
(Dec 1996)

\bibitem[Sou92]{51}
I.A. D'Souza. C.S. Kalman, Preons, Models of Leptons , 
Quarks and Gauge bosons
as Composite Objects (World Scientific, Singapore) 1992



\bibitem[Sta90]{29}
A. Staudt, K.Muto, H.V. Klapdor--Kleingrothaus, 
Europhys. Lett. {\bf 13} (1990) 31



\bibitem[Ste91]{67}
J. Steinberger, Phys. Rep. {\bf 203} (1991) 345





\bibitem[Sto96]{38} 
S. Stoica, in \cite{tren}

\bibitem[Suh96]{39}
J. Suhonen, in \cite{tren}

\bibitem[Suz97]{Suz97}
A. Suzuki, priv. comm. 1997, and KAMLAND proposal, (in Japanese)


\bibitem[Tak96]{9}
E.Takasugi, in \cite{tren}  

\bibitem[Tak97]{Tak97}
E. Takasugi, in \cite{Kla98}

\bibitem[Tom87]{32}
T.Tomoda, A. Faessler, Phys. Lett. {\bf B199} (1987) 475


\bibitem[Tre95]{83}
V.I. Tretyak, Yu. Zdesenko, At. Data Nucl. Data Tables {\bf 61} (1995) 43


\bibitem[Val96]{13}
J.W.F. Valle, in \cite{tren}

\bibitem[Vog86]{35}
P.Vogel, M.R. Zirnbauer, Phys. Rev. Lett. {\bf 57} (1986)3148


\bibitem[Vog96]{36}
P.Vogel, in \cite{tren}

\bibitem[Vui93]{92} J.-C. Vuilleumier et al., Phys. Rev. {\bf D 48} (1993) 1009

\bibitem[Wol81]{Wol81} L. Wolfenstein, Phys. Lett {\bf 107 B} (1981) 77

\bibitem[You95]{87}
Ke You et al., Phys. Lett. {\bf B 265} (1995) 53 


\end{thebibliography}
\end{document}